\documentclass[10pt]{article}
\usepackage{hyperref}
\usepackage{geometry}
\usepackage{graphicx}
\usepackage{slashed}
\usepackage{cite}
\usepackage{amssymb}
\usepackage{amsmath}
\usepackage{color}
\usepackage{indentfirst}
\usepackage{multirow}
\allowdisplaybreaks[4]

\begin{document}

\title{\huge\bf Double pion photoproduction off nucleons in covariant chiral perturbation theory}

\author{Kai-Ge Kang\textsuperscript{1}\footnote{kaige@stu.pku.edu.cn}, Xiong-Hui Cao\textsuperscript{2}\footnote{xhcao@itp.ac.cn}, De-Liang Yao\textsuperscript{3,2}\footnote{yaodeliang@hnu.edu.cn (corresponding author)}  and Han-Qing Zheng\textsuperscript{4}\footnote{zhenghq@scu.edu.cn (corresponding author)}\\
\small \textsuperscript{1} School of Physics, Peking University, Beijing 100871, China\\
\small \textsuperscript{2} CAS Key Laboratory of Theoretical Physics, Institute of Theoretical Physics,\\
\small Chinese Academy of Sciences, Beijing 100190, China\\
\small \textsuperscript{3} School of Physics and Electronics, Hunan University, Changsha 410082, China\\
\small \textsuperscript{4} Institute of Particle and Nuclear Physics, 
Sichuan University, Chengdu, Sichuan 610065, China}
\maketitle

\begin{abstract}
 The double pion photoproduction off nucleons near threshold is analyzed in a covariant baryon chiral perturbation theory up to next to leading order, where the $\Delta(1232)$, $N^*(1400)$ and $\rho(770)$ resonances are included as explicit degrees of freedom. For the process $\gamma p \to \pi^+ \pi^0 n$, the chiral results of total cross sections, invariant-mass distributions and beam-helicity asymmetry are in good agreement with the experimental data within uncertainties. For the process $\gamma p \to \pi^0 \pi^0 p$, the prediction of total cross section deviates from the existing experimental data. Once the final-state interaction of $\pi \pi$ in the isoscalar S-wave channel is taken into account, a good description of the cross section is achieved. The effect of the Roper resonance always turns out be negligible, and hence can be thrown away in future study of this process. 
\end{abstract}

\section{Introduction}\label{sec:i}
Single and double pion photoproduction has been studied for decades, mainly aiming at solving the so-called missing resonance problem. That is, some of the resonances predicted by the quark model and lattice QCD were not found in $\pi N$ scattering~\cite{refprd341986,refepja102001,refprd842011}. An explanation is that these resonances are weakly coupled to $\pi N$, and need to be further explored through the photoproduction processes. The photoproduction processes also play an important role in deciphering the electromagnetic properties of resonances from experimental data. For instance, the single $\pi$ photo- and electroproduction off the nucleon near threshold are investigated in Refs~\cite{Ma:2020hpe,Cao:2021kvs} to extract the $\gamma^{(*)}N$ coupling to the sub-threshold resonance~\cite{Wang:2017agd,Wang:2018gul,Wang:2018nwi,Cao:2022zhn} in the $S_{11}$ channel.

High-energy nucleon intermediate excitations prefer sequential decays into many meson states, where the double pion production is considered to dominate~\cite{refppnp512003,refarxiv2207}. Plenty of experimental results for double pion photoproduction on the nucleon have been accumulated so far, including total cross sections, invariant-mass distributions, angle differential cross sections and the asymmetry related to the polarization of photons and initial nucleons~\cite{refplb3631995,refplb5782004,refplb6242005,refepja342007,refprl1032009,refepja512015,refplb5512003,refplb7882019,refplb8472023,refprl1252020,refepja482012,ref2207.14079,refprl872001}. However, data near the threshold are scarce. It is therefore challenging to decode resonant information in the low-energy region.

Various effective models have been proposed in previous theoretical works in order to analyze the contributions of different resonances. The Bonn-Gatchina model~\cite{refplb6592008} performed partial wave analysis to fit experimental cross sections of $\gamma p \to \pi^0 \pi^0 p$, finding that the $D_{33}(1700)$ wave makes the most sizeable contribution and $D_{13}(1520)$ accounts for a large part of the first bump. The Valencia model~\cite{refnpa6952001} employed phenomenological Lagrangian approach, where the Born term and the resonance contributions of $P_{33}(1232)$, $P_{11}(1440)$, $D_{13}(1520)$, $D_{33}(1700)$ and $\rho(770)$ are considered at tree level. A good description is achieved for cross sections of $\gamma p \to \pi^0 \pi^0 p, \gamma p \to \pi^+ \pi^- p, \gamma p \to \pi^+ \pi^0 n$. It is also found that the $D_{13}(1520)$ state is the dominant contribution amongst the resonant terms and a strong $P_{11}(1440)$ contribution was ruled out therein. The MAID model~\cite{refepja252005} is very similar to the Valencia model, but more higher lying states, such as $F_{15}(1680), D_{15}(1570)$, are also taken into account. Although it yields a satisfactory description for total cross sections of $\gamma p \to \pi^+ \pi^- p, \gamma p \to \pi^+ \pi^0 n$, the $\pi^0 \pi^0$ production on the proton near the threshold is underestimated. In the MAID model, the $D_{13}(1520)$ excitation turns out to be dominant in resonant region and at higher energies $F_{15}(1680)$ plays an important role. The influence owing to the rescattering mechanism of $ \pi^+ \pi^- \to \pi^0 \pi^0$ in the neutral channel has been investigated in Ref.~\cite{refrpj602017}.

Besides, hitting longitudinally polarized nuclei by circularly polarized photons, the beam-target helicity asymmetry $\sigma_{3 /2}-\sigma_{1/2}$ can be obtained. It is the difference between the total cross sections that the spins of photons and nucleons are oriented parallel and anti-parallel. The beam-target helicity asymmetry can be related to the anomalous magnetic moment of the nucleon through the Gerasimov-Drell-Hearn (GDH) sum rule~\cite{refprl161966}. It is pointed out that the double pion photoproduction process has a non-negligible contribution to the overall GDH sum rule~\cite{ refplb5512003}.

Double pion photoproduction amplitudes at threshold have been calculated within chiral perturbation theory (ChPT) in the 1990s~\cite {refprl731994,refnpa5801994}. Among them, Ref.~\cite {refprl731994} employed covariant formalism and calculated the leading order (LO) amplitude with explicit inclusion of the $\Delta(1232)$ resonance. Therein, the neutral channels $\gamma p \to \pi^0 \pi^0 p,\gamma n \to \pi^0 \pi^0 n$ are zero in the chiral limit, while the $\Delta(1232)$ does not contribute to $\gamma n \to \pi^0 \pi^0 n$ channel and its contribution to other channels is not as pronounced as expected. On the other hand, calculation was done in heavy baryon (HB) formalism up to $\mathcal{O}(p^3)$ in Ref.~\cite{refnpa5801994}. The amplitudes are derived exactly at threshold, where most of the Feynman diagrams do not contribute. It is found that, for the neutral channel, the loop diagrams play a major role and $\Delta(1232)$ has no contribution. For the channels with one charged pion, i.e., $\gamma p \to \pi^+ \pi^0 n,\gamma n \to \pi^- \pi^0 p$, the $\mathcal{O}(p^2)$ diagrams are of crucial importance, while the loop contribution is negligible. For the double-charged channel $\gamma p \to \pi^+ \pi^- p,\gamma n \to \pi^+ \pi^- n$, major contribution stems from the loop diagrams. The obtained HB results of cross sections of $\gamma p \to \pi^0 \pi^0 p$ are in good agreement with the experimental data~\cite{refplb5782004}, however, only up to the energy point $\sim 40~{\rm MeV}$ above the threshold. In addition, for the $\gamma p \to \pi^+ \pi^0 n$ channel, it is found that the HB ChPT predictions are much lower than existing data.

The main purpose of this work is to extend the calculation of $\gamma N\to \pi \pi N$ process beyond the threshold by imposing a covariant chiral perturbation theory scheme, in which the resonances $\Delta(1232),N^*(1440)$ and $\rho(770)$ are considered as explicit degrees of freedom. The importance of various contributions from the Born terms and resonances is discussed. The cross sections for $\gamma p \to \pi^0 \pi^0 p$ and $\gamma p \to \pi^+ \pi^0 n$ channel are calculated, which are then confronted with the available data so as to test the validity of chiral perturbation theory. For the $\gamma p \to \pi^0 \pi^0 p$ channel, the effect of final-state interaction (FSI) of $\pi \pi$ with $IJ=00$ is taken into account. In consequence, a good description of the existing experimental data can be obtained, even though the contribution of chiral loop diagrams is absent.

The structure of this manuscript is as follows. Section~\ref{sec:tf} represents the theoretical formalism, comprising the required chiral effective Lagrangian, basic formulae for the amplitude structure as well as observables, and the obtained chiral amplitudes. Section~\ref{sec:cwd} shows the numerical results of total cross sections, the $\pi\pi$ invariant mass distribution and the beam-target helicity asymmetry. Comparison with the available experimental data is also discussed. We summarize in Section~\ref{sec:s}. Correspondence between the theories with and without the $\rho$ meson is discussed in Appendix~\ref{sec:rhotheory}. The influence of pion mass is discussed in Appendix~\ref{sec.mpis}. Explicit expressions of the $\mathcal{O}(p)$ and $\mathcal{O}(p^2)$ non-resonant chiral amplitudes are relegated to Appendix~\ref{sec.amp.non}.

\section{Theoretical formalism}\label{sec:tf}
\subsection{Covariant chiral effective Lagrangian}
ChPT is a low-energy effective theory of Quantum Chromodynamics (QCD) describing the interaction between Goldstone-bosons and matter~\cite{refweinberg,refgasserandleutwyler,refscherer}. Its calculations are organized as a perturbative expansion in terms of the Goldstone-boson masses and external momentum, denoted collectively by $p$ hereafter. The chiral order of a given diagram, with $L$ loops, $I_{\phi}$ internal pion lines, $I_{N}$ internal nucleon lines and $N^{(k)}$ vertices from $\mathcal{O}(p^k)$ Lagrangian, is assigned to be~\cite{refnpb3631991}
\begin{equation}
    D=4L-2I_{\phi}-I_N+\sum_{k}^{\infty}kN^{(k)}\ .
\end{equation}
The leading order chiral Lagrangian for purely pionic interaction reads~\cite{refap1581984}
\begin{equation}
    \mathcal{L}_{\pi \pi}^{(2)}=\frac{F^2}{4}{\rm Tr}\left[ \nabla_\mu U \left(\nabla^\mu U\right)^\dagger \right]+\frac{F^2}{4}{\rm Tr}\left[ \chi U^\dagger+U \chi^\dagger  \right],
\end{equation}
where the superscript represents the chiral order and $F$ is the meson decay constant in chiral limit. The pion field in this Lagrangian is parameterized in the SU(2) matrix
\begin{equation}
U={\rm exp} \left( i\frac{\phi}{F} \right),\ \phi=\vec{\phi}\cdot \vec{\tau}=\begin{pmatrix}
\pi_0&\sqrt{2}\pi^+\\\sqrt{2}\pi^-&-\pi_0 
      \end{pmatrix}\ ,
\end{equation}
with $\tau_i(i=1,2,3)$ being the Pauli matrices. The covariant derivative acting on the pion fields is defined as $\nabla_\mu U=\partial_\mu U- i r_\mu \cdot U+i U\cdot l_\mu$. The photon field $A_\mu$ enters through the external fields by setting $r_\mu=l_\mu=-e A_\mu Q$, where $e>0$ is the elementary charge and $Q=(\mathbf{1}+\tau_3 )/2$. In addition, $\chi=M^2 \mathbf{1}$ in the isospin symmetric limit, with $M$ the mass of pions. The required Lagrangian describing the interaction with the nucleon is~\cite{refap2832000}
\begin{equation}
\begin{aligned}
    \mathcal{L}_{\pi N}^{(1)}=&\overline{\Psi}\left( i \slashed{D}-m +\frac{g}{2}\gamma^\mu \gamma_5 u_\mu \right)\Psi\ ,\\
    \mathcal{L}_{\pi N}^{(2)}=&c_1 {\rm Tr}[\chi_+]\overline{\Psi}\Psi-\frac{c_2}{4m^2}{\rm Tr}[u_\mu u_\nu]\left( \overline{\Psi}D^\mu D^\nu \Psi+\mathrm{h.c.} \right) +\frac{c_3}{2}{\rm Tr}[u^\mu u_\mu]\overline{\Psi}\Psi\\
   &-\frac{c_4}{4}\overline{\Psi}\gamma^\mu \gamma^\nu[u_\mu,u_\nu]\Psi+c_5 \overline{\Psi}\left[ \chi_+ -\frac{1}{2}{\rm Tr}[\chi_+] \right]\Psi+\overline{\Psi}\sigma^{\mu \nu}\left[ \frac{c_6}{8m}f^+_{\mu \nu}+\frac{c_7}{8m}{\rm Tr}\left[f^+_{\mu \nu}\right] \right]\Psi\ .
\end{aligned}
\end{equation}
The nucleon is contained in the isospin doublet: $\Psi=\begin{pmatrix}
    p\\n 
   \end{pmatrix}$. 
Here $m$ is the nucleon mass, $g$ is the axial coupling constant, and $c_i$'s are $\mathcal{O}(p^2)$ low-energy constants (LECs). The covariant derivative with respect to the nucleon field is given by:
\begin{equation}
    \begin{aligned}
        &D_\mu =\partial_\mu +\Gamma_\mu\ ,\\
        &\Gamma_\mu=\frac{1}{2}\left[ u^\dagger \left( \partial_\mu -i r_\mu \right)u+u\left( \partial_\mu -i l_\mu \right)u^\dagger \right]\ ,\\
        &u=\sqrt{U}={\rm exp}\left(\frac{i\phi}{2F}\right)\ .
    \end{aligned}
\end{equation}
Furthermore, the chiral
vielbein is defined as $u_\mu=i \left[ u^\dagger \left( \partial_\mu -i r_\mu \right)u-u\left( \partial_\mu -i l_\mu \right)u^\dagger \right]$ and the chiral operator $\chi_{+}$ reads $\chi_{+}=u^\dagger \chi u^\dagger + u \chi^\dagger u$. For convenience, we also need the electromagnetic tensor $f^{+}_{\mu \nu}=uf_{L\mu \nu}u^\dagger + u^\dagger f_{R\mu \nu}u$ with $f_{L\mu \nu}=f_{R\mu \nu }=-eQ(\partial_\mu A_\nu-\partial_\nu A_\mu)$.
Due to the closeness of the $\Delta(1232)$ to the nucleon, including it in ChPT will lead to a faster convergence property and generate more accurate predictions in the description of the $\gamma N$ and $\pi N$ processes~\cite{refap3362013,refprd1002019}. Hence, we ought to take the $\Delta(1232)$ excitations as explicit degrees freedom. The spin of $\Delta(1232)$ is $3/2$, which can be described using the vector-spinor field of the Rarita-Schwinger formalism~\cite{refpr601941}. Its isospin is also $3/ 2$, and can be described in the form of isovector-isospinor. With explicit $\Delta(1232)$ in ChPT, the Lagrangian for its interaction with pions can be constructed as~\cite{refjpg241998}
\begin{equation}
\begin{aligned}
    \mathcal{L}_{\pi \Delta}^{(1)}=&-\overline{\Psi}_\mu^i \xi^{\frac{3}{2}}_{ij}\left[ (i\slashed{D}_{jk}-m_\Delta \delta_{jk})g_{\mu \nu}+iA(\gamma^\mu D^\nu_{jk}+\gamma^\nu D^\mu_{jk}) \right. \\
    &+\frac{i}{2}(3A^2+2A+1)\gamma^\mu \slashed{D}_{jk}\gamma^\nu+m_\Delta \gamma^\mu\delta_{jk}\gamma^\nu \\
    &\left.+\frac{g_1}{2}\slashed{u}_{jk}\gamma_5 g^{\mu \nu}+\frac{g_2}{2}(\gamma^\mu u^\nu_{jk}+\gamma^\nu u^\mu_{jk})\gamma_5+\frac{g_3}{2}\gamma^\mu \slashed{u}_{jk}\gamma_5 \gamma^\nu  \right]\xi^{\frac{3}{2}}_{kl}\Psi_{\nu,l}\ ,
\end{aligned}
\end{equation}
where $m_\Delta$ and $g_1, g_2, g_3$ are the mass and coupling constants of the $\Delta(1232)$,  respectively. Here, $A\neq -\frac{1}{2}$ is an arbitrary real parameter and we take $A=-1$ in the calculation throughout this work. $\Psi_\mu^i$ represents the field  corresponding to $\Delta(1232)$, where $\mu=0,1,2,3$ are the Lorentz vector indices and  $i=1,2,3$ are the isovector indices. Furthermore, $\xi_{ij}^ {\frac{3}{2}}=\delta_{ij}-\frac{1}{3}\tau_i \tau_j$ is the matrix form of the projection operator of the isospin-$3/2$ component, e.g., $u_{\mu,ij}=\xi^{\frac{3}{2}}_{ik}u_\mu \xi^{\frac{3}{2}}_{kj}$. The covariant derivative acting on $\Psi_\mu^i$ is defined by $D_{\mu,ij}=\delta_{ij}\partial_{\mu}-2i\epsilon_{ijk}\Gamma_{\mu,k}+\delta_{ij}\Gamma_\mu$. The leading order Lagrangian describing the $\pi N \Delta$ interaction is
\begin{equation}
    \mathcal{L}_{\pi N \Delta}^{(1)}=h \overline{\Psi}u_j^\mu \Theta_{\mu \alpha}(z_0) \xi^{\frac{3}{2}}_{ji} \Psi^{\alpha}_i+\mathrm{h.c.}
\end{equation}
where $h$ is the pion-nucleon-delta coupling constant. $\Theta_{\mu \alpha}(z_0)=g_{\mu \alpha}+z_0\gamma_\mu \gamma_\alpha$, here $z_0$ is an off-shell parameter. The parameters of $g_2,g_3,z_0$ can be absorbed by a redefinition of other LECs as done in Ref.~\cite{refplb6832010}, therefore we can simply set $g_2=g_3=z_0=0$ for convenience. 

In addition to $\Delta(1232)$, we also consider the lowest-order contribution of the $N^*(1440)$ resonance, i.e. the Roper resonance. The relevant Lagrangian reads~\cite{refplb7602016},
\begin{equation}\mathcal{L}^{(1)}_{\pi N R}=\frac{g_{\pi N R}}{2} \overline{\Psi}_{\rm R} \gamma^\mu \gamma_5 u_\mu \Psi+\mathrm{h.c.}
\end{equation}
As for the vector meson resonance $\rho(770)$, following Ref.~\cite{refprc862012}, we adopt the so-called massive Yang-Mills approach~\cite{refpr1611988}, which was suggested as a general one by demanding the self-consistency in the sense of constraints and perturbative renormalizability~\cite{refijmpa252010}.
\begin{equation}
\begin{aligned}
    &\mathcal{L}_{\pi \rho}=-\frac{1}{2}{\rm Tr}\left( \rho_{\mu \nu}\rho^{\mu \nu} \right)+M_{\rho}^2 {\rm Tr}\left[ \left( \rho_\mu-\frac{i}{g_\rho}\tilde{\Gamma}_\mu \right)\left( \rho^\mu-\frac{i}{g_\rho}\tilde{\Gamma}^\mu \right) \right]\\
    &\mathcal{L}_{\pi \rho N}= g_\rho \overline{\Psi} \left[ \rho_\mu-\frac{i}{g_\rho}\tilde{\Gamma}_\mu \right]\gamma^\mu \Psi+\frac{G_\rho}{2}\overline{\Psi}\rho_{\mu \nu}\sigma^{\mu \nu}\Psi  ,
\end{aligned}
\label{eq.rho}
\end{equation}
where $\tilde{\Gamma}_\mu=\Gamma_\mu-\frac{1}{2}{\rm Tr}(\Gamma_\mu),\ \rho_\mu=\rho_{i \mu}\frac{\tau_i}{2},\  \rho_{\mu \nu}=\partial_\mu \rho_\nu-\partial_\nu \rho_\mu -i g_\rho \left[ \rho_\mu,\rho_\nu \right],\ g_\rho^2=\frac{M_\rho^2}{2F^2}$~\cite{refprl932004}.\footnote{In fact, this equation is the so-called (Kawarabayashi-Suzuki-Riadzuddin-Fayyazuddin) KSRF relation~\cite{Kawarabayashi:1966kd, Riazuddin:1966sw}. The KSRF relation can be generalized by including the contributions of the crossed channel resonance exchange, as pointed out, e.g. in Ref.~\cite{Basdevant:1970hm}, and discussed in details in Ref.~\cite{Guo:2007ff}. }   We can get $\mathcal{O}(p^0)$ three-$\rho$ and four-$\rho$ interaction terms, and $\mathcal{O}(p^1)$ $\rho$-photon, $\rho$-$\pi$-$\pi$ and $\rho$-photon-$\pi$-$\pi$ interaction terms from $\mathcal{L}_{\pi \rho}$. $\mathcal{L}_{\pi \rho N}$ will give $\mathcal{O}(p^0)$ and $\mathcal{O}(p^1)$ interactions between $\rho$ and nucleons.

\subsection{Amplitude structure and observables}
The amplitude of the $\gamma(k) N(p_1) \to \pi^a(q_1) \pi^b(q_2) N(p_2)$ process can be decomposed into 6 parts according to the isospin structure:
\begin{equation}
\mathcal{M}_{ab}=\mathcal{M}^1\delta_{ab}+\mathcal{M}^2\delta_{ab}\tau_3+\mathcal{M}^3 \delta_{a3}\tau_b+\mathcal{M}^4 \delta_{b3}\tau_a+\mathcal{M}^5 \epsilon_{abc}\tau_c+\mathcal{M}^6 \epsilon_{3ab}\mathbf{1}\ .\label{eq:Mijiso}
\end{equation}
The amplitude should satisfy the Bose symmetry due to the presence of two $\pi$ in the final states. That is, by exchanging $a, b$ and $q_1, q_2$ simultaneously, the amplitude should remain unchanged. Consequently, the exchange of $q_1 \leftrightarrow q_2$ leads to the interchanging between the isospin amplitudes $\mathcal{M}^i$'s: \footnote{Note that, in the isospin limit, one furter has $q_1=q_2$ at threshold. In this case, the threshold amplitude is symmetric in $q_{1}$ and $q_{2}$, which leads to $\mathcal{M}^3=\mathcal{M}^4,\mathcal {M}^5=\mathcal{M}^6=0$.}
\begin{equation}
{\mathcal{M}}^1 \leftrightarrow \mathcal{M}^1, \;{\mathcal{M}}^2\leftrightarrow\mathcal{M}^2, \;{\mathcal{M}}^3\leftrightarrow\mathcal{M}^4, \;{\mathcal{M}}^4\leftrightarrow\mathcal{M}^3, \;{\mathcal{M}}^5\leftrightarrow-\mathcal{M}^5, \;{\mathcal{M}}^6\leftrightarrow-\mathcal{M}^6
\end{equation}

The isospin amplitudes can be further decomposed according to the Lorentz structure as
\begin{equation}
    \mathcal{M}^i= \epsilon_\mu(k) \overline{u}(p_2) \bigg[ \sum_{j=1}^{12} \mathcal{M}^i_{j} l_j^\mu \bigg]  u(p_1)\equiv\epsilon_\mu(k) \overline{u}(p_2)  \mathcal{M}^{i,\mu}  u(p_1)\ ,\label{eq.lorentz}
\end{equation}
where the Lorentz operators are defined by
\begin{equation}
    \begin{aligned}
        &l_1^\mu=(P\cdot k) \gamma^\mu-P^\mu \slashed{k}, \ \ l_2^\mu=\gamma^\mu \slashed{k}, \ \ l_3^\mu=(P\cdot k) \gamma^\mu \slashed{Q}_- -P^\mu \slashed{k} \slashed{Q}_-, \ \ l_4^\mu=\gamma^\mu \slashed{k} \slashed{Q}_-,\\
        &l_5^\mu=(P\cdot k) Q_+^\mu-P^\mu (Q_+\cdot k),\ \ l_6^\mu=Q_+^\mu \slashed{k}-(Q_+\cdot k) \gamma^\mu, \ \ l_7^\mu=(P\cdot k) Q_+^\mu \slashed{Q}_- -P^\mu (Q_+\cdot k) \slashed{Q}_-,\\
        &l_8^\mu=Q_+^\mu \slashed{k} \slashed{Q}_- -(Q_+\cdot k) \gamma^\mu \slashed{Q}_-,\ \ l_9^\mu=(P\cdot k) Q_-^\mu-P^\mu (Q_-\cdot k), \ \ l_{10}^\mu=Q_-^\mu \slashed{k}-(Q_-\cdot k) \gamma^\mu,\\
        &l_{11}^\mu=(P\cdot k) Q_-^\mu \slashed{Q}_--P^\mu (Q_-\cdot k) \slashed{Q}_-,\ \ l_{12}^\mu=Q_-^\mu \slashed{k}\slashed{Q}_- -(Q_-\cdot k) \gamma^\mu \slashed{Q}_-,
    \end{aligned}
\end{equation}
with $Q_+=(q_1+q_2)/2,\ Q_-=(q_1-q_2)/2,\ P=(p_1+p_2)/2$. It should be stressed that each Lorentz structure above obeys gauge invariance separately. In Eq.~\eqref{eq.lorentz},  $\mathcal{M}^i_{j}$ are scalar functions of Mandelstam variables. For a process with 2 initial and 3 final states, there are 5 independent Mandelstam variables in total. In this work, they are chosen to be:
\begin{equation}
\begin{aligned}
    s=(k +p_1 )^2\ ,\;
    s_1 =&(p_2 +q_1 )^2 \ ,\;\ s_2 =(p_2 +q_2 )^2\ ,\;
    t_1 =(k -q_1 )^2 \ ,\; t_2 =(k -q_2 )^2\ .
\end{aligned}
\end{equation} 
At threshold, the 3-momenta of final states in the center-of-mass (CM) frame are zero, so $s^{{\rm thr}}=(m+2M)^2$ with the superscript denoting the threshold for short. In the laboratory reference frame, the energy of the incident photon is $E_\gamma=({s-m^2})/({2m})$. Taking $m=938.27~{\rm MeV}, M=139.57~{\rm MeV}$, we can get for the threshold $\sqrt{s}^{{\rm thr}}=1217.5~{\rm MeV}$ and $E_\gamma^{{\rm thr}}=320.7~{\rm MeV}$.
\begin{figure}[h!]
    \centering
   \includegraphics[scale=0.4]{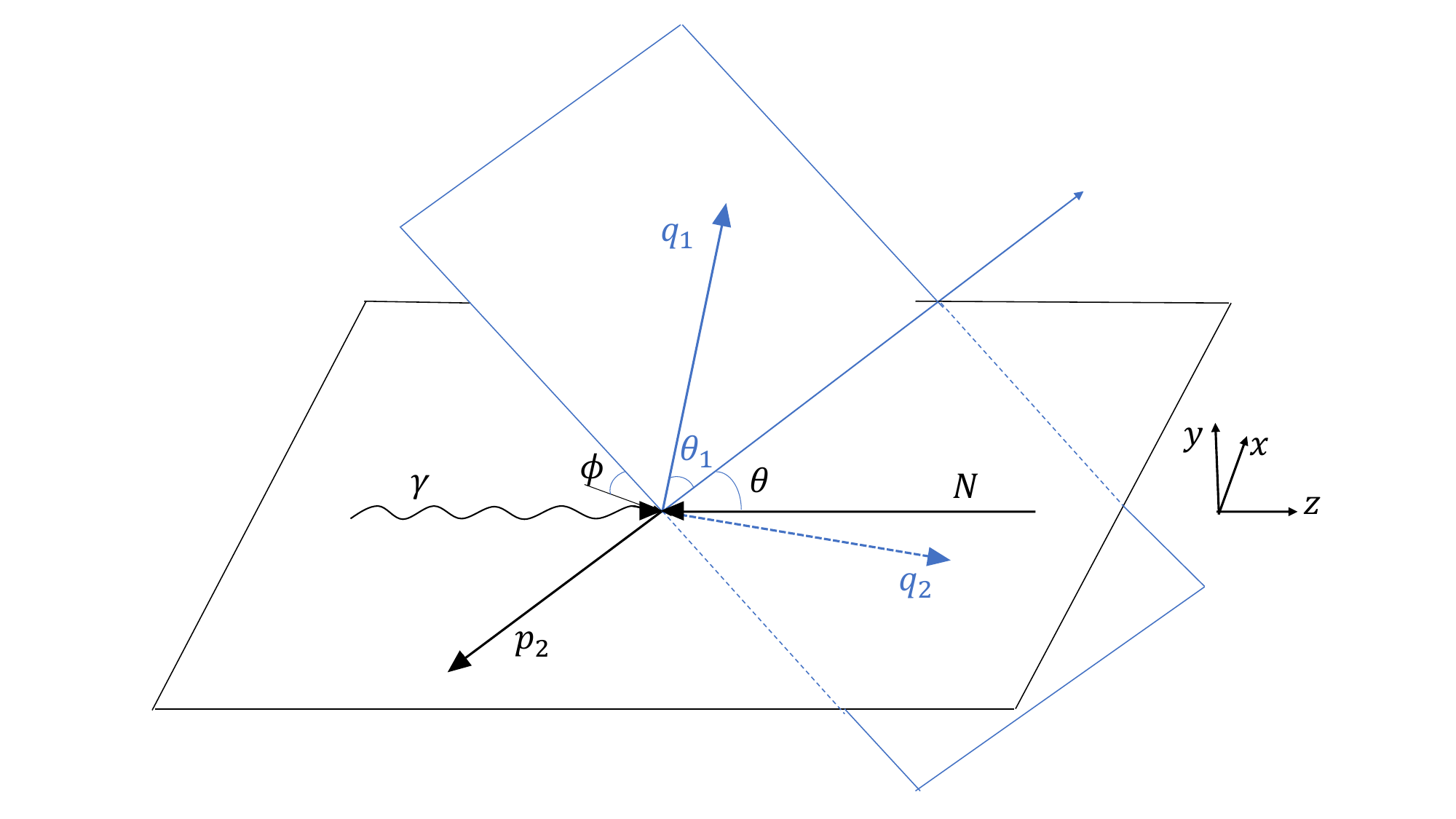}
    \caption{Kinematics in the CM frame of the intial $\gamma N$ system. The $z$-axis is oriented in the direction of the three-momentum of the incoming photon.}
    \label{tuydx}
\end{figure}

We denote the amplitude for the physical process ${\rm P}_i$ by $\mathcal{N}({\rm P}_i)$, which can be expressed in terms of the above isospin amplitudes $\mathcal{M}^i$ via
\begin{equation}
\begin{array}{cll }
{\rm P}_1 & \gamma p \to \pi^+ \pi^0 n: & \mathcal{N}({\rm P}_1)=\sqrt{2}\mathcal{M}^4-i\sqrt{2}\mathcal{M}^5 \ , \\
{\rm P}_2 &  \gamma n \to \pi^- \pi^0 p:& \mathcal{N}({\rm P}_2)=\sqrt{2}\mathcal{M}^4+i\sqrt{2}\mathcal{M}^5 \ ,\\
{\rm P}_3 & \gamma p \to \pi^0 \pi^0 p:&\mathcal{N}({\rm P}_3)=\mathcal{M}^1+\mathcal{M}^2+\mathcal{M}^3+\mathcal{M}^4\ ,\\
{\rm P}_4 & \gamma n \to \pi^0 \pi^0 n:&\mathcal{N}({\rm P}_4)=\mathcal{M}^1-\mathcal{M}^2-\mathcal{M}^3-\mathcal{M}^4\ ,\\
{\rm P}_5 & \gamma p \to \pi^+ \pi^- p: & \mathcal{N}({\rm P}_5)=\mathcal{M}^1+\mathcal{M}^2+i\mathcal{M}^5+i\mathcal{M}^6 \ ,\\
{\rm P}_6 & \gamma n \to \pi^+ \pi^- n: & \mathcal{N}({\rm P}_6)=\mathcal{M}^1-\mathcal{M}^2-i\mathcal{M}^5+i\mathcal{M}^6 \ .
\end{array}
\label{eq.physical}
\end{equation}
It can be checked that at threshold one has 
\begin{align}
 \mathcal{N}({\rm P}_3)-\mathcal{N}({\rm P}_5)=\mathcal{N}({\rm P}_6)-\mathcal{N}({\rm P}_4)=\sqrt{2}\mathcal{N}({\rm P}_1)=\sqrt{2}\mathcal{N}({\rm P}_2)\ .   
\end{align}

The modular square of the amplitude for the physical process, that is averaged for the initial state polarizations and summed for the final state polarizations, is denoted by $\left| \overline{\mathcal{N}} \right|^{2}$. Then, the unpolarized cross section can be obtained by performing the three-body phase space integration over $\Phi_3$,
\begin{equation}
    \sigma= \frac{S}{4 p_1 \cdot k}\int \left| \overline{\mathcal{N}} \right|^{2} {\rm d}\Phi_3\ .
\label{gssihi}
\end{equation}
Here $S$ is a symmetry factor for the identical particles in the final states, which is $1/2$ for the neutral channels and $1$ for the charged channels. The kinematics of double-pion photoproduction on the nucleon in the CM frame is shown in Fig.~\ref{tuydx}. The polar and azimuth angles of $\vec{p_2}$ are denoted by $(\pi-\theta)$ and $\varphi$, respectively. The dihedral angle $\phi$ is spanned by the ($\vec{k},\vec{p_2}$) plane and the ($\vec{q_1},\vec{q_2}$) plane. The phase space integration can be written as a multidimensional integral over the angles $\theta,\varphi,\phi$ and the Dalitz variables $s_1,s_{23}$ via
\begin{equation}
    {\rm d}\Phi_3 = \frac{{\rm d}s_{1}{\rm d}s_{23}}{(2\pi)^5 \cdot 32s}{\rm d} \cos \theta\, {\rm d} \varphi\, {\rm d} \phi\ .\label{eq.phi3}
\end{equation}
Note that $s_{23}$ is related to the Mandelstam variables through $s_{23}=m^2+2M^2+s-s_1-s_2$. The integration range for $\theta$ is $[0, \pi]$, while for $\varphi$ and $\phi$ one has $(0,2\pi]$ . The upper and lower limits of the Dalitz variables $s_1,s_{23}$ are given by
\begin{equation}
\begin{aligned}
    (s_{23})_{{\rm max}}&=(\sqrt{s}-m)^2,\quad (s_{23})_{{\rm min}}=4M^2\ ,\\
    (s_{1})_{{\rm max}}&=(\hat{E}_{p2}+\hat{E}_{q1})^2-\left( \sqrt{(\hat{E}_{p2})^2-m^2}-\sqrt{(\hat{E}_{q1})^2-M^2} \right)^2\ ,\\
    (s_{1})_{{\rm min}}&=(\hat{E}_{p2}+\hat{E}_{q1})^2-\left( \sqrt{(\hat{E}_{p2})^2-m^2}+\sqrt{(\hat{E}_{q1})^2-M^2} \right)^2\ ,
\end{aligned}
\label{gssmqm}
\end{equation}
where $\hat{E}_{p2}$ and $\hat{E}_{q1}$ are the energies of the nucleon and the pion (with momentum $q_1$) in the two-pion CM frame, respectively:
\begin{equation}
   \hat{E}_{p2}=\frac{s-m^2-s_{23}}{2\sqrt{s_{23}}},\quad \hat{E}_{q1}=\frac{\sqrt{s_{23}}\ ,}{2}
\end{equation}
with $\sqrt{s_{23}}$ being the invariant mass of the pion pair. By substituting Eq.~\eqref{eq.phi3} into Eq.~\eqref{gssihi} and performing the integration over all the involved variables except $s_{23}$, one can derive the differential cross section with respect to $\sqrt{s_{23}}$ :
\begin{equation}
    \frac{{\rm d}\sigma}{{\rm d} \sqrt{s_{23}}}=\frac{S}{2(s-m_N^2)}\int \left| \overline{\mathcal{N}} \right|^{2} 2 \sqrt{s_{23}} \frac{{\rm d}s_1}{(2\pi)^5 \cdot 32s}{\rm d} \cos \theta {\rm d} \varphi\, {\rm d} \phi
    \label{gsdddd}
\end{equation}
The spinor for a nucleon polarized along the momentum $\vec{p}=\left| \vec{p} \right|(\sin \theta \cos \varphi,\sin \theta \sin \varphi,\cos \theta)$ reads
\begin{equation}
    u_\lambda (p)=\begin{pmatrix} \sqrt{p\cdot \sigma} \xi_\lambda \\ \sqrt{p\cdot \overline{\sigma}}\xi_\lambda  \end{pmatrix}.
    \label{eq.spinor}
\end{equation}
where $\sigma^\mu=(\mathbf{1},\vec{\tau}),\overline{\sigma}^\mu=(\mathbf{1},-\vec{\tau})$. $\xi_+=\begin{pmatrix} \cos \frac{\theta}{2}\\ e^{i \varphi}\sin \frac{\theta}{2}  \end{pmatrix}$ and $\xi_-=\begin{pmatrix}  -e^{-i \varphi}\sin \frac{\theta}{2}\\ \cos \frac{\theta}{2}  \end{pmatrix}$ represent the helicity $+\frac{1}{2}$ and $-\frac{1}{2}$ respectively.
On the other hand, the polarization vector for a circularly polarized photon with helicity $+,-$ are
\begin{align}
    &\epsilon^\mu_+(k,n)=(0,\frac{1}{\sqrt{2}},\frac{i }{\sqrt{2}},0)\ ,\\
    &\epsilon^\mu_-(k,n)=(0,\frac{1}{\sqrt{2}},-\frac{i}{\sqrt{ 2}},0)\ ,
\end{align}
respectively. Consequently, the helicity amplitude for the physical process ${\rm P}^i$ of $\gamma N \rightarrow \pi \pi N$ can be defined through
\begin{align}
\mathcal{N}_{\lambda_1 \lambda_2 \lambda_3}=\epsilon^\mu_{\lambda_1}(k) \overline{u}_{\lambda_3}(p_2)\mathcal{N}_\mu u_{\lambda_2}(p_1) 
\label{eq.helicity.amp}
\end{align}
where $\mathcal{N}_\mu$ can be expressed in terms of $\mathcal{M}_\mu^i$ with the help of Eq.~\eqref{eq.physical} and Eq.~\eqref{eq.lorentz}. Ignoring the influence of resonance width, some relations among $\mathcal{N}_{\lambda_1 \lambda_2 \lambda_3}^i$'s can be found
\begin{equation}
    \begin{aligned}
        \mathcal{N}_{+++}=\left( \mathcal{N}_{---} \right)^\ast\ , &\quad \mathcal{N}_{+--}=\left( \mathcal{N}_{-++} \right)^\ast\ ,\\
        \mathcal{N}_{++-}=-\left( \mathcal{N}_{--+} \right)^\ast\ , &\quad \mathcal{N}_{-+-}=-\left( \mathcal{N}_{+-+} \right)^\ast\ .
    \end{aligned}
\end{equation}
And at threshold, we have 
\begin{align}
 &\mathcal{N}_{+++}=\mathcal{N}_{---}\ , \notag\\
 &\mathcal{N}_{+--}=\mathcal{N}_{++-}^i=\mathcal{N}_{-+-}=0\ .
\end{align} 
Eventually, the total cross section for the scattering of a circularly polarized photon by a proton polarized with its spin parallel ($\sigma_{3/2}$) and anti-parallel ($\sigma_{1/2}$) to the photon spin are
\begin{equation}
    \begin{aligned}
        \sigma_{\frac{3}{2}}\equiv& \frac{S}{4\sqrt{(p_1 \cdot k)^2-m_{p_1}^2 \cdot m_{k}^2}}\int \sum_{\lambda_3} \left| \mathcal{N}_{+-\lambda_3} \right|^{2} {\rm d}\Phi_3 \ ,\\
        \sigma_{\frac{1}{2}}\equiv& \frac{S}{4\sqrt{(p_1 \cdot k)^2-m_{p_1}^2 \cdot m_{k}^2}}\int \sum_{\lambda_3} \left| \mathcal{N}_{++\lambda_3} \right|^{2} {\rm d}\Phi_3\ ,
    \end{aligned}
    \label{gsspdm}
\end{equation}
respectively. The beam-target helicity asymmetry of the total cross section is defined to be 
\begin{align}
\Delta \sigma=\sigma_{3/2}-\sigma_{1/2}\ .
\end{align}

\subsection{Chiral amplitudes}\label{sec:rat}
\begin{figure}[htbp]
    \centering
   \includegraphics[scale=0.5]{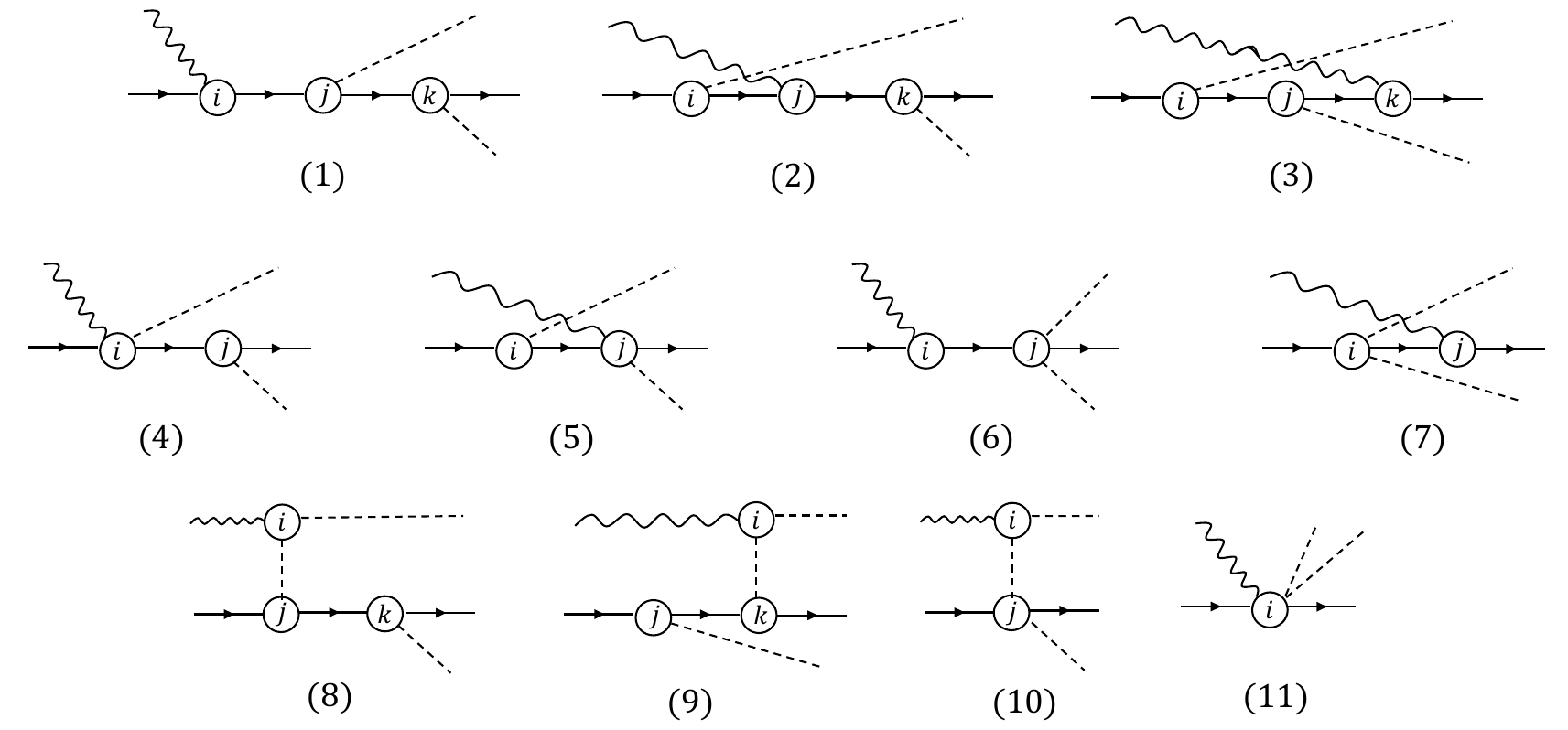}
    \caption{Topologies of tree-level diagrams of $\gamma N \to \pi \pi N$ without resonances. The wavy, dashed, and solid lines represent photons, pions and nucleons, respectively. Circled numbers mark the chiral orders of the vertices. Crossed diagrams are not shown.}
    \label{tutree}
\end{figure}
\begin{figure}[htbp]
    \centering
   \includegraphics[scale=0.5]{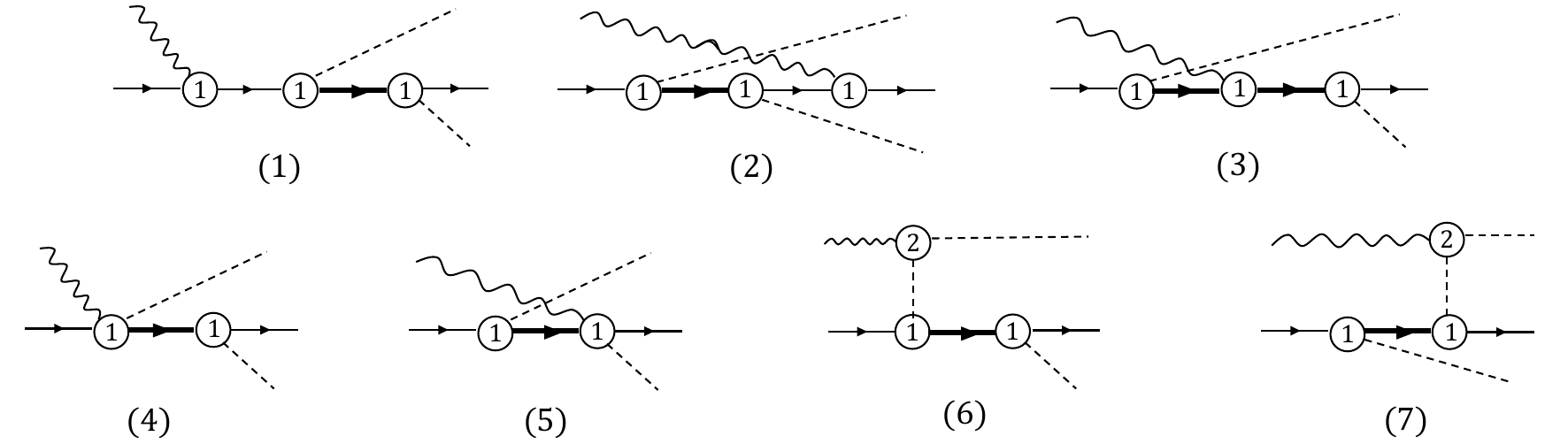}
    \caption{Tree diagrams of $\gamma N \to \pi \pi N$ with $\Delta(1232)$. The wavy, dashed, solid and thick lines represent photons, pions, nucleons and deltas, in order. Circled numbers mark the chiral orders of the vertices. Crossed diagrams are not shown.}
    \label{tuop1delta}
\end{figure}
\begin{figure}[htbp]
    \centering
   \includegraphics[scale=0.5]{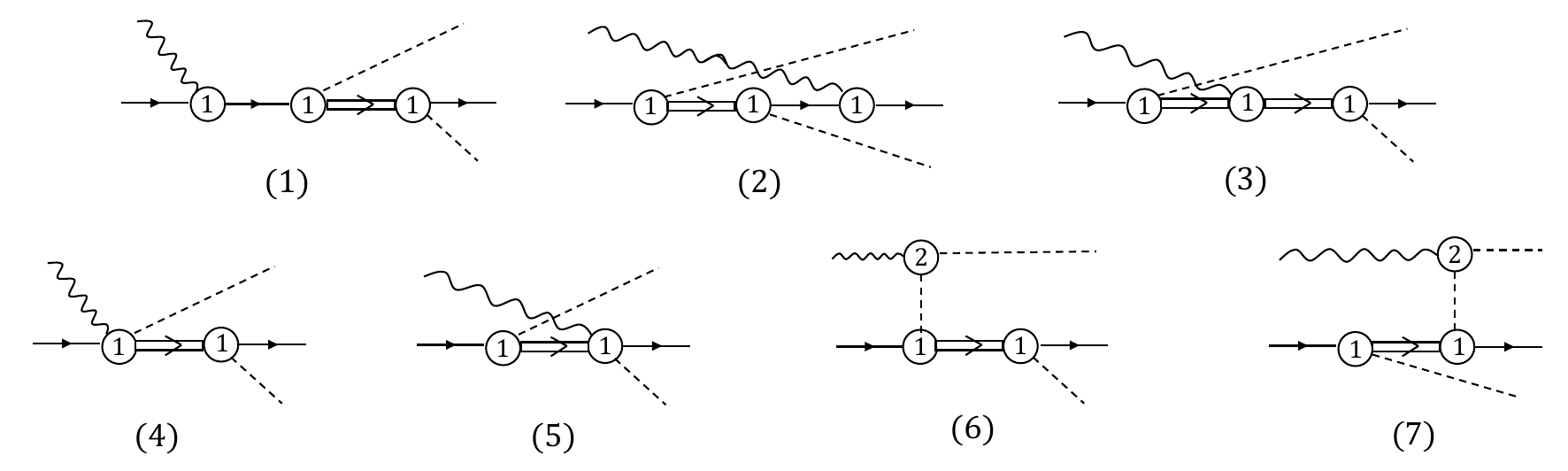}
    \caption{Tree diagrams of $\gamma N \to \pi \pi N$ with Roper resonances. The wavy, dashed, solid and double lines represent photons, pions, nucleons and Ropers, in order. Circled numbers mark the chiral orders of the vertices. Crossed diagrams are not shown.}
    \label{tuop1roper}
\end{figure}
\begin{figure}[htbp]
    \centering
   \includegraphics[scale=0.5]{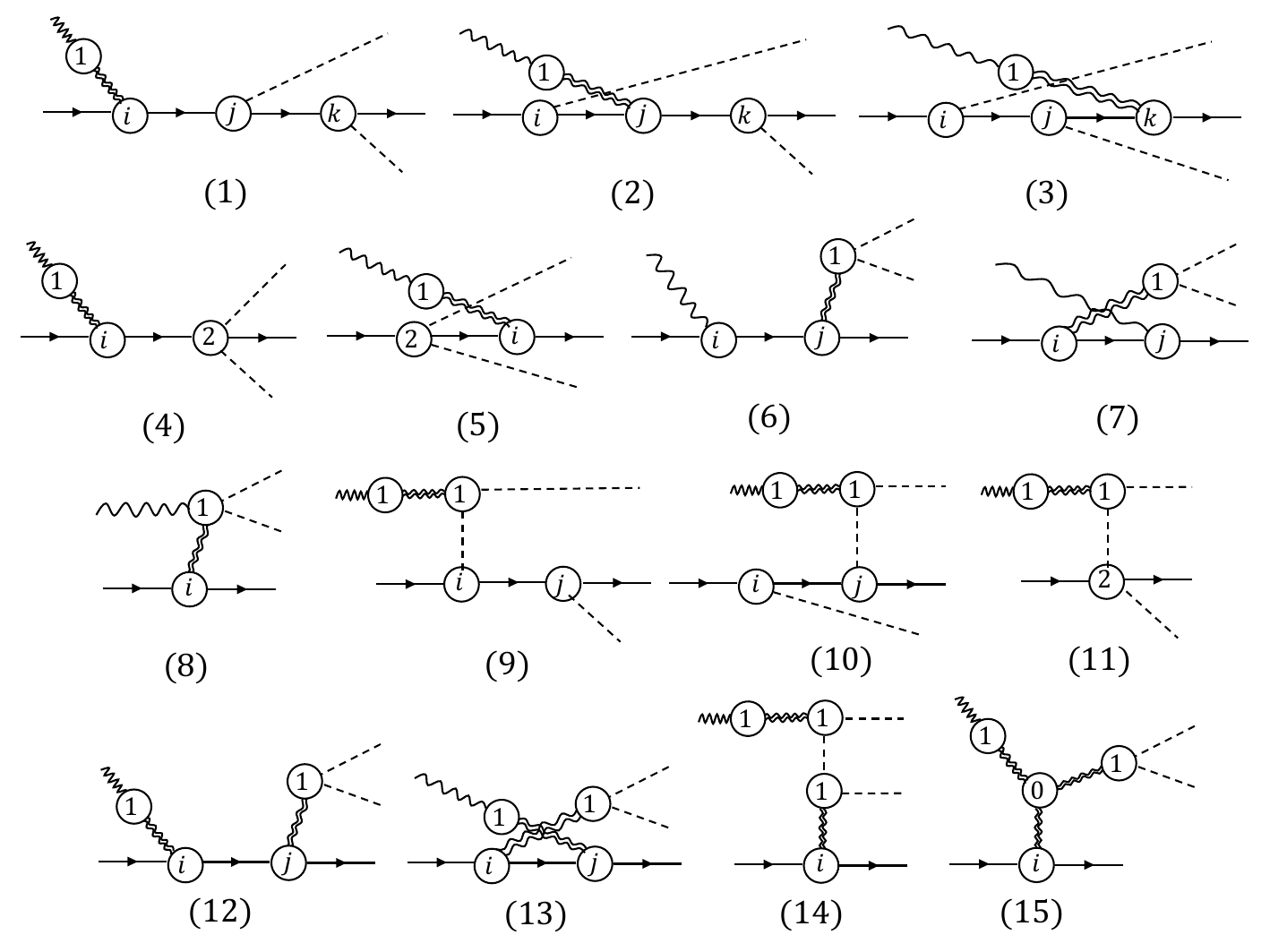}
    \caption{Tree diagrams of $\gamma N \to \pi \pi N$ with $\rho$ meson. The wavy, dashed, solid and double wavy lines represent photons, pions, nucleons and $\rho$ mesons, in order. Circled numbers mark the chiral order of the vertices. Crossed diagrams are not shown.}
    \label{turho}
\end{figure}
\begin{figure}[htbp]
    \centering
   \includegraphics[scale=0.5]{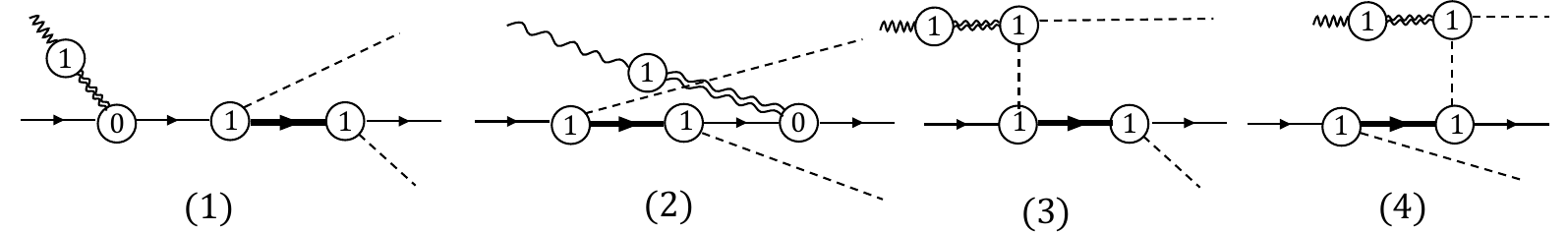}
    \caption{Tree diagrams of $\gamma N \to \pi \pi N$ with both $\rho$ meson and baryon resonance. The wavy, dashed, solid, double wavy and thick lines represent photons, pions, nucleons, $\rho$ mesons and baryon resonance ($\Delta(1232)$ or Roper) in order. Circled numbers mark the chiral orders of the vertices. Crossed diagrams are not shown.}
    \label{turhodelta}
\end{figure}

Topologies of tree-level diagrams for the double pion photoproduction without resonances are displayed in Fig.~\ref{tutree}, where $i,j,k$ refer to the chiral orders of the vertices. For the calculation up to $\mathcal{O}(p^2)$, it is worthy noting that the vertex for the interaction between photon and mesons must be $\mathcal{O}(p^2)$. In consequence, one can get all the $\mathcal{O}(p)$ diagrams by uniformly setting the chiral orders of the vertices involving nucleons to be 1. Similarly, all the $\mathcal{O}(p^2)$ diagrams  can be obtained by merely setting one of the vertices containing nucleons to $\mathcal{O}(p^2)$, while the remaining ones to $\mathcal{O}(p)$. Expressions of the $\mathcal{O}(p)$ and $\mathcal{O}(p^2)$ non-resonant chiral amplitudes are shown in Appendix~\ref{sec.amp.non}.

The tree diagrams with explicit $\Delta(1232),N^*(1440),\rho(770)$ resonances are shown in Figs.~\ref{tuop1delta},~\ref{tuop1roper},~\ref{turho} and \ref{turhodelta}, in order. Note that, if the $\rho$ mesons are explicitly included via the Yang-Mills approach, those diagrams of Fig~\ref{tutree}, containing the LO photon-$\pi$-$\pi$, LO $\pi$-$\pi$-N-N, LO photon-$\pi$-$\pi$-N-N vertices and $e\overline{\Psi}\slashed{A}\frac{\tau_3}{2}\Psi$, do not show up. As illustrated in Appendix~\ref{sec:rhotheory}, their contributions are actually incorporated by the exchange of $\rho$ meson. The expressions of the amplitudes with explicit resonant contributions are too lengthy to be shown here.\footnote{The expressions of the amplitudes are obtainable from the authors.}

\section{Numerical results and discussions}\label{sec:cwd}

At present, there are few experimental data for double pion photoproduction near threshold with the photon energy $E_\gamma\leq 450~{\rm MeV}$. 
The available data we can use are total cross sections of $\gamma p \to \pi^0 \pi^0 p$. For $\gamma p \to \pi^+ \pi^0 n$,  one has data on total cross sections, pion-pair invariant-mass distributions and beam-target helicity asymmetry. In this section, comparisons between our model and the above-mentioned experimental results will be discussed. Predictions for the other channels will be shown as well.

\subsection{Parameters}
In our numerical computation, the parameters $m$, $g$, $F$ and $M$, stemming from the chiral Lagrangians, can be replaced by the corresponding physical ones $m_N$, $g_A$, $F_\pi$ and $M_\pi$, respectively. The error caused by such an replacement is of higher order beyond the accuracy we are considering. In the SU(2) isospin limit, the values of the physical quantities are taken to be
\begin{align}
 m_N=938.27~{\rm MeV}\ ,
 \quad
 F_\pi=92.42~{\rm MeV}\ ,\quad
 g_A=1.267\ ,\quad
 e=0.303\ .
\end{align}
For processes ${\rm P}_1$ and ${\rm P}_2$, the pion mass is set as the average of the $\pi^+$ and $\pi^0$ masses, i.e. $M_\pi = 137.3~{\rm MeV}$. For processes ${\rm P}_5$ and ${\rm P}_6$, $M_\pi=139.57~{\rm MeV}$, and for the neutral ones $M_\pi=135~{\rm MeV}$. For the $O(p^2)$ $\pi N$ LECs, we choose the determinations given in Refs.~\cite{refjhep052016}: 
\begin{align}
& c_1 = -0.99\pm 0.02~{\rm GeV}^{-1},\quad c_2 = 1.38\pm 0.03~{\rm GeV}^{-1},\notag\\ 
&c_3 = -2.33\pm 0.03~{\rm GeV}^{-1},\quad c_4=1.71\pm 0.02 ~{\rm GeV}^{-1}  \ .
\end{align}
The LECs $c_6$ and $c_7$ can be determined by the anomalous magnetic moment of the nucleon $\kappa_p = 1.793, \kappa_n = -1.913$ with the help of the following relations~\cite{refprc862012,refpdg}
\begin{align}
c_6=\kappa_p-\kappa_n+2m_N \frac{G_\rho}{g_\rho}, \quad c_7=\kappa_n-m_N \frac{G_\rho}{g_\rho}\ .
\end{align}
For the $\Delta(1232)$, we use $m_\Delta=1210~{\rm MeV}$ and $h=1.28\pm 0.01$~\cite{refjhep052016}. For the Roper, we take $m_{\rm R}=1396~{\rm MeV}$, $g_{\pi N R}=\pm 0.47$, following Ref.~\cite{refplb7602016}. As for the $\rho$ parameters, $M_\rho=775~{\rm MeV}$, $g_\rho=\frac{M_\rho}{\sqrt{2}F_\pi}=5.93$, and $G_\rho=-16.9~{\rm GeV^{-1}}$ are used~\cite{refprc862012}.

To compare the contributions of different types, we compute the moduli of the helicity amplitude $ \left| \mathcal{N}^i_{+++}\right|$, defined in Eq.~\eqref{eq.helicity.amp}, at threshold. Results are shown in Table~\ref{btreeyu}. The resonant width effect have been incorporated by replacing the propagator in the following way, 
\begin{equation}
    \frac{1}{s-m_\xi^2}\to \frac{1}{s-m_\xi^2+i m_\xi \Gamma_\xi}
\end{equation}
where $\xi$ represents $\Delta(1232)$, $N^*(1440)$ or $\rho(770)$. The width of Roper is taken as constant $\Gamma_{R}=175$~MeV, but for $\Delta$ and $\rho$ we use the energy-dependent results of Ref.~\cite{refplb7632016,refplb4121997}.
\begin{equation}
\begin{aligned}
   \Gamma_\Delta(s)=&\frac{h^2 \lambda^\frac{3}{2}(s,M^2,m^2)}{192\pi F^2 s^3}\left[ (s-M^2+m^2)m_\Delta+2sm \right]\ ,
   \\
   \Gamma_\rho(s)=&\frac{M_\rho s}{96\pi F^2}\left( 1-\frac{4M^2}{s} \right)^{\frac{3}{2}}
   \ ,
\end{aligned}
\end{equation}
where $\lambda(a,b,c)=a^2+b^2+c^2-2ab-2ac-2bc$ is the K\"all\'en function.
 
The effects of the resonances are as follows. It can be seen from the fourth column of Table~\ref{btreeyu} that, the $\Delta(1232)$ does not contribute to the $\gamma n \to \pi^0 \pi^0 n$ channel, but makes a small contribution to the $\gamma p \to \pi^0 \pi^0 p$ channel. As for the other channels, the inclusion of $\Delta(1232)$ leads to sizeable changes in the results. It can also be found that the contributions of Roper are negligible in all the channels. The impact of $\rho$ on the neutral channel is also negligible, but it changes the amplitude of other channels greatly at threshold.

As for non-resonant contributions, the HB ChPT results from Ref.~\cite{refnpa5801994} are shown for comparison in Table~\ref{btreeyu}. It should be noted that, in the HB formalsim, all the $\mathcal{O}(p^1)$ amplitudes at threshold are zero due to the selection rules. In fact, the $\mathcal{O}(p^1)$ amplitudes are sizable in covariant ChPT for almost all the channels except $\gamma n \to \pi^0 \pi^0 n$, as can be seen from the second column of Table~\ref{btreeyu}. The $\mathcal{O}(p^2)$ HB results are comparable with our predictions (third column) expect for the $\gamma p \to \pi^0 \pi^0 p$ channel and $\gamma p \to \pi^+ \pi^- p$ channel. 
\begin{table}[h!]
   \centering
   \begin{tabular}{|cc|c|c|c|c|c|c|}
      \hline
   $ \left| \mathcal{N}_{+++}({\rm P}_i)\right|$&   Process & $\mathcal{O}(p^1)$ & $+\mathcal{O}(p^2)$ & $+\Delta$ & $+{\rm Roper}$ & $+\rho$ & $\mathcal{O}(p^2)$ in HB ChPT~\cite{refnpa5801994}\\
      \hline

    ${\rm P}_1:$ & $\gamma p \to \pi^+ \pi^0 n$ & $9.39$ & $4.55$ & $2.03$ & $1.58$ & $45.47$ & $5.10$\\
      \hline

  ${\rm P}_2:$ &    $\gamma n \to \pi^- \pi^0 p$ & $9.39$ & $4.55$ & $2.03$ & $1.58$ & $45.47$ & $5.10$\\
      \hline

 ${\rm P}_3:$ &    $\gamma p \to \pi^0 \pi^0 p$ & $2.62$ & $4.75$ & $5.16$ & $5.24$ & $6.74$ & $0$\\
     \hline
     
  ${\rm P}_4:$ &   $\gamma n \to \pi^0 \pi^0 n$ & $0$ & $0.54$ & $0.54$ & $0.54$ & $0.75$ & $0.97$\\
     \hline

   ${\rm P}_5:$ &   $\gamma p \to \pi^+ \pi^- p$ & $16.07$ & $1.30$ & $2.71$ & $3.44$ & $62.04$ & $7.22$\\
      \hline
      
 ${\rm P}_6:$ &    $\gamma n \to \pi^+ \pi^- n$ & $13.29$ & $7.03$ & $3.47$ & $2.82$ & $65.36$ & $7.22$\\
      \hline
   \end{tabular}
   \caption{The moduli of chiral amplitudes at threshold for all the physical production processes in unit of ${\rm GeV}^{-1}$.}
   \label{btreeyu}
\end{table}

\subsection{The $\gamma p \to \pi^0 \pi^0 p$ process}

\begin{figure}[htbp]
    \centering
\includegraphics[width=0.7\textwidth]{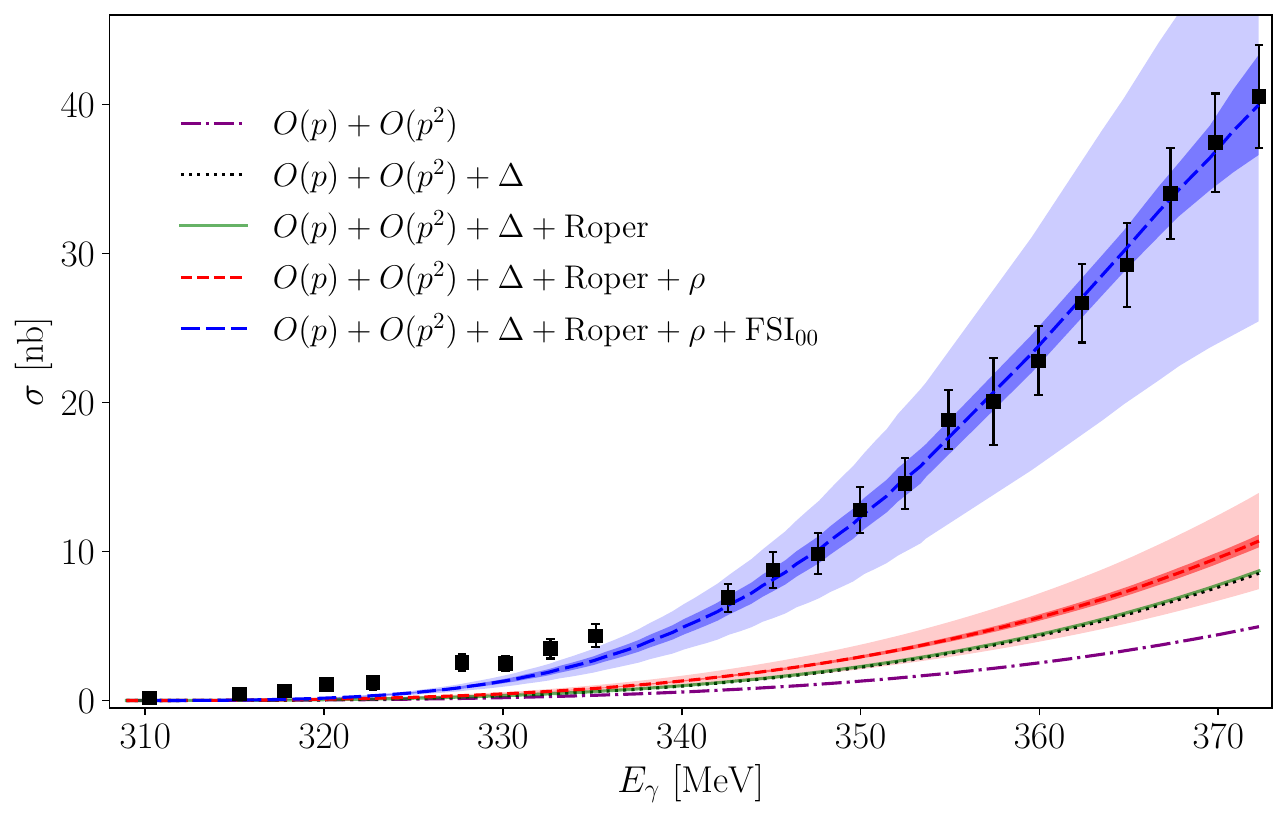}
    \caption{Total cross sections for $\gamma p\to \pi^0 \pi^0 p$. Data are taken from Ref.~\cite{refepja482012}. Red short-dashed line represents the sum of various contributions: non-resonant terms up to $\mathcal{O}(p^2)$, $\Delta$, Roper and $\rho$ resonances. The inner red band stands for the statistical uncertainty propagating from the errors of LECs. The outer red band is obtained by quadratically adding the theoretical error, due to the chiral truncation, to the statistical one. Blue dashed line with bands stands for the result obtained by taking the FSI effect into account. The meaning of the green solid, black dotted and purple dash-dotted lines can be read from the legends. }
    \label{tusigman3}
\end{figure}

For $\gamma p \to \pi^0 \pi^0 p$, we 
take $M_\pi=135.0~{\rm MeV}$ for the neutral pion in the numerical computation. Results of the cross section are shown in Fig.~\ref{tusigman3}. In Fig.~\ref{tusigman3}, the inner bands represent the uncertainties propagated from the errors of the parameters. Specifically, 1000 groups of random numbers for the LECs, $c_i$ and $h$, are generated with a normal distribution. Then, observables such as cross sections are computed for 1000 times, from which the standard deviations can be determined. The outer bands are obtained by quadratically adding the systematic errors, which are responsible for higher-order contribution due to the truncation in the chiral series. We follow the method of Ref.~\cite{refepja512015} to estimate such errors due to the chiral truncation at $\mathcal{O}(p^2)$. Namely, for a given observable $X$, one has
\begin{equation}
    \delta X^{(2)}={\rm max}\left( |X^{(1)}|Q^2,|X^{(2)}-X^{(1)}|Q \right)
    \label{eqxmqx}
\end{equation}
with $Q=E_\gamma^{\rm CM}/\Lambda_b$. Here $E_\gamma^{\rm CM}$ and $\Lambda_b$ are the photon energy in the CM frame and the breakdown scale of the chiral expansion, respectively. We set $\Lambda_b\sim 4\pi F_\pi\sim 1~{\rm GeV}$. Appropriate powering counting rules for the resonances are needed for a reasonable estimation with Eq.~\eqref{eqxmqx}. The inclusion of resonances introduces relevant extra mass scales. Following the treatment in Ref.~\cite{refplb7602016}, here we assign $(m_\Delta-m_N)\sim \delta $, $m_{\rm R}-m_N \sim \delta$ and $\delta^2\sim M_\pi \sim O(p^1)$. This counting rule is actually a generalization of the so-called $\delta$ counting proposed in Ref.~\cite{Pascalutsa:2002pi}, where only the $\Delta$ resonance is incorporated. As a result, the contributions of diagrams with explicit Delta and Roper are either $\mathcal{O}(p^{1.5})$ or $\mathcal{O}(p^{2})$. As for $\rho$, we employ the power counting scheme in Ref.~\cite{refplb5752003} and count the $\rho$ meson propagator as $\mathcal{O}(p^{0})$.

It can be seen from Fig.~\ref{tusigman3} that the ChPT predictions are far away from the experimental data. Even in the vicinity of the threshold, large deviation is observed. We have further checked the influence of the $\mathcal{O}(p^3)$ trees, by using the values of the involved $O(p^3)$ LECs from Ref.~\cite{refprd1002019,refjhep052016}. We found that the inclusion of $\mathcal{O}(p^3)$ tree can not make the situation better. On the other hand, it is pointed out in Ref.~\cite{refprl611988} that there exits strong FSI between the final di-pions in the isospin-zero S-wave channel, corresponding to $IJ=00$. Since the isospin-zero S-wave FSI is expected to yield sizeable contribution to the $\gamma p \to \pi^0 \pi^0 p$ process, we proceed with the discussion of the effect of FSI of $\pi^0\pi^0$. 

According
 to Watson's final-state theorem~\cite{Watson:1952ji}, we incorporate the $\pi\pi$ FSI by means of~\cite{refprd351987}
\begin{equation}
    \mathcal{N}(\gamma p\to \pi^0\pi^0 p)= \mathcal{N}_{\rm tree}\,\mathcal{P}(s_{23})\,\Omega_{00}(s_{23})\ .
    \label{gsnnst}
\end{equation}
We refer the readers to, e.g. Ref.~\cite{Yao:2020bxx}, for review of various approaches of implementing FSI effects. Here,
$\mathcal{N}_\text{tree}$ denotes the tree-level chiral amplitude, which has already been derived above. The $\Omega_{00}(s_{23})$ is the Omn\` es function~\cite{Omnes:1958hv} in the $IJ=00$ channel. Its explicit expression reads
\begin{align}
  \Omega_{00}(s_{23})=\exp \left[\frac{s_{23}-s_{23}^{\rm th}}{\pi} \int_{s_{23}^{\rm th}}^{\infty} \frac{\mathrm{d} s_{23}^{\prime}}{s_{23}^{\prime}-s_{23}^{\rm th}} \frac{\delta_{00}\left(s_{23}^{\prime}\right)}{s_{23}^{\prime}-s_{23}}\right]\ ,
\end{align}
where $s_{23}^{\rm th}=4M_\pi^2$ and the phase shift $\delta_{00}$ is taken from the Roy-type equations presented in Ref.~\cite{rfprd832011}. It is sufficient to take $\mathcal{P}(s_{23})$ in Eq.~\eqref{gsnnst} to be a second-order polynomial, i.e.,
\begin{align}
\mathcal{P}(s_{23})=1+ \alpha_1(s_{23}-4M_\pi^2)+\alpha_2(s_{23}-4M_\pi^2)^2  \ .
\end{align}
The coefficients $\alpha_1$ and $\alpha_2$ are free parameters to be determined by fitting experimental data. The fit results are 
\begin{align}
\alpha_1=(2.16\pm 0.23)\times 10^2{\rm GeV}^{-2}\ ,\quad\alpha_2=(-1.05 \pm 0.18)\times 10^4{\rm GeV}^{-4}\ ,
\label{eq:agga}
\end{align}
with $\chi^2/{\rm d.o.f}=1.96$. The FSI-improved ChPT prediction of cross section is shown by the blue line in Fig.~\ref{tusigman3}. The blue narrow band stands for the statistical errors due to the uncertainties of the parameters including the $\alpha_i$'s, while the broad band is obtained by taking into account the influence of the theoretical uncertainties due to chiral truncation.  Generally speaking, a good description of data is now achieved for $E_\gamma>340{\rm MeV}$, indicating the $IJ=00$ FSI effect is indeed sizeable. For the energy $E_\gamma<340{\rm MeV}$, deviation still exists. Such deviation is attributed to the absence of the chiral loop contributions, as concluded in Ref.~\cite{refnpa5801994}.

\subsection{The $\gamma p \to \pi^+ \pi^0 n$ process}

\begin{figure}[htbp]
    \centering
   \includegraphics[width=0.7\textwidth]{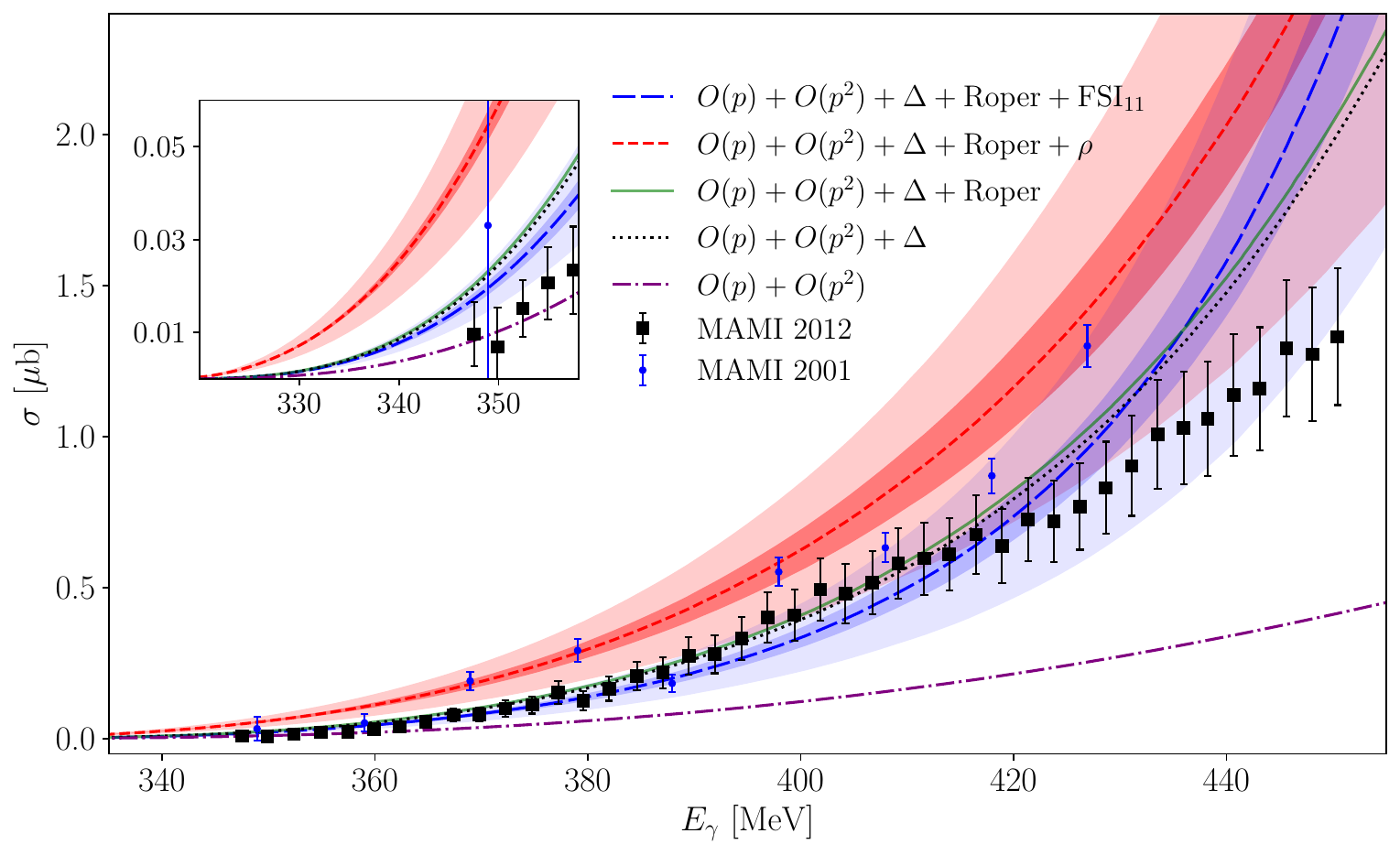}
    \caption{Cross sections for $\gamma p\to \pi^+ \pi^0 n$. Data are taken from Refs.~\cite{refepja482012,refprl872001}. The red short-dashed line with bands represents the full ChPT prediction, while the blue long-dashed one with bands stands for the result obtained by implementing the FSI effect of $\pi^+\pi^0$ pair via Omn\`es approach.  }
    \label{tusigman1}
\end{figure}

For $\gamma p \to \pi^+ \pi^0 n$,  where both the charged and neutral pions are involved, we take the average of the $\pi^+$ and $\pi^0$ masses in our numerical computation.  That is $M_\pi = (M_{\pi^+} + M_{\pi^0}) / 2 = 137.3~{\rm MeV}$.\footnote{The effect caused by using a different pion mass value, e.g. $M_\pi=M_\pi^+$, is subleading in the sense that the resultant outcome is mostly within the 1-$\sigma$ error band of the result obtained with the average pion mass. We refer the readers to  App.~\ref{sec.mpis} for more discussions. } Results of total cross section are shown in Fig.~\ref{tusigman1}. It can be seen that the ChPT prediction with only non-resonant $\mathcal{O}(p)+\mathcal{O}(p^2)$ contributions is not adequate to describe the data from Ref.~\cite{refepja482012,refprl872001}; see the purple dash-dotted curve in Fig.~\ref{tusigman1}. As the incident photon energy increases, the deviation becomes larger and larger. The $\Delta(1232)$ resonance contributes largely, which is necessary for a good agreement with the data in the energy region below $440$~MeV. The contribution of the Roper resonance is negligible. The $\rho$ meson increases the modulus of cross section obviously, especially near the threshold. The incorporation of the $\rho$ meson makes the theoretical result well consistent with the experimental data of Ref.~\cite{refprl872001}, but not with the one of Ref.~\cite{refepja482012}. For the $\gamma p \to \pi^+ \pi^0 n$ process, there is no need to consider higher-order tree and loop contributions, in the sense that it is already sufficient to establish a good description of data by using the $\mathcal{O}(p^2)$ ChPT result with explicit $\Delta$, Roper and $\rho$ resonances. It is also worthy noting that the FSI of the $\pi^+\pi^0$ system is dominated by the $IJ=11$ partial wave, which is actually compensated by the explicit inclusion of the $\rho$ meson. 

\begin{figure}[h!]
    \centering
   \includegraphics[width=0.7\textwidth]{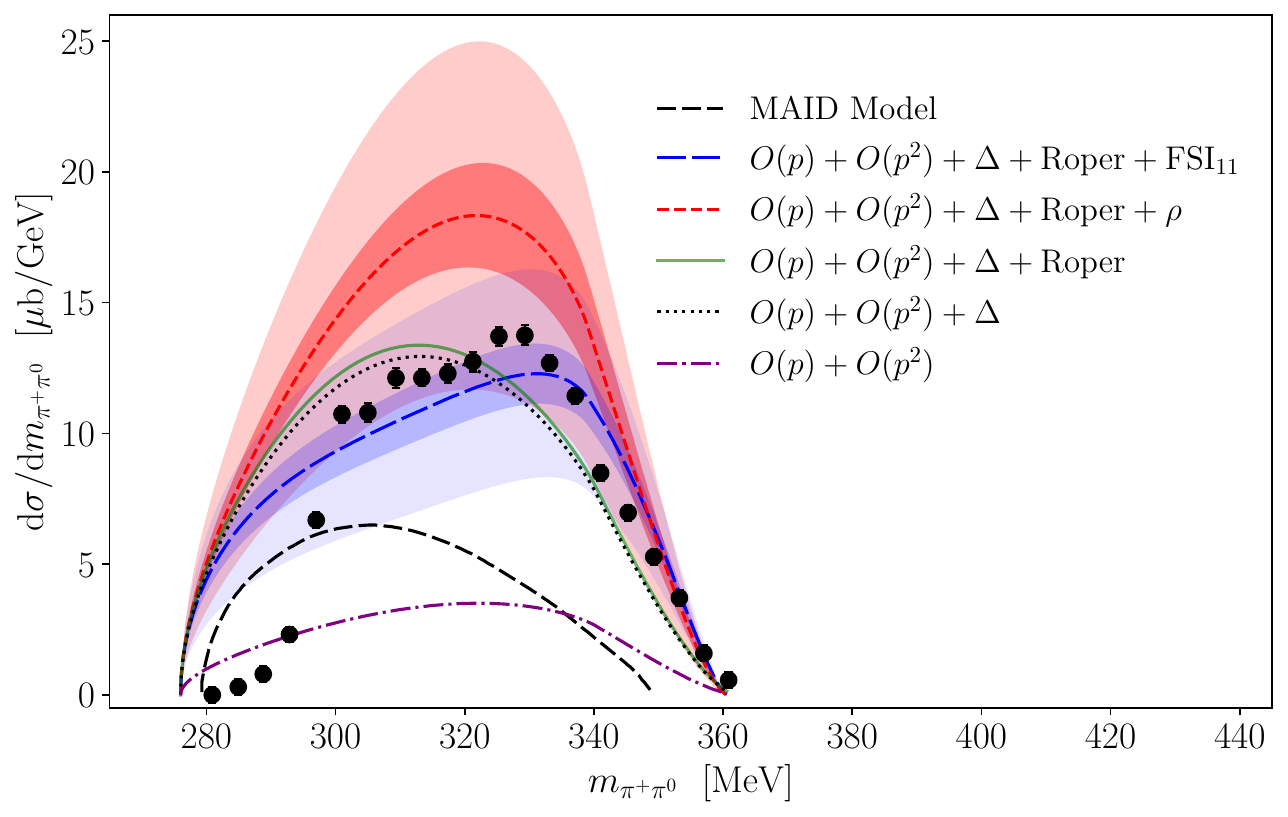}
    \caption{Pion-pion invariant-mass distribution for $\gamma p\to \pi^+ \pi^0 n$ with incident photon energy bin of $400-430~{\rm MeV}$. Data are taken from~\cite{refepja482012}.  The red short-dashed line with bands represents the full ChPT prediction, while the blue long-dashed one with bands stands for the result obtained by implementing the FSI effect of $\pi^+\pi^0$ pair via Omn\`es approach. For comparison, the result by the MAID model~\cite{refepja252005} is shown as the black long-dashed line.   }
    \label{tudsigman1}
\end{figure}

Besides the total cross section, invariant-mass distribution of pion pair can also be computed with the help of Eq.~\eqref{gsdddd}. To confront our model with the experimental measurement in a given energy bin $\left[E_{\gamma {\rm min}},E_{\gamma {\rm max}}\right]$, we perform a convolution on the differential cross section in the following way, 
\begin{equation}
    \frac{d\sigma}{d \sqrt{s_{23}}}=\frac{1}{E_{\gamma {\rm max}}-E_{\gamma {\rm min}}}\int_{E_{\gamma {\rm min}}}^{E_{\gamma {\rm max}}}\frac{d\sigma(E_\gamma)}{d \sqrt{s_{23}}}dE_\gamma
\end{equation}
For the photon energy interval $[400,430]~{\rm MeV}$, our result of the di-pion invariant mass distribution is shown in Fig.~\ref{tudsigman1}.
The $\mathcal{O}(p)+\mathcal{O}(p^2)$ non-resonant contribution does not agree with the data, even near the pion-pair production threshold. The $\Delta(1232)$ resonance improves the result significantly, and the theoretical curve with $\Delta$ contribution tends to be comparable with some of the data points. Similar to the case of total cross section, the impact of the Roper resonance remains negligible. The effect of the $\rho$ meson is to deform the shape of the curve, leading to a good agreement with data within $1$-$\sigma$ uncertainty. 
For comparison, we also show the result obtained by an effective model (i.e. MAID model) calculation~\cite{refepja252005}. It can be seen from the figure that the chiral prediction, up to $\mathcal{O}(p^2)$ with explicit resonances, is much better than the MAID-model determination. However, it should be mentioned that inconsistency occurs for low di-pion invariant masses close to the threshold. That is, the experimental data rise slowly, while the ChPT result grows sharply. The MAID-model calculation suffers from the same problem~\cite{refepja252005}. This problem is expected to be addressed, in future, by further incorporating leading one-loop contribution with more complicated analytical structure. 

\begin{figure}[h!]
\centering
\includegraphics[width=0.7\textwidth]{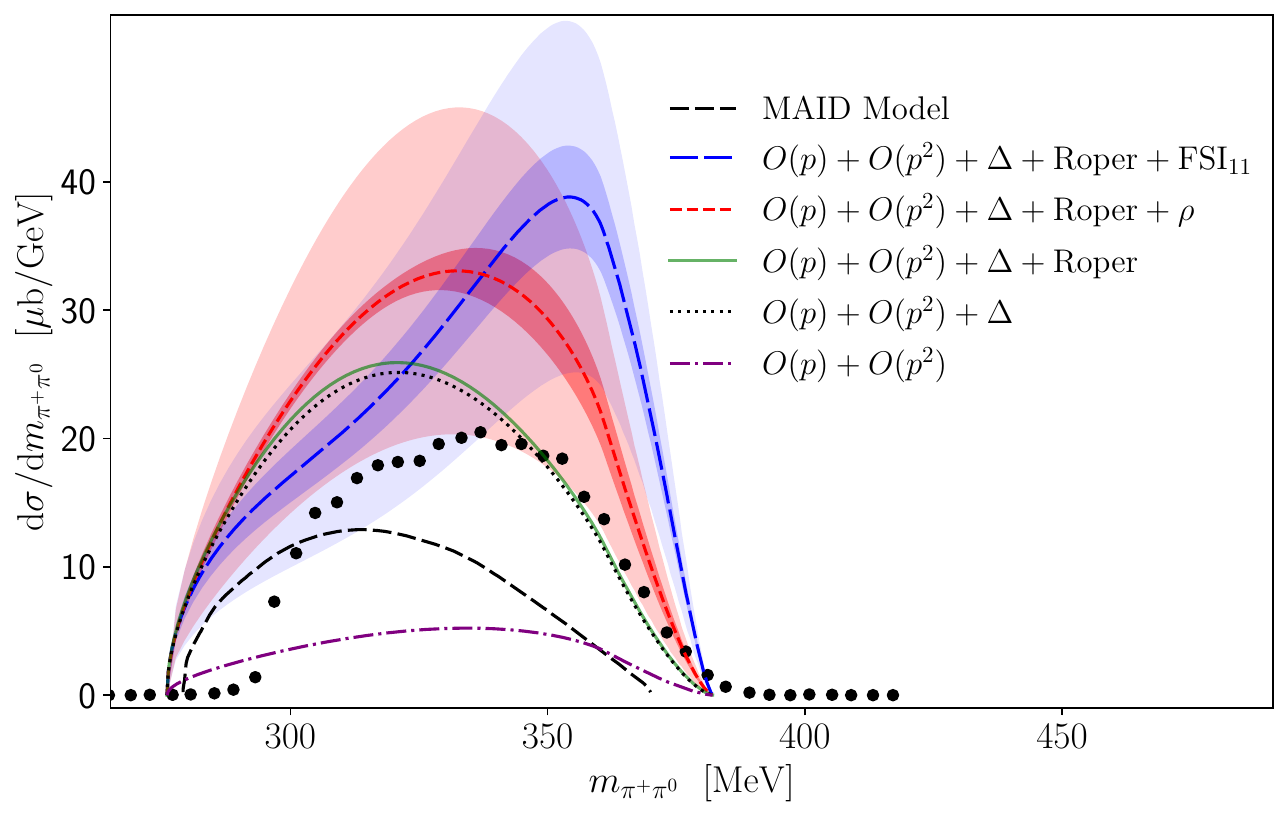}
\caption{Pion-pion invariant-mass distribution for $\gamma p\to \pi^+ \pi^0 n$ with incident photon energy bin of $430-460$~MeV. Data are taken from~\cite{refepja482012}. The description of the curves is the same as in Fig.~\ref{tudsigman1}.}
\label{tudsigman1445}
\end{figure}
\begin{figure}[h!]
    \centering
   \includegraphics[width=0.7\textwidth]{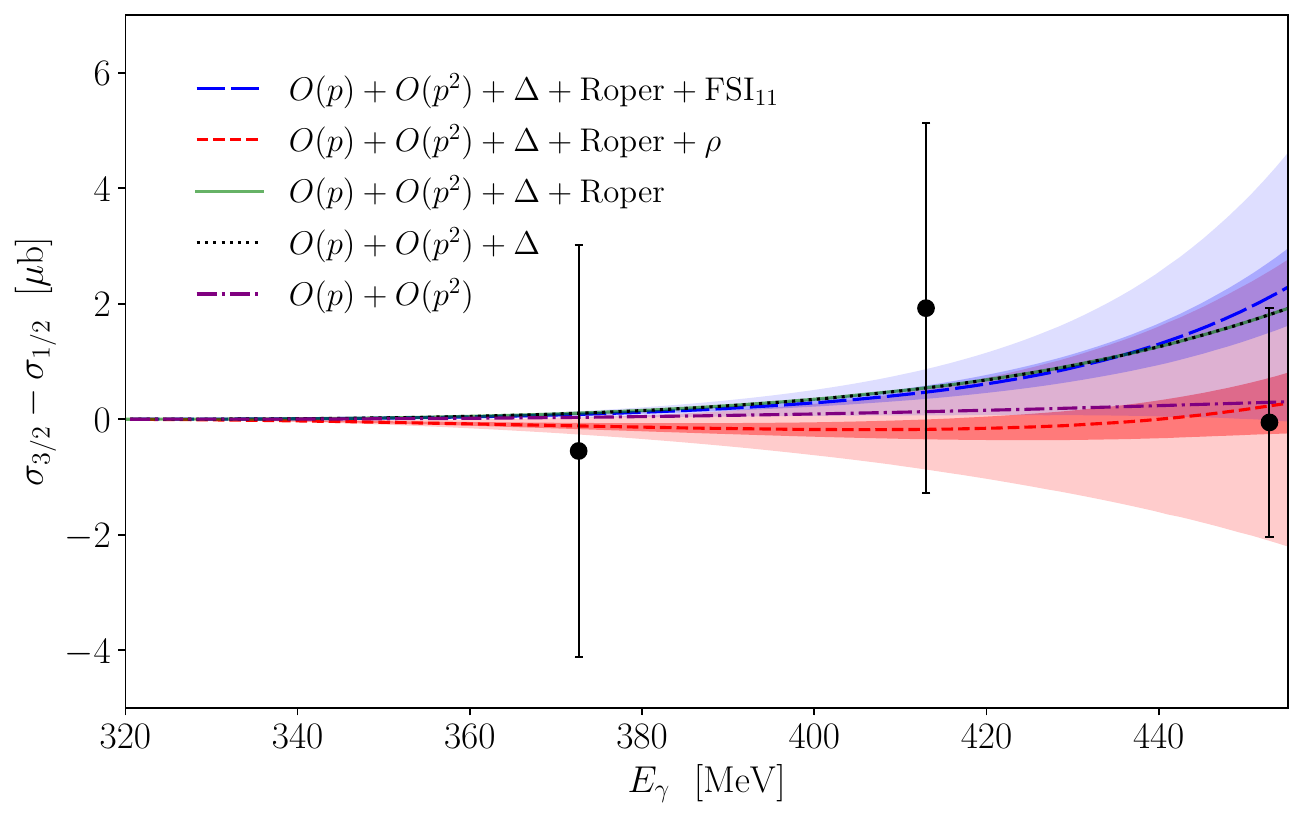}
    \caption{Beam-target helicity asymmetry for $\gamma p\to \pi^+ \pi^0 n$. Data are taken from Ref.~\cite{refplb5512003}. The description of the curves is the same as in Fig.~\ref{tudsigman1}.}
    \label{tusigma32n1}
\end{figure}

We have also calculated the invariant-mass distribution for the energy bin of $[430,460]$~MeV, which is shown in Fig.~\ref{tudsigman1445}. Effect of each resonance is similar to the case with energy bin $[400,430]~{\rm MeV}$. However, it can be seen from Fig.~\ref{tudsigman1445} that the ChPT result starts to fail in describing the data, since the $1$-$\sigma$ error band merely overlaps with the experimental data marginally.

Finally, by making use of Eq.~\eqref{gsspdm}, the beam-target helicity asymmetry $\sigma_{3/2}-\sigma_{1/2}$ for $\gamma p\to \pi^+ \pi^0 n$ is predicted as well. Result is shown in Fig.~\ref{tusigma32n1}. The available experimental data are close to zero with large error bars~\cite{refplb5512003}. The contribution from the non-resonant terms is identical to zero, so is the Roper resonance. The $\Delta(1232)$ yields a positive value of the asymmetry, due to the fact that it mainly contributes to the total cross section $\sigma_{3/2}$ rather than $\sigma_{1/2}$. On the contrary, the influence of the $\rho$ meson is opposite to that of the $\Delta(1232)$. It contributes much more to $\sigma_{1/2}$ than to $\sigma_{3/2}$, leading to a negative asymmetry eventually.

Before ending this subsection, it would be interesting to discuss the inclusion of FSI effect via the Omn\`es-function approach, as done for the $\gamma p\to \pi^0\pi^0 p$ channel. To that end,  the effect of $IJ=11$ FSI for $\gamma p \to \pi^+ \pi^0 n$ is implemented by, 
\begin{equation}
    \mathcal{N}(\gamma p\to \pi^+\pi^0 n)= \mathcal{N}_{\rm tree}\,\mathcal{P}(s_{23})\,\Omega_{11}(s_{23})\ .
    \label{gsnnstappen}
\end{equation}
which is similar to Eq.~\eqref{gsnnst}. Here,  the contribution of the $\rho$ meson  must be excluded from the tree-level amplitude $\mathcal{N}_{\rm tree}$, in order to avoid double counting problem.  The $P$-wave phase shift $\delta_{11}$ is taken from Ref.~\cite{rfprd832011}. We fit the total cross sections and the pion-pion invariant mass distribution with incident photon energy bin $400-430$ MeV. The fit results of the parameters in $\mathcal{P}(s_{23})$ are
\begin{align}
\alpha_1=(-15.62\pm 0.58)~{\rm GeV}^{-2}\ ,\quad\alpha_2=(5.29 \pm 0.17)\times 10^2~{\rm GeV}^{-4}\ ,
\end{align}
with $\chi^2/{\rm d.o.f}=27.07$. The unexpected large number is caused by the failure of description of the di-pion mass distribution data, as one can see from the blue long-dased lines in Fig~\ref{tudsigman1} and Fig.~\ref{tudsigman1445}.  Based on the fitted values of the parameters, the total cross sections, the invariant mass distribution and the beam-target helicity asymmetry are shown as blue long-dashed lines in Figs.~\ref{tusigman1},~\ref{tudsigman1},~\ref{tudsigman1445} and ~\ref{tusigma32n1}. For the total cross section, the experimental data of Ref.~\cite{refepja482012} are now well described up to the photon energy $\sim E_\gamma=420$~MeV.  Compared to the case with explicit $\rho$ meson, the inclusion of FSI in the Omn\`es manner does not improve the description of the data of di-pion invariant mass distribution. Besides, the former is more consistent with the beam-target helicity asymmetry than the latter, as can be seen from Fig.~\ref{tusigma32n1}. We have checked that, by tuning $\alpha_1$ and $\alpha_2$, one can obtain similar results of cross sections and di-pion invariant mass distributions as the ones with explicit inclusion of the $\rho$ meson. Therefore, it can be concluded that the two ways of implementation of the two-pion FSI effects, i.e., via the $\rho$-meson or through the Omn\`es approach, are more or less equivalent.

\subsection{ The other four physical processes}

\begin{figure}[htbp]
    \centering
    \includegraphics[width=0.8\textwidth]{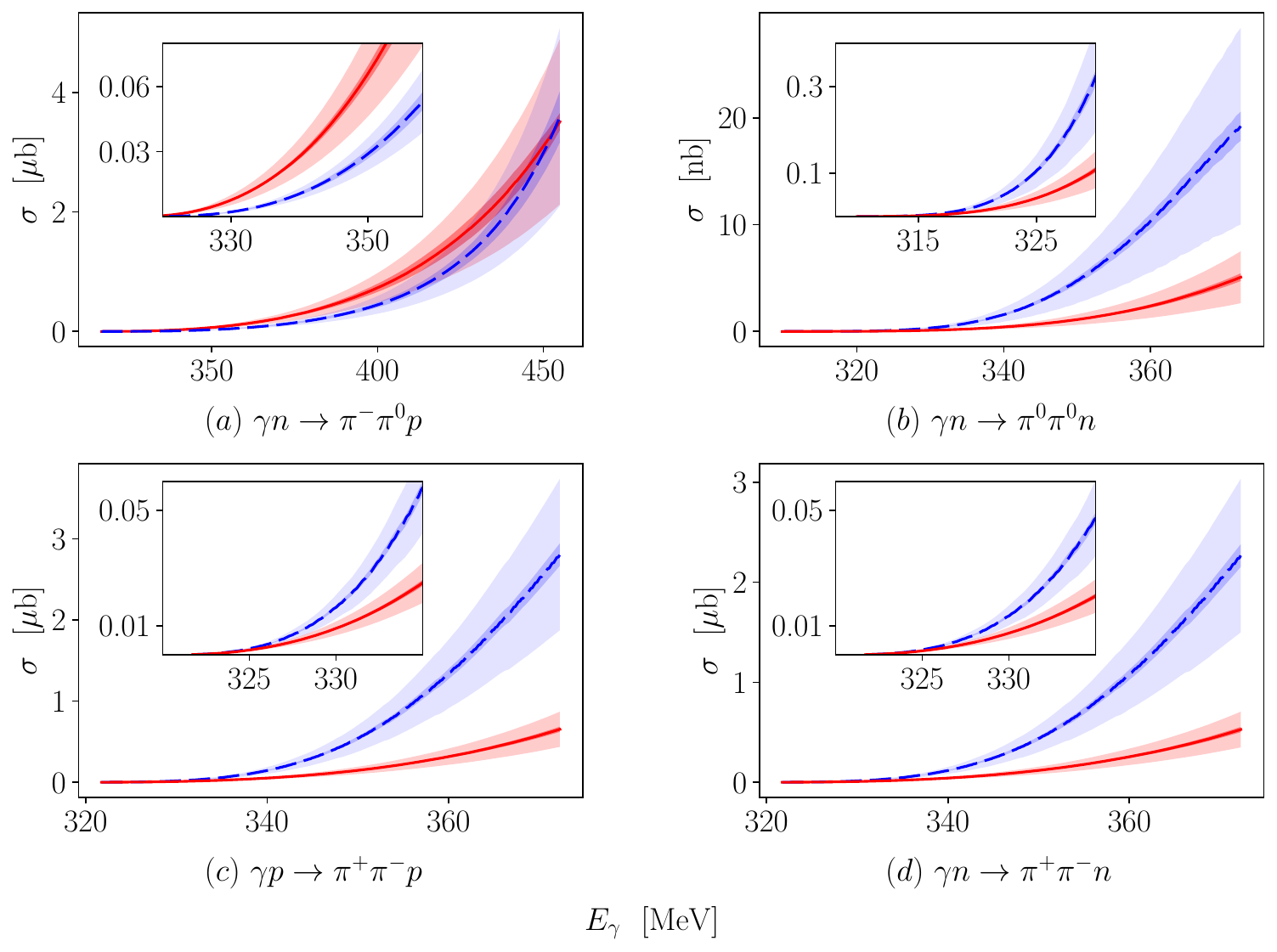}
    \caption{Predictions of total cross sections. The red solid and  blue dashed lines with bands represent the pure ChPT and Omn\`es-\lq\lq improved" ChPT results, respectively.}
    \label{fig.sigma_predictions}
\end{figure}
\begin{figure}[htbp]
    \centering
    \includegraphics[width=0.8\textwidth]{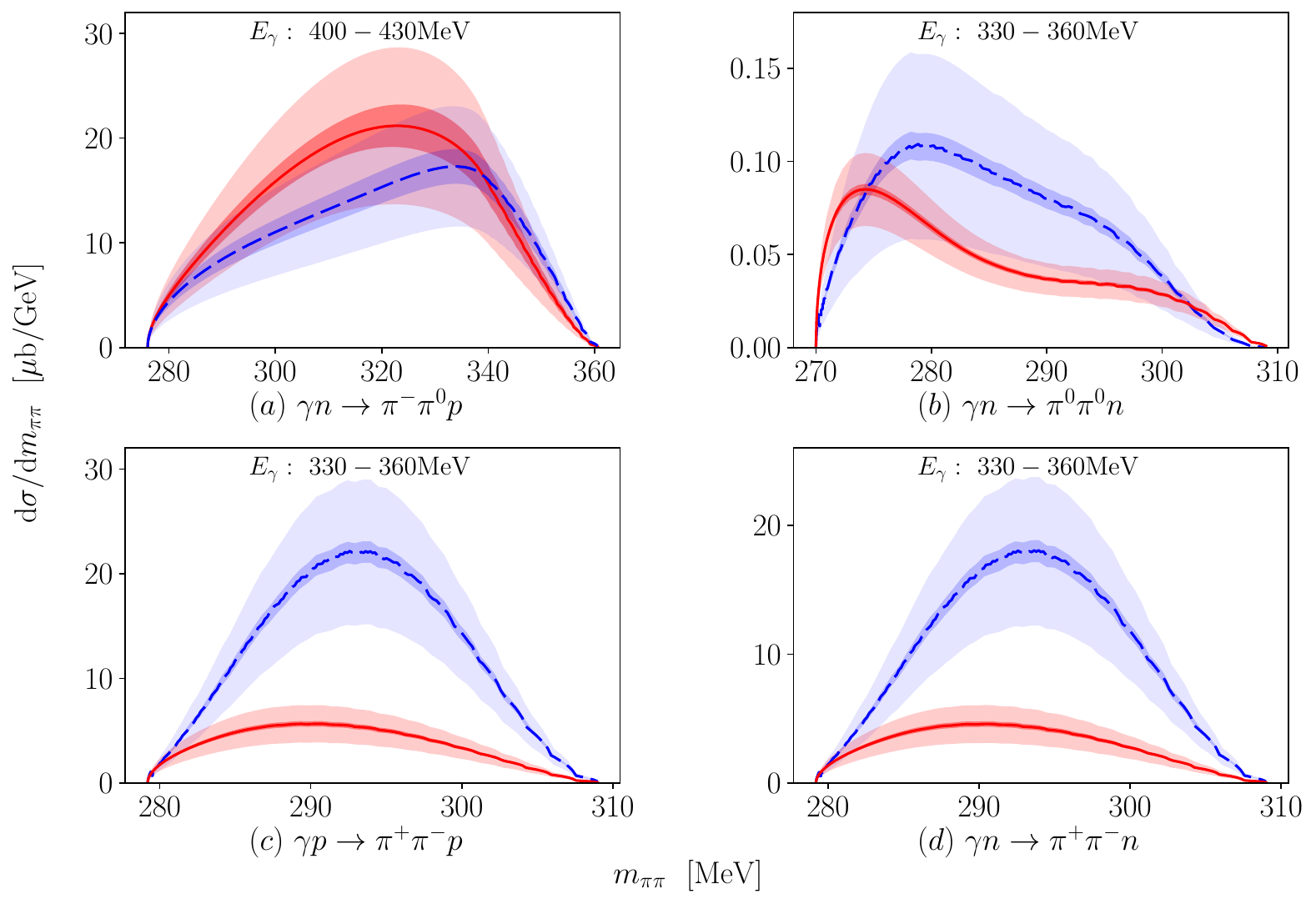}
    \caption{Predictions of di-pion invariant mass distributions. The red solid and  blue dashed lines with bands represent the pure ChPT and Omn\`es-\lq\lq improved" ChPT results, respectively.}
    \label{fig.deltasigma_predictions}
\end{figure}
Based on the above results, we can make predictions of the cross sections for the other four physical channels: $(a)$ $\gamma n \to \pi^- \pi^0 p$, $(b)$ $\gamma n \to \pi^0 \pi^0 n$, $(c)$ $\gamma p \to \pi^+ \pi^- p$ and $(d)$ $\gamma n \to \pi^+ \pi^- n$.   So far, those processes have not been experimentally measured for incident photon energies $\le 430~{\rm MeV}$ in the laboratory reference frame. Our predictions of total cross sections and invariant mass distributions are shown in Fig.~\ref{fig.sigma_predictions} and Fig.~\ref{fig.deltasigma_predictions}. The red solid and blue dashed lines with bands in Fig.~\ref{fig.sigma_predictions} and Fig.~\ref{fig.deltasigma_predictions} represent the pure ChPT and Omn\`es-\lq\lq improved" ChPT predictions, respectively.

The situation of  $\gamma n \to \pi^- \pi^0 p$ channel is similar to that of the $\gamma p \to \pi^+ \pi^0 n$ channel, where the FSI of the di-pion system is dominated by the $P$-wave contribution with $IJ=11$. Such a contribution can be equivalently implemented by explicit inclusion of the $\rho$ meson or by the Omn\`es function, as concluded in the previous subsection. Results of $\gamma n \to \pi^- \pi^0 p$ keep in line with this conclusion. As one can see from the left-top panels of Fig.~\ref{fig.sigma_predictions} and Fig.~\ref{fig.deltasigma_predictions}, the pure BChPT and Omn\`es-\lq\lq improved" ChPT predictions are comparable with each other.

As for the $\gamma n \to \pi^0 \pi^0 n$, $\gamma p \to \pi^+ \pi^- p$ and $\gamma n \to \pi^+ \pi^- n$ channels, the loop contributions are as important as that in the $\gamma p \to \pi^0 \pi^0 p$ channel. Assuming the effect of loop diagrams can be compensated by the $IJ=00$ $\pi \pi $ FSI, just like the $\gamma p \to \pi^0 \pi^0 p$ channel, the predictions of the total cross sections and invariant mass distributions for those channels are shown as blue long-dashed lines in the right-top, left-bottom and right-bottom panels of Figs.~\ref{fig.sigma_predictions} and~\ref{fig.deltasigma_predictions}. 
Here, the values of the FSI parameters $\alpha_1$ and $\alpha_2$ in Eq.~\eqref{eq:agga} are used. It can be seen that the incorporation of $IJ=00$ FSI effect leads to larger predictions than pure ChPT. Note that all the above predicted results are obtained by considering only the dominant partial-wave contribution in the FSI.  Therefore, one should be cautious when comparing them with future experimental data.

\section{Summary}\label{sec:s}
The double pion photoproduction off nucleons is analyzed in a covariant chiral perturbation theory with pions, nucleons, the $\Delta(1232)$, the $N^*(1440)$ and the $\rho$ resonances as explicit degrees of freedoms. The production amplitudes for all the 6 physical processes are obtained up to $\mathcal{O}(p^2)$. All the parameters involved in the chiral amplitudes have been determined elsewhere, which allows us to make numerical predictions. The findings are as follows.
 
\begin{itemize}
\item It is found that, at threshold, all the considered resonances have little contribution to the neutral channels $\gamma N\to \pi^0\pi^0 N$ with $N$ being proton or neutron, while the $\Delta(1232)$ and $\rho$ resonances contribute largely to the other channels. The effect of the Roper resonance is negligible for all the channels.
\item
For $\gamma p\to \pi^0 \pi^0 p$, our ChPT prediction of the total cross section is far below the corresponding experimental data. Even if the $\mathcal{O}(p^3)$ tree contribution is taken into account, the large discrepancy persists. The deviation can be remedied by implementing the isoscalar S-wave FSI of the $\pi^0\pi^0$ system, and a good description of data is established.  

\item
For $\gamma p\to \pi^+ \pi^0 n$, our results of total cross section, pion-pair invariant-mass distribution and beam-target helicity asymmetry are in good agreement with the existing experimental data. Sizeable contributions are from the $\mathcal{O}(p)+\mathcal{O}(p^2)$ non-resonant terms, the $\Delta$ and the $\rho$ resonances. The impact of the Roper resonance is slight and can be neglected. The explicit inclusion of the $\rho$ meson actually accounts for the FSI of the $\pi^+\pi^0$ system in the $IJ=11$ channel.

\end{itemize}

A more comprehensive ChPT analysis can be done at loop-level in the future to achieve better agreement with experimental data. The Roper resonance can be omitted due to its negligible contribution. On the other hand, more data near threshold are required from experiments to pin down the involved LECs that will show up in higher-order ChPT calculations.
\\
\\

{\bf\large Acknowlegement}

We would like to thank Feng-Kun Guo for a careful reading of the manuscript. 
This work is supported by National Nature Science Foundations of China (NSFC) under Contract Nos. 12275076, 11905258, 12335002, 12125507, 12047503, 12347120; by the Chinese Academy of Sciences under Grant No. YSBR-101; by the Science Fund for Distinguished Young Scholars of Hunan Province under Grant No. 2024JJ2007; by the Fundamental Research Funds for the Central Universities; and by the Postdoctoral Fellowship Program of China Postdoctoral Science Foundation (CPSF) under Grant Nos. GZC20232773, 2023M743601. DLY appreciates the support of Peng Huan-Wu visiting professorship and the hospitality of Institute of Theoretical Physics at Chinese Academy of Sciences, where this work was finalized.

\appendix
\section{Correspondence between theories with and without the $\rho$ meson\label{sec:rhotheory}}

\begin{figure}[ht]
 \centering
\includegraphics[scale=0.5]{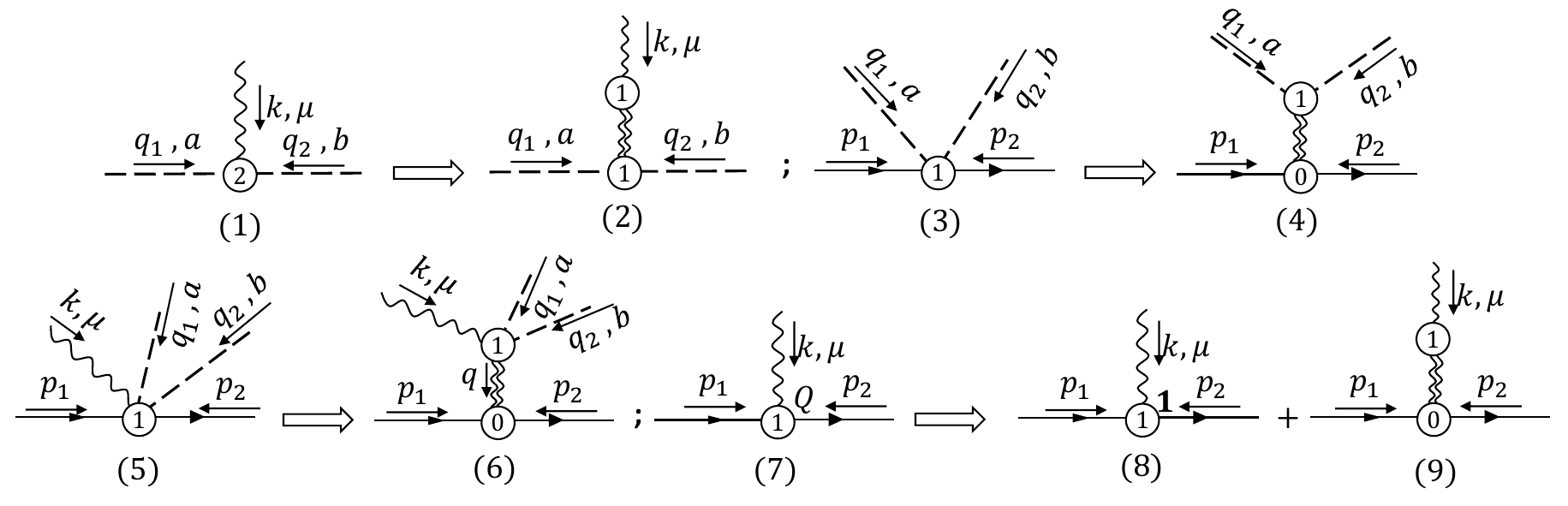}
    \caption{Correspondence between the vertices with and without $\rho(770)$. The dashed, solid, wavy, and double wavy lines represent pions, nucleons, photons, and $\rho$, respectively. Circled numbers mark the chiral orders of the vertices. $Q$ in diagram(g) and $\mathbf{1}$ in diagram(i) represent terms proportional to matrices $Q$ and $\mathbf{1}$, respectively.}
    \label{turhover}
\end{figure}

The leading order photon-$\pi$-$\pi$, $\pi$-$\pi$-N-N, photon-$\pi$-$\pi$-N-N interaction terms and $e\overline{\Psi}\slashed{A}\frac{\tau_3}{2}\Psi$ term from $\mathcal{L}^{(2)}_{\pi \pi}$ and $\mathcal{L}^{(1)}_{\pi N}$ will be canceled by the corresponding terms generated from Eq.~\eqref{eq.rho}. Their vertices will be replaced by the ones with an internal $\rho$ line inserted, as shown in Fig.~\ref{turhover}. The effective theory with $\rho$ as explicit degree of freedom reduces to the $\rho$-less one by integrating out the $\rho$ meson. Specifically, for the vertices in Fig.~\ref{turhover}, one can expand the $\rho$ propagators and keep the leading term, which leads to the following correspondences:
\begin{equation}
\begin{aligned}
{\rm Vertex}_{(2)}&=\left( \frac{ie}{g_\rho}M_\rho^2 \delta^{i3}g^{\mu \alpha} \right) \left[ \frac{i\delta^{ij}}{k^2-M_\rho^2}\left( -g_{\alpha \beta}+\frac{k_\alpha k_\beta}{M_\rho^2} \right) \right] \left[ g_\rho \varepsilon^{abj}\left(q_2^\beta-q_1^\beta \right) \right]\\
&=e \varepsilon^{ab3}(q_1^\mu-q_2^\mu)+\mathcal{O}(p^3)={\rm Vertex}_{(1)}+\mathcal{O}(p^3)\ ,\\
{\rm Vertex}_{(4)}&=\left[ g_\rho \varepsilon^{abi}\left( q_2^\mu-q_1^\mu \right) \right] \left[ \frac{i\delta^{ij}}{(q_1+q_2)^2-M_\rho^2}\left( -g_{\mu \nu}+\frac{(q_{1\mu}+q_{2\mu}) (q_{1\nu}+q_{2\nu})}{M_\rho^2} \right) \right]\left( ig_\rho \gamma^\nu \frac{\tau_j}{2} \right)\\
&=\frac{1}{2F^2}\varepsilon^{abi}\frac{\tau_i}{2}\left( \slashed{q}_1-\slashed{q}_2 \right)+\mathcal{O}(p^3)={\rm Vertex}_{(3)}+\mathcal{O}(p^3)\ ,\\
{\rm Vertex}_{(6)}&=\left[ -\frac{ieM_\rho^2}{2F^2 g_\rho} g^{\mu \alpha}\left( 2\delta_{ab}\delta_{i3}-\delta_{b3}\delta_{ia}-\delta_{a3}\delta_{ib} \right) \right] \left[ \frac{i\delta^{ij}}{q^2-M_\rho^2}\left( -g_{\alpha \beta}+\frac{q_\alpha q_\beta}{M_\rho^2} \right) \right] \left( ig_\rho \gamma^\beta \frac{\tau_j}{2} \right)\\
&=\frac{ie}{4F^2}\gamma^\mu \left( 2\delta_{ab}\tau_3-\delta_{a3}\tau_b-\delta_{b3}\tau_a \right)+\mathcal{O}(p^2)={\rm Vertex}_{(5)}+\mathcal{O}(p^2)\ ,\\
{\rm Vertex}_{(8)}+{\rm Vertex}_{(9)}&=-ie\frac{\mathbf{1}}{2}\gamma^\mu+\left( \frac{ie}{g_\rho}M_\rho^2 \delta^{i3}g^{\mu \alpha} \right) \left[ \frac{i\delta^{ij}}{k^2-M_\rho^2}\left( -g_{\alpha \beta}+\frac{k_\alpha k_\beta}{M_\rho^2} \right) \right] \left( ig_\rho \gamma^\beta \frac{\tau_j}{2} \right)\\
&=-ieQ\gamma^\mu+\mathcal{O}(p^2)={\rm Vertex}_{(7)}+\mathcal{O}(p^2)\ .
\end{aligned}
\end{equation}
The difference between the left- and right-hand sides (with and without $\rho$, respectively) of each equation is of higher order. In another word, the inclusion of the $\rho$ meson accounts for a resummation of higher-order contributions.

\section{The influence of the pion mass \label{sec.mpis}}

\begin{figure}[htbp]
    \centering
    \includegraphics[width=0.85\textwidth]{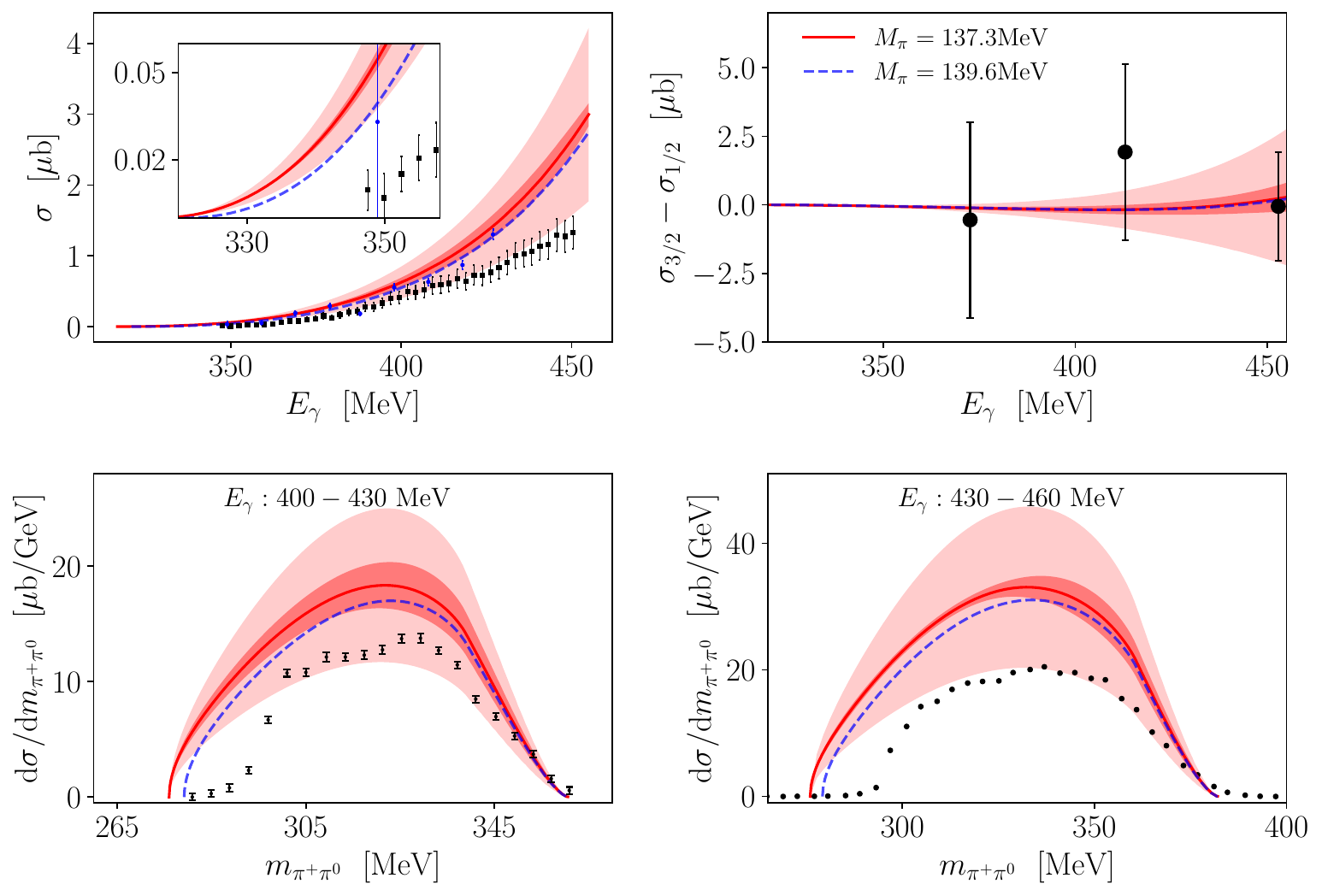}
    \caption{Comparison of results obtained with different pion masses.}
    \label{fig.deltasigma_predictions_mpi}
\end{figure}

For the calculation within the framework of ChPT constructed in the isospin limit, we need to assume that $\pi^+$ and $\pi^0$ have the same mass. In reality,  there is a mass difference of about $4.6~{\rm MeV}$ between the charged and neutral pions. Choosing different pion masses will result in a change in the energy threshold of the incident photon in the laboratory frame, since $E_\gamma^{{\rm thr}} = 2M_\pi(m_N+M_\pi)/m_N$. The obtained cross sections will also differ. To assess the impact of the isospin-breaking effect in the $\gamma p \to \pi^+ \pi^0 n$ channel, we have calculated the total cross sections, the di-pion invariant mass distribution with incident photon energy bin of $400-430~{\rm MeV}$ and $430-460~{\rm MeV}$, and the beam-target helicity asymmetry by employing $M_\pi = M_{\pi^+}= 139.6~{\rm MeV}$. Results are shown as the blue dashed lines in Fig.~\ref{fig.deltasigma_predictions_mpi}.  It can be seen from Fig.~\ref{fig.deltasigma_predictions_mpi} that an increase of the pion mass by $\sim 2.28$~MeV leads to smaller results of total cross sections and invariant mass distributions compared to the $M_\pi=137.3$~MeV case.  Meanwhile, the beam-target helicity asymmetry is almost unchanged. The effect of the small change of the pion mass due to isospin breaking is actually of higher order beyond our working accuracy. The influence is covered by the systematical uncertainties which are estimated by using Eq.~\eqref{eqxmqx}. More specifically,  we can see that the curves of the new results lie within the error bands almost in the whole energy region. However, the mass difference may lead to relatively significant impact in the vicinity of thresholds, especially for the di-pion invariant mass distributions.

\section{Expressions of non-resonant amplitudes \label{sec.amp.non}}
In this appendix, the $O(p^1)$ and $O(p^2)$ non-resonant amplitudes $\mathcal{M}_j^i$ are shown for easy reference. The definition of $\mathcal{M}_j^i$ can be found in Eq.~\eqref{eq:Mijiso} and Eq.~\eqref{eq.lorentz}. The expressions in subsection~\ref{sec.p1.exp} (\ref{sec.p2.exp}) correspond to the sum of the contributions of the $\mathcal{O}(p^1)$ ($\mathcal{O}(p^2)$) Feynman diagrams in Fig.~\ref{tutree} and their crossed partners. 

\subsection{At $\mathcal{O}(p^1)$\label{sec.p1.exp}}
\begin{itemize}
\item Amplitudes $\mathcal{M}^1_j$:
\begin{align}
\mathcal{M}^1_{1}=&\mathcal{M}^1_{3}=\mathcal{M}^1_{8}=\mathcal{M}^1_{9}=\mathcal{M}^1_{12}=0\ ,\notag\\
\mathcal{M}^1_{2}=&-\frac{i e g^2 m_{N} \left(2 M^2-t_{1}-t_{2}\right)}{2 F^2 \left(m_{N}^2-s\right) \left(-2 M^2-m_{N}^2+s+t_{1}+t_{2}\right)}\ ,\notag\\ 
\mathcal{M}^1_{4}=&\frac{i e g^2}{8 F^2} \left(\frac{3 m_{N}^2+s_{1}}{\left(m_{N}^2-s\right) \left(m_{N}^2-s_{1}\right)}+\frac{M^2+4 m_{N}^2-s+s_{2}-t_{1}}{\left(M^2-s+s_{2}-t_{1}\right) \left(-2 M^2-m_{N}^2+s+t_{1}+t_{2}\right)} \right. \notag \\
&\left. +\frac{4 m_{N}^2}{\left(m_{N}^2-s_{1}\right) \left(-M^2+s-s_{1}+t_{2}\right)} \right)-(s_1 \leftrightarrow s_2,t_1 \leftrightarrow t_2)\ , \notag\\
\mathcal{M}^1_{5}=&-\frac{4 i e g^2 m_{N}}{F^2 \left(m_{N}^2-s\right) \left(-2 M^2-m_{N}^2+s+t_{1}+t_{2}\right)}\ ,\notag\\
\mathcal{M}^1_{6}=&\frac{i e g^2 m_{N}^2}{2 F^2} \frac{1}{\left(m_{N}^2-s_{1}\right) \left(-M^2+s-s_{1}+t_{2}\right)}+(s_1 \leftrightarrow s_2,t_1 \leftrightarrow t_2)\ , \notag\\
\mathcal{M}^1_{7}=&-\frac{i e g^2}{F^2 \left(-2 M^2-2 m_{N}^2+2 s+t_{1}+t_{2}\right)} \left(\frac{M^2+4 m_{N}^2-s+s_{1}-t_{2}}{\left(M^2-s+s_{1}-t_{2}\right) \left(-2 M^2-m_{N}^2+s+t_{1}+t_{2}\right)}\right.\notag \\
&\left. +\frac{3 m_{N}^2+s_{1}}{\left(m_{N}^2-s\right) \left(m_{N}^2-s_{1}\right)}\right)-(s_1 \leftrightarrow s_2,t_1 \leftrightarrow t_2)\ ,\notag \\
\mathcal{M}^1_{10}=&\frac{i e g^2}{4 F^2} \left(\frac{2 m_{N}^2}{\left(m_{N}^2-s_{1}\right) \left(-M^2+s-s_{1}+t_{2}\right)}+\frac{3 m_{N}^2+s_{1}}{\left(m_{N}^2-s\right) \left(m_{N}^2-s_{1}\right)}\right)-(s_1 \leftrightarrow s_2,t_1 \leftrightarrow t_2)\ , \notag\\
\mathcal{M}^1_{11}=&-\frac{4 i e g^2 m_{N}^2}{F^2 \left(-2 M^2-2 m_{N}^2+2 s+t_{1}+t_{2}\right)} \frac{1}{\left(m_{N}^2-s_{1}\right) \left(-M^2+s-s_{1}+t_{2}\right)}+(s_1 \leftrightarrow s_2,t_1 \leftrightarrow t_2)\ .
\end{align}
\item Amplitudes $\mathcal{M}^2_j$:
\begin{align} \mathcal{M}^2_{1}=&\mathcal{M}^2_{3}=\mathcal{M}^2_{8}=\mathcal{M}^2_{9}=\mathcal{M}^2_{12}=0,\mathcal{M}^2_{2}=\mathcal{M}^1_{2},\ \mathcal{M}^2_{5}=\mathcal{M}^1_{5}
\ ,\notag\\
\mathcal{M}^2_{4}=&\frac{i e g^2}{8 F^2} \left(\frac{3 m_{N}^2+s_{1}}{\left(m_{N}^2-s\right) \left(m_{N}^2-s_{1}\right)}+\frac{M^2+4 m_{N}^2-s+s_{2}-t_{1}}{\left(M^2-s+s_{2}-t_{1}\right) \left(-2 M^2-m_{N}^2+s+t_{1}+t_{2}\right)} \right. \notag \\
&\left. -\frac{4 m_{N}^2}{\left(m_{N}^2-s_{1}\right) \left(-M^2+s-s_{1}+t_{2}\right)} \right)-(s_1 \leftrightarrow s_2,t_1 \leftrightarrow t_2)\ ,\notag\\
\mathcal{M}^2_{6}=&-\frac{i e}{4 F^2} \left(-\frac{g^2 \left(M^2+4 m_{N}^2-s+s_{1}-t_{2}\right)}{\left(M^2-t_{1}\right) \left(M^2-s+s_{1}-t_{2}\right)}+\frac{2 g^2 m_{N}^2}{\left(m_{N}^2-s_{1}\right) \left(-M^2+s-s_{1}+t_{2}\right)}\right.\notag\\
&\left.+\frac{g^2 \left(3 m_{N}^2+s_{1}\right)}{\left(M^2-t_{2}\right) \left(m_{N}^2-s_{1}\right)}+\frac{2}{M^2-t_{1}}\right)+(s_1 \leftrightarrow s_2,t_1 \leftrightarrow t_2)\ , \notag\\
\mathcal{M}^2_{7}=&-\frac{i e}{F^2 \left(-2 M^2-2 m_{N}^2+2 s+t_{1}+t_{2}\right)} \left(\frac{2 g^2 \left(M^2+4 m_{N}^2-s+s_{1}-t_{2}\right)}{\left(M^2-t_{1}\right) \left(M^2-s+s_{1}-t_{2}\right)}\right.\notag\\
&+\frac{g^2 \left(M^2+4 m_{N}^2-s+s_{1}-t_{2}\right)}{\left(M^2-s+s_{1}-t_{2}\right) \left(-2 M^2-m_{N}^2+s+t_{1}+t_{2}\right)}+\frac{2 g^2 \left(3 m_{N}^2+s_{1}\right)}{\left(M^2-t_{2}\right) \left(m_{N}^2-s_{1}\right)}\notag\\
&\left. +\frac{g^2 \left(3 m_{N}^2+s_{1}\right)}{\left(m_{N}^2-s\right) \left(m_{N}^2-s_{1}\right)}-\frac{4}{M^2-t_{1}}\right)-(s_1 \leftrightarrow s_2,t_1 \leftrightarrow t_2)\ , \notag\\
\mathcal{M}^2_{10}=&-\frac{i e}{4 F^2} \left(\frac{2}{M^2-t_{1}}+\frac{4 g^2 m_{N}^2}{\left(M^2-t_{2}\right) \left(M^2-s+s_{2}-t_{1}\right)}+\frac{2 g^2 m_{N}^2}{\left(m_{N}^2-s_{2}\right) \left(M^2-s+s_{2}-t_{1}\right)}\right.\notag\\
&\left. +\frac{4 g^2 m_{N}^2}{\left(M^2-t_{1}\right) \left(m_{N}^2-s_{2}\right)}-\frac{2 g^2}{M^2-t_{1}}-\frac{4 g^2 m_{N}^2}{\left(m_{N}^2-s\right) \left(m_{N}^2-s_{1}\right)}\right)-(s_1 \leftrightarrow s_2,t_1 \leftrightarrow t_2)\ ,\notag\\
\mathcal{M}^2_{11}=&\frac{2 i e}{F^2 \left(-2 M^2-2 m_{N}^2+2 s+t_{1}+t_{2}\right)} \left(\frac{2}{M^2-t_{1}}-\frac{g^2 \left(M^2+4 m_{N}^2-s+s_{1}-t_{2}\right)}{\left(M^2-t_{1}\right)\left(M^2-s+s_{1}-t_{2}\right)} \right.\notag\\
&\left.+\frac{2 g^2 m_{N}^2}{\left(m_{N}^2-s_{1}\right)\left(-M^2+s-s_{1}+t_{2}\right)}+\frac{g^2 \left(3 m_{N}^2+s_{1}\right)}{\left(M^2-t_{2}\right)\left(m_{N}^2-s_{1}\right)}\right)+(s_1 \leftrightarrow s_2,t_1 \leftrightarrow t_2)\ .
\end{align}
\item Amplitudes $\mathcal{M}^3_j$:
\begin{align}
\mathcal{M}^3_{1}=&\mathcal{M}^3_{2}=\mathcal{M}^3_{3}=\mathcal{M}^3_{5}=\mathcal{M}^3_{8}=\mathcal{M}^3_{9}=\mathcal{M}^3_{12}=0\ ,\notag\\
\mathcal{M}^3_{4}=&\frac{i e}{8 F^2} \left(-\frac{g^2 \left(M^2+4 m_N^2-s+s_1-t_2\right)}{\left(M^2-s+s_1-t_2\right) \left(-2 M^2-m_N^2+s+t_1+t_2\right)}+\frac{4 g^2 m_N^2}{\left(m_N^2-s_1\right) \left(-M^2+s-s_1+t_2\right)}\right.\notag\\
&-\frac{g^2 \left(M^2+4 m_N^2-s+s_2-t_1\right)}{\left(M^2-s+s_2-t_1\right) \left(-2 M^2-m_N^2+s+t_1+t_2\right)}+\frac{4 g^2 m_N^2}{\left(m_N^2-s_2\right) \left(M^2-s+s_2-t_1\right)}\notag\\
&\left.-\frac{g^2 \left(3 m_N^2+s_1\right)}{\left(m_N^2-s\right) \left(m_N^2-s_1\right)}-\frac{g^2 \left(3 m_N^2+s_2\right)}{\left(m_N^2-s\right) \left(m_N^2-s_2\right)}+\frac{2}{-2 M^2-m_N^2+s+t_1+t_2}-\frac{2}{m_N^2-s}\right)\ , \notag\\
\mathcal{M}^3_{6}=&\frac{i e}{2 F^2} \left(\frac{1}{M^2-t_2}+\frac{g^2 m_N^2}{\left(m_N^2-s_1\right) \left(-M^2+s-s_1+t_2\right)}-\frac{2 g^2m_N^2}{\left(M^2-t_2\right) \left(M^2-s+s_2-t_1\right)}\right.\notag\\
&\left.+\frac{g^2 m_N^2}{\left(m_N^2-s_2\right)\left(-M^2+s-s_2+t_1\right)}+\frac{g^2 (s_1+m_N^2)}{\left(M^2-t_2\right) \left(m_N^2-s_1\right)}\right)\ , \notag\\
\mathcal{M}^3_{7}=&\frac{i e}{F^2 \left(-2 M^2-2 m_N^2+2 s+t_1+t_2\right)} \left(-\frac{g^2 \left(M^2+4 m_N^2-s+s_1-t_2\right)}{\left(M^2-s+s_1-t_2\right) \left(-2 M^2-m_N^2+s+t_1+t_2\right)}\right.\notag\\
&-\frac{2 g^2 \left(M^2+4 m_N^2-s+s_2-t_1\right)}{\left(M^2-t_2\right) \left(M^2-s+s_2-t_1\right)}-\frac{g^2 \left(M^2+4 m_N^2-s+s_2-t_1\right)}{\left(M^2-s+s_2-t_1\right) \left(-2 M^2-m_N^2+s+t_1+t_2\right)}\notag\\
&+\frac{2 g^2 \left(3 m_N^2+s_1\right)}{\left(M^2-t_2\right) \left(m_N^2-s_1\right)}+\frac{g^2 \left(3 m_N^2+s_1\right)}{\left(m_N^2-s\right) \left(m_N^2-s_1\right)}+\frac{g^2 \left(3 m_N^2+s_2\right)}{\left(m_N^2-s\right) \left(m_N^2-s_2\right)}\notag\\
&\left.+\frac{2}{-2 M^2-m_N^2+s+t_1+t_2}+\frac{4}{M^2-t_2}+\frac{2}{m_N^2-s}\right)\ ,\notag\\
\mathcal{M}^3_{10}=&\frac{i e}{4 F^2} \left(\frac{g^2 \left(M^2+4 m_N^2-s+s_1-t_2\right)}{\left(M^2-t_1\right) \left(M^2-s+s_1-t_2\right)}-\frac{4g^2 m_N^2}{\left(M^2-t_1\right) \left(M^2-s+s_1-t_2\right)}-\frac{g^2}{M^2-t_1}\right.\notag\\
&-\frac{2 g^2 m_N^2}{\left(m_N^2-s_1\right)\left(M^2-s+s_1-t_2\right)}+\frac{g^2 \left(M^2+4 m_N^2-s+s_2-t_1\right)}{\left(M^2-t_2\right)\left(M^2-s+s_2-t_1\right)}+\frac{g^2}{M^2-t_2}\notag\\
&+\frac{2 g^2 m_N^2}{\left(m_N^2-s_2\right) \left(M^2-s+s_2-t_1\right)}-\frac{4g^2 m_N^2}{\left(M^2-t_2\right)\left(m_N^2-s_1\right)}-\frac{4 g^2m_N^2}{\left(m_N^2-s\right) \left(m_N^2-s_1\right)}\notag\\
&\left.-\frac{4 g^2 m_N^2}{\left(m_N^2-s\right)\left(m_N^2-s_2\right)}+\frac{2 g^2}{m_N^2-s}+\frac{2}{t_2-M^2}-\frac{2}{m_N^2-s}\right)\ ,\notag\\
\mathcal{M}^3_{11}=&-\frac{4 i e}{F^2 \left(-2 M^2-2 m_N^2+2s+t_1+t_2\right)} \left(\frac{g^2 m_N^2}{\left(m_N^2-s_1\right) \left(-M^2+s-s_1+t_2\right)}\right.\notag\\
&-\frac{2 g^2m_N^2}{\left(M^2-t_2\right) \left(M^2-s+s_2-t_1\right)}+\frac{g^2 m_N^2}{\left(m_N^2-s_2\right)\left(-M^2+s-s_2+t_1\right)}+\frac{g^2 s_1}{\left(M^2-t_2\right) \left(m_N^2-s_1\right)}\notag\\
&\left.+\frac{g^2m_N^2}{\left(M^2-t_2\right) \left(m_N^2-s_1\right)}+\frac{1}{M^2-t_2}\right)\ .
\end{align}
\item Amplitudes $\mathcal{M}^4_j$:
\begin{align}  \mathcal{M}^4_{1}=&\mathcal{M}^4_{2}=\mathcal{M}^4_{3}=\mathcal{M}^4_{5}=\mathcal{M}^4_{8}=\mathcal{M}^4_{9}=\mathcal{M}^4_{12}=0\ ,\notag\\
\mathcal{M}^4_{4}=&-\mathcal{M}^3_{4}(s_1 \leftrightarrow s_2,t_1 \leftrightarrow t_2),\ \mathcal{M}^4_{6}=\mathcal{M}^3_{6}(s_1 \leftrightarrow s_2,t_1 \leftrightarrow t_2),\ \mathcal{M}^4_{7}=-\mathcal{M}^3_{7}(s_1 \leftrightarrow s_2,t_1 \leftrightarrow t_2)\ ,\notag\\
\mathcal{M}^4_{10}=&-\mathcal{M}^3_{10}(s_1 \leftrightarrow s_2,t_1 \leftrightarrow t_2),\ \mathcal{M}^4_{11}=\mathcal{M}^3_{11}(s_1 \leftrightarrow s_2,t_1 \leftrightarrow t_2)\ . 
\end{align}
\item Amplitudes $\mathcal{M}^5_j$:
\begin{align}
\mathcal{M}^5_{1}=&\mathcal{M}^5_{2}=\mathcal{M}^5_{3}=\mathcal{M}^5_{5}=\mathcal{M}^5_{8}=\mathcal{M}^5_{9}=\mathcal{M}^5_{12}=0\ ,\notag\\
\mathcal{M}^5_{4}=&\frac{e}{8 F^2} \left(\frac{g^2 \left(M^2+4 m_N^2-s+s_1-t_2\right)}{\left(M^2-s+s_1-t_2\right) \left(-2 M^2-m_N^2+s+t_1+t_2\right)}+\frac{4 g^2 m_N^2}{\left(m_N^2-s_1\right) \left(M^2-s+s_1-t_2\right)}\right.\notag\\
&+\frac{g^2 \left(M^2+4 m_N^2-s+s_2-t_1\right)}{\left(M^2-s+s_2-t_1\right) \left(-2 M^2-m_N^2+s+t_1+t_2\right)}+\frac{4 g^2 m_N^2}{\left(m_N^2-s_2\right) \left(M^2-s+s_2-t_1\right)}\notag\\
&\left.-\frac{g^2 \left(3 m_N^2+s_1\right)}{\left(m_N^2-s\right) \left(m_N^2-s_1\right)}-\frac{g^2 \left(3 m_N^2+s_2\right)}{\left(m_N^2-s\right) \left(m_N^2-s_2\right)}-\frac{2}{-2 M^2-m_N^2+s+t_1+t_2}-\frac{2}{m_N^2-s}\right)\ ,\notag\\
\mathcal{M}^5_{6}=&\frac{e g^2 m_N^2}{2 F^2 \left(m_N^2-s_1\right) \left(M^2-s+s_1-t_2\right)}-(s_1 \leftrightarrow s_2,t_1 \leftrightarrow t_2)\ , \notag\\
\mathcal{M}^5_{7}=&\frac{e}{F^2 \left(-2 M^2-2 m_N^2+2 s+t_1+t_2\right)} \left(\frac{g^2 \left(M^2+4 m_N^2-s+s_1-t_2\right)}{\left(M^2-s+s_1-t_2\right) \left(-2 M^2-m_N^2+s+t_1+t_2\right)}\right.\notag\\
&+\frac{g^2 \left(M^2+4 m_N^2-s+s_2-t_1\right)}{\left(M^2-s+s_2-t_1\right) \left(-2 M^2-m_N^2+s+t_1+t_2\right)}+\frac{g^2 \left(3 m_N^2+s_1\right)}{\left(m_N^2-s\right) \left(m_N^2-s_1\right)}\notag\\
&\left.+\frac{g^2 \left(3 m_N^2+s_2\right)}{\left(m_N^2-s\right) \left(m_N^2-s_2\right)}-\frac{2}{-2 M^2-m_N^2+s+t_1+t_2}+\frac{2}{m_N^2-s}\right)\ ,\notag\\
\mathcal{M}^5_{10}=&\frac{e}{4 F^2} \left(\frac{2 g^2 m_N^2}{\left(m_N^2-s_1\right) \left(M^2-s+s_1-t_2\right)}+\frac{2 g^2m_N^2}{\left(m_N^2-s_2\right) \left(M^2-s+s_2-t_1\right)}\right.\notag\\
&\left.-\frac{4 g^2 m_N^2}{\left(m_N^2-s\right)\left(m_N^2-s_1\right)}-\frac{4 g^2 m_N^2}{\left(m_N^2-s\right) \left(m_N^2-s_2\right)}-\frac{2\left(1-g^2\right)}{m_N^2-s}\right)\ ,\notag\\
\mathcal{M}^5_{11}=&\frac{4 e g^2 m_N^2 \left(M^2 (s_2-s_1)+m_N^2 (s_1-s_2+t_1-t_2)+s s_1-s s_2-s_1^2+s_1 t_2+s_2^2-s_2 t_1\right)}{F^2 \left(m_N^2-s_1\right) \left(m_N^2-s_2\right) \left(M^2-s+s_1-t_2\right) \left(M^2-s+s_2-t_1\right) \left(-2 M^2-2 m_N^2+2 s+t_1+t_2\right)}\ .
\end{align}
\item Amplitudes $\mathcal{M}^6_j$:
\begin{align}
\mathcal{M}^6_{1}=&\mathcal{M}^6_{2}=\mathcal{M}^6_{3}=\mathcal{M}^6_{8}=\mathcal{M}^6_{12}=0\ ,\notag\\
\mathcal{M}^6_{4}=&\frac{e}{8 F^2} \left(\frac{g^2 \left(M^2+4 m_N^2-s+s_1-t_2\right)}{\left(M^2-s+s_1-t_2\right) \left(-2 M^2-m_N^2+s+t_1+t_2\right)}+\frac{4 g^2 m_N^2}{\left(m_N^2-s_1\right) \left(-M^2+s-s_1+t_2\right)}\right.\notag\\
&+\frac{g^2 \left(M^2+4 m_N^2-s+s_2-t_1\right)}{\left(M^2-s+s_2-t_1\right) \left(-2 M^2-m_N^2+s+t_1+t_2\right)}-\frac{4 g^2 m_N^2}{\left(m_N^2-s_2\right) \left(M^2-s+s_2-t_1\right)}\notag\\
&\left.-\frac{g^2 \left(3 m_N^2+s_1\right)}{\left(m_N^2-s\right) \left(m_N^2-s_1\right)}-\frac{g^2 \left(3 m_N^2+s_2\right)}{\left(m_N^2-s\right) \left(m_N^2-s_2\right)}-\frac{2}{-2 M^2-m_N^2+s+t_1+t_2}-\frac{2}{m_N^2-s}\right)\ ,\notag\\
\mathcal{M}^6_{5}=&-\frac{8 e g^2 m_N (t_1-t_2)}{F^2 \left(M^2-t_1\right) \left(M^2-t_2\right) \left(-2 M^2-2 m_N^2+2 s+t_1+t_2\right)}\ ,\notag\\
\mathcal{M}^6_{6}=&\frac{e g^2 m_N^2}{4 F^2} \left(\frac{2}{\left(m_N^2-s_1\right) \left(-M^2+s-s_1+t_2\right)}+\frac{4}{\left(M^2-t_2\right)\left(m_N^2-s_1\right)}\right.\notag\\
&\left.+\frac{4}{\left(M^2-t_2\right) \left(M^2-s+s_2-t_1\right)}\right)-(s_1 \leftrightarrow s_2,t_1 \leftrightarrow t_2)\ ,\notag\\
\mathcal{M}^6_{7}=&\frac{e }{F^2 \left(-2 M^2-2 m_N^2+2 s+t_1+t_2\right)}\left(\frac{2 g^2 \left(M^2+4 m_N^2-s+s_1-t_2\right)}{\left(M^2-t_1\right) \left(M^2-s+s_1-t_2\right)}+\frac{2 g^2 \left(3 m_N^2+s_2\right)}{\left(M^2-t_1\right) \left(m_N^2-s_2\right)}\right.\notag\\
&+\frac{g^2 \left(M^2+4 m_N^2-s+s_1-t_2\right)}{\left(M^2-s+s_1-t_2\right) \left(-2 M^2-m_N^2+s+t_1+t_2\right)}+\frac{2 g^2 \left(M^2+4 m_N^2-s+s_2-t_1\right)}{\left(M^2-t_2\right) \left(M^2-s+s_2-t_1\right)}\notag\\
&+\frac{g^2 \left(M^2+4 m_N^2-s+s_2-t_1\right)}{\left(M^2-s+s_2-t_1\right) \left(-2 M^2-m_N^2+s+t_1+t_2\right)}+\frac{2 g^2 \left(3 m_N^2+s_1\right)}{\left(M^2-t_2\right) \left(m_N^2-s_1\right)}\notag\\
&\left.+\frac{g^2 \left(3 m_N^2+s_1\right)}{\left(m_N^2-s\right) \left(m_N^2-s_1\right)}+\frac{g^2 \left(3 m_N^2+s_2\right)}{\left(m_N^2-s\right) \left(m_N^2-s_2\right)}-\frac{2}{-2 M^2-m_N^2+s+t_1+t_2}+\frac{2}{m_N^2-s}\right)\ ,\notag\\
\mathcal{M}^6_{9}=&-\frac{8 e g^2 m_N \left(2 M^2-t_1-t_2\right)}{F^2 \left(M^2-t_1\right) \left(M^2-t_2\right) \left(-2 M^2-2 m_N^2+2 s+t_1+t_2\right)}\ ,\notag\\
\mathcal{M}^6_{10}=&\frac{e}{4 F^2} \left(\frac{4 g^2 m_N^2 \left(-M^2-m_N^2+s+s_1-s_2+t_1\right)}{\left(M^2-t_2\right)\left(m_N^2-s_1\right) \left(M^2-s+s_2-t_1\right)}+\frac{2 g^2 m_N^2 \left(M^2+2 m_N^2-2s_1-t_1\right)}{\left(M^2-t_1\right) \left(m_N^2-s_1\right) \left(-M^2+s-s_1+t_2\right)}\right.\notag\\
&+\frac{2 g^2m_N^2}{\left(m_N^2-s_2\right) \left(-M^2+s-s_2+t_1\right)}+\frac{g^2 \left(3m_N^2+s_2\right)}{\left(t_1-M^2\right) \left(m_N^2-s_2\right)}+\frac{g^2}{t_1-M^2}-\frac{g^2 \left(3m_N^2+s_1\right)}{\left(m_N^2-s\right) \left(m_N^2-s_1\right)}\notag\\
&\left.-\frac{g^2 \left(3m_N^2+s_2\right)}{\left(m_N^2-s\right) \left(m_N^2-s_2\right)}+\frac{2}{s-m_N^2}\right)\ ,\notag\\
\mathcal{M}^6_{11}=&\frac{2 e g^2}{F^2 \left(-2 M^2-2 m_N^2+2s+t_1+t_2\right)} \left(\frac{M^2+4 m_N^2-s+s_1-t_2}{\left(M^2-t_1\right)\left(M^2-s+s_1-t_2\right)}\right. \notag\\
&\left.+\frac{1}{\left(m_N^2-s_1\right) \left(M^2-s+s_1-t_2\right)}-\frac{3m_N^2+s_1}{\left(M^2-t_2\right) \left(m_N^2-s_1\right)}\right)-(s_1 \leftrightarrow s_2,t_1 \leftrightarrow t_2)\ .
\end{align}
\end{itemize}
\subsection{At $\mathcal{O}(p^2)$\label{sec.p2.exp}}
\begin{itemize}
\item Amplitudes $\mathcal{M}^1_j$:
\begin{align}
\mathcal{M}^1_{1}=&-\frac{i e g^2 (c_6+2 c_7)}{8 F^2}\left(\frac{m_N^2+s-2s_1-t_1+t_2}{\left(M^2-s+s_1-t_2\right) \left(-2M^2-m_N^2+s+t_1+t_2\right)}\right.\notag\\
&\left.+\frac{-M^2-m_N^2+s+t_2}{\left(m_N^2-s_1\right)\left(M^2-s+s_1-t_2\right)}+\frac{m_N^2+s-2s_1}{\left(m_N^2-s\right)\left(m_N^2-s_1\right)}\right)+(s_1 \leftrightarrow s_2,t_1 \leftrightarrow t_2)\ ,\notag\\
\mathcal{M}^1_{2}=&-\frac{2 i e M^2 c_1}{F^2 \left(-2 M^2-m_N^2+s+t_1+t_2\right)}-\frac{2 i e M^2 c_1}{F^2 \left(m_N^2-s\right)}-\frac{i e c_3 \left(m_N^2+s-s_1-s_2\right)}{2 F^2 \left(-2 M^2-m_N^2+s+t_1+t_2\right)}\notag\\
&-\frac{i e c_3 \left(m_N^2+s-s_1-s_2\right)}{2 F^2 \left(m_N^2-s\right)}+\frac{i e (c_6+2c_7)g^2 m_N}{2 F^2 \left(-2 M^2-m_N^2+s+t_1+t_2\right)}+\frac{i e (c_6+2c_7)g^2 \left(m_N^2+s\right)}{4F^2 m_N \left(m_N^2-s\right)}\notag\\
&+\frac{i e c_2 \left(-M^4+s \left(-2 M^2-m_N^2+s_1\right)+s_2\left(M^2+m_N^2-s_1\right)+M^2 s_1\right)}{4 F^2 m_N^2 \left(m_N^2-s\right)}-\frac{i e c_2 t_1}{4 F^2m_N^2}\notag\\
&-\frac{i e c_2 (s-s_2+t_1) \left(2 M^2+m_N^2-s_1-t_1\right)}{4 F^2 m_N^2 \left(-2M^2-m_N^2+s+t_1+t_2\right)}\ ,\notag\\
\mathcal{M}^1_{3}=&\frac{i e (c_6+2c_7)g^2 m_N \left(2 M^2+m_N^2-s-s_1+s_2-2 t_1\right)}{2 F^2 \left(M^2+m_N^2-s_1-t_1\right)\left(M^2-s+s_2-t_1\right) \left(-2 M^2-m_N^2+s+t_1+t_2\right)}\notag\\
&+\frac{i e (c_6+2c_7)g^2 m_N \left(M^2-t_1\right)}{2F^2 \left(m_N^2-s_1\right) \left(M^2+m_N^2-s_1-t_1\right) \left(-M^2+s-s_1+t_2\right)}\notag\\
&-\frac{i e (c_6+2c_7)g^2m_N}{2 F^2 \left(m_N^2-s_2\right) \left(-M^2+s-s_2+t_1\right)}+\frac{i e(c_6+2c_7) g^2 m_N (s_1-s_2)}{2F^2 \left(m_N^2-s\right) \left(m_N^2-s_1\right) \left(m_N^2-s_2\right)}\ ,\notag\\
\mathcal{M}^1_{4}=&\frac{i e g^2 (c_6+2 c_7) }{8 F^2}\left(\frac{3 M^2+5 m_N^2-2 s+s_2-2 t_1-t_2}{\left(M^2-s+s_2-t_1\right) \left(-2M^2-m_N^2+s+t_1+t_2\right)}+\frac{M^2+3 m_N^2-s+2 s_2-t_1}{\left(m_N^2-s_2\right)\left(M^2-s+s_2-t_1\right)}\right.\notag\\
&\left.+\frac{2 m_N^2+s+s_1}{\left(m_N^2-s\right) \left(m_N^2-s_1\right)}\right)-(s_1 \leftrightarrow s_2,t_1 \leftrightarrow t_2)\ ,\notag\\
\mathcal{M}^1_{5}=&\frac{16 i e c_1 M^2}{F^2 \left(m_N^2-s\right) \left(-2 M^2-m_N^2+s+t_1+t_2\right)}+\frac{2 i ec_2 (s-s_2+t_1) \left(2 M^2+m_N^2-s_1-t_1\right)}{F^2 m_N^2 \left(m_N^2-s\right) \left(-2M^2-m_N^2+s+t_1+t_2\right)}\notag\\
&+\frac{2 i ec_2 \left(M^2+m_N^2-s-t_1\right)\left(M^2+m_N^2-s-s_1+s_2-t_1\right)}{F^2 m_N^2 \left(m_N^2-s\right) \left(-2 M^2-2 m_N^2+2s+t_1+t_2\right)}+\frac{i e c_2}{F^2 m_N^2}\notag\\
&+\frac{4 i e c_3\left(m_N^2+s-s_1-s_2\right)}{F^2 \left(m_N^2-s\right) \left(-2 M^2-m_N^2+s+t_1+t_2\right)}\notag\\
&+\frac{i e (c_6+2c_7)g^2 m_N \left(2 M^2-2 s+s_1+s_2-t_1-t_2\right)}{F^2 \left(M^2-s+s_1-t_2\right)\left(M^2-s+s_2-t_1\right) \left(2 M^2+2 m_N^2-2 s-t_1-t_2\right)}\notag\\
&+\frac{i e(c_6+2c_7) g^2 m_N\left(\left(m_N^2-s\right) \left(2 m_N^2-s_1-s_2\right)+(s_1-s_2) (t_1-t_2)\right)}{F^2\left(m_N^2-s\right) \left(m_N^2-s_1\right) \left(m_N^2-s_2\right) \left(2 M^2+2 m_N^2-2 s-t_1-t_2\right)}\ ,\notag\\
\mathcal{M}^1_{6}=&\frac{i e g^2}{8 F^2} (c_6+2 c_7) \left(-\frac{m_N^2+s-2 s_1-t_1+t_2}{\left(M^2-s+s_1-t_2\right) \left(-2M^2-m_N^2+s+t_1+t_2\right)}\right.\notag\\
&\left.+\frac{-M^2-3 m_N^2+s-2 s_1+t_2}{\left(m_N^2-s_1\right)\left(M^2-s+s_1-t_2\right)}+\frac{m_N^2+s-2 s_1}{\left(m_N^2-s\right) \left(m_N^2-s_1\right)}\right)+(s_1 \leftrightarrow s_2,t_1 \leftrightarrow t_2)\ ,\notag\\
\mathcal{M}^1_{7}=&0\ ,\notag\\
\mathcal{M}^1_{8}=&-\frac{i e g^2 (c_6+2 c_7)}{8F^2 m_N} \left(\frac{M^2+4 m_N^2-s+s_1-t_2}{\left(M^2-s+s_1-t_2\right) \left(-2M^2-m_N^2+s+t_1+t_2\right)}+\frac{3 m_N^2+s_1}{\left(m_N^2-s\right) \left(m_N^2-s_1\right)}\right)\notag\\
&-(s_1 \leftrightarrow s_2,t_1 \leftrightarrow t_2)\ ,\notag\\
\mathcal{M}^1_{9}=&\frac{i e g^2 (c_6+2 c_7)}{2 F^2 \left(-2 M^2-2 m_N^2+2s+t_1+t_2\right)} \left(\frac{2 m_N \left(-2 M^2-m_N^2+s+t_1+t_2\right)}{\left(m_N^2-s_1\right)\left(M^2-s+s_1-t_2\right)}+\frac{3 m_N^2+s_1}{m_N^3-m_N s_1}\right)\notag\\
&-\frac{i c_2 e (2 s_1+t_1)}{F^2 m_N^2 \left(-2 M^2-2 m_N^2+2 s+t_1+t_2\right)}-(s_1 \leftrightarrow s_2,t_1 \leftrightarrow t_2)\ ,\notag\\
\mathcal{M}^1_{10}=&\frac{i e(c_6+2 c_7) g^2 \left(-m_N^2+s+2 s_1\right)}{2 F^2 \left(m_N^2-s\right) \left(m_N^2-s_1\right)}-\frac{i e (c_6+2 c_7)g^2m_N^2}{2 F^2 \left(m_N^2-s_1\right) \left(M^2-s+s_1-t_2\right)}-(s_1 \leftrightarrow s_2,t_1 \leftrightarrow t_2)\ ,\notag\\
\mathcal{M}^1_{11}=&0 \ ,\notag\\
\mathcal{M}^1_{12}=&\frac{i e g^2 m_N (c_6+2 c_7)}{2 F^2 \left(m_N^2-s_1\right) \left(M^2-s+s_1-t_2\right)}+(s_1 \leftrightarrow s_2,t_1 \leftrightarrow t_2)\ .
\end{align}
\item Amplitudes $\mathcal{M}^2_j$:
\begin{align}
\mathcal{M}^2_{1}=&\frac{i c_6 e g^2}{2 F^2 \left(-2 M^2-m_N^2+s+t_1+t_2\right)}-\frac{i c_6 e g^2}{2 F^2\left(m_N^2-s\right)}\ ,\notag\\
\mathcal{M}^2_{2}=&-\frac{2 i ec_1 M^2}{F^2 \left(-2 M^2-m_N^2+s+t_1+t_2\right)}-\frac{2 i ec_1 M^2}{F^2 \left(m_N^2-s\right)}+\frac{i ec_2}{4 F^2} \left(\frac{2 M^2+s-s_1-s_2}{m_N^2}\right.\notag\\
&+\frac{s^2-s s_1-s s_2+s t_1+s t_2+s_1 s_2-s_1 t_1-s_2 t_2+t_1 t_2}{m_N^2 \left(2M^2+m_N^2-s-t_1-t_2\right)}+\frac{-2M^2-s+s_1+s_2}{m_N^2-s}\notag\\
&\left.+\frac{-M^4+M^2 s_1+M^2 s_2-s_1 s_2}{m_N^2 \left(m_N^2-s\right)}\right)+\frac{i e c_3\left(2 M^2-t_1-t_2\right) \left(m_N^2+s-s_1-s_2\right)}{2 F^2 \left(m_N^2-s\right) \left(-2M^2-m_N^2+s+t_1+t_2\right)}+\frac{i ec_4}{F^2}\notag\\
&-\frac{i e c_6g^2 m_N}{2 F^2 \left(2 M^2+m_N^2-s-t_1-t_2\right)}+\frac{i e c_6g^2 m_N}{4 F^2\left(M^2-s+s_1-t_2\right)}+\frac{i e c_6g^2 m_N}{4 F^2 \left(M^2-s+s_2-t_1\right)}\notag\\
&+\frac{i ec_6 g^2m_N}{2 F^2 \left(m_N^2-s\right)}-\frac{i e g^2 m_N}{4 F^2 \left(m_N^2-s_1\right)}-\frac{i ec_6g^2 m_N}{4 F^2 \left(m_N^2-s_2\right)}-\frac{i ec_6}{4 F^2 m_N}\ ,\notag\\
\mathcal{M}^2_{3}=&\mathcal{M}^2_{4}= \frac{i e c_6 m_N g^2}{2F^2} \left(\frac{1}{\left(M^2-s+s_1-t_2\right) \left(2 M^2+m_N^2-s-t_1-t_2\right)}+\frac{1}{\left(m_N^2-s_1\right)\left(M^2-s+s_1-t_2\right)}\right.\notag\\
&\left.+\frac{1}{\left(m_N^2-s\right) \left(m_N^2-s_1\right)}\right)-(s_1 \leftrightarrow s_2,t_1 \leftrightarrow t_2)\ ,\notag\\
\mathcal{M}^2_{5}=&\frac{16 i e c_1 M^2}{F^2 \left(m_N^2-s\right) \left(-2 M^2-m_N^2+s+t_1+t_2\right)}+\frac{2 i e c_2 (s-s_2+t_1) \left(2 M^2+m_N^2-s_1-t_1\right)}{F^2 m_N^2 \left(m_N^2-s\right) \left(-2M^2-m_N^2+s+t_1+t_2\right)}\notag\\
&+\frac{2 i e c_2\left(M^2+m_N^2-s-t_1\right)\left(M^2+m_N^2-s-s_1+s_2-t_1\right)}{F^2 m_N^2 \left(m_N^2-s\right) \left(-2 M^2-2 m_N^2+2s+t_1+t_2\right)}+\frac{i e c_2 }{F^2 m_N^2}\notag\\
&+\frac{4 i e c_3\left(m_N^2+s-s_1-s_2\right)}{F^2 \left(m_N^2-s\right) \left(-2 M^2-m_N^2+s+t_1+t_2\right)}-\frac{4 i e c_4 \left(M^2+m_N^2-s-s_1+s_2-t_1\right)}{F^2 \left(t_2-M^2\right) \left(M^2+2 m_N^2-2s-t_1\right)}\notag\\
&+\frac{8 i ec_4 \left(M^2+m_N^2-s-t_1\right) \left(M^2+m_N^2-s-s_1+s_2-t_1\right)}{F^2\left(M^2-t_1\right) \left(M^2+2 m_N^2-2 s-t_1\right) \left(-2 M^2-2 m_N^2+2 s+t_1+t_2\right)}+\frac{4 i ec_4}{F^2\left(M^2-t_1\right)}\notag\\
&-\frac{i e c_6 g^2 m_N}{F^2 \left(m_N^2-s_1\right) \left(2 M^2+2 m_N^2-2 s-t_1-t_2\right)}-\frac{i ec_6 g^2 m_N}{F^2 \left(m_N^2-s_2\right) \left(2 M^2+2 m_N^2-2s-t_1-t_2\right)}\notag\\
&-\frac{i e c_6 g^2m_N}{F^2 \left(M^2-s+s_1-t_2\right) \left(2 M^2+2 m_N^2-2 s-t_1-t_2\right)}\notag\\
&+\frac{i c_6 e g^2 m_N\left(t_1- t_2\right)}{F^2 \left(m_N^2-s\right) \left(m_N^2-s_1\right) \left(2 M^2+2 m_N^2-2s-t_1-t_2\right)}\notag\\
&-\frac{i e c_6 g^2 m_N}{F^2 \left(M^2-s+s_2-t_1\right) \left(2 M^2+2 m_N^2-2s-t_1-t_2\right)}\notag\\
&-\frac{i c_6 e g^2 m_N \left(t_1-t_2\right)}{F^2\left(m_N^2-s\right) \left(m_N^2-s_2\right) \left(2 M^2+2 m_N^2-2 s-t_1-t_2\right)}\ ,\notag\\
\mathcal{M}^2_{6}=&\frac{2 i e c_4m_N}{F^2 \left(t_2-M^2\right)}-\frac{2 i e c_4m_N}{F^2 \left(M^2-t_1\right)}+\frac{i e c_6g^2 s_1}{2 F^2 \left(m_N^2-s_1\right) \left(M^2-s+s_1-t_2\right)}+\frac{i e c_6 g^2}{2 F^2 \left(m_N^2-s\right)}\notag\\
&+\frac{i e c_6g^2 s_2}{2F^2 \left(m_N^2-s_2\right) \left(M^2-s+s_2-t_1\right)}+\frac{i e c_6 g^2}{2 F^2 \left(M^2-s+s_1-t_2\right)}+\frac{i e c_6 g^2}{2F^2 \left(M^2-s+s_2-t_1\right)}\notag\\
&-\frac{i e c_6g^2}{2 F^2 \left(2M^2+m_N^2-s-t_1-t_2\right)}\ ,\notag\\
\mathcal{M}^2_{7}=&\frac{16 i c_4 e m_N (t_1-t_2)}{F^2 \left(M^2-t_1\right) \left(M^2-t_2\right) \left(-2 M^2-2 m_N^2+2 s+t_1+t_2\right)}\ ,\notag\\
\mathcal{M}^2_{8}=&\frac{i c_6 e g^2 m_N \left(2 M^2+m_N^2-s-s_1+s_2-2 t_1\right)}{2 F^2\left(M^2+m_N^2-s_1-t_1\right) \left(M^2-s+s_2-t_1\right) \left(-2 M^2-m_N^2+s+t_1+t_2\right)}\notag\\
&-\frac{ic_6 e g^2 m_N}{2 F^2 \left(M^2+m_N^2-s_1-t_1\right) \left(-M^2+s-s_1+t_2\right)}-\frac{i c_6e g^2 m_N (s_1-s_2)}{2 F^2 \left(m_N^2-s\right) \left(m_N^2-s_1\right)\left(m_N^2-s_2\right)}\ ,\notag\\
\mathcal{M}^2_{9}=&\frac{2 i e c_2 \left(M^2+m_N^2-s-s_1+s_2-t_1\right)}{F^2 m_N^2 \left(-2 M^2-2 m_N^2+2s+t_1+t_2\right)}+\frac{i ec_2}{F^2 m_N^2}+\frac{4 i e c_4\left(M^2+m_N^2-s-s_1+s_2-t_1\right)}{F^2 \left(t_2-M^2\right) \left(M^2+2 m_N^2-2s-t_1\right)}\notag\\
&+\frac{8 i ec_4 \left(m_N^2-s\right) \left(M^2+m_N^2-s-s_1+s_2-t_1\right)}{F^2\left(M^2-t_1\right) \left(M^2+2 m_N^2-2 s-t_1\right) \left(-2 M^2-2 m_N^2+2 s+t_1+t_2\right)}+\frac{4 i ec_4}{F^2\left(M^2-t_1\right)}\notag\\
&-\frac{2 i ec_6 g^2 m_N}{F^2 \left(m_N^2-s_1\right) \left(2 M^2+2 m_N^2-2 s-t_1-t_2\right)}+\frac{i ec_6 g^2m_N}{F^2 \left(m_N^2-s_2\right) \left(M^2-s+s_2-t_1\right)}\notag\\
&+\frac{i e c_6g^2 m_N}{F^2 \left(M^2-s+s_1-t_2\right) \left(2 M^2+2 m_N^2-2 s-t_1-t_2\right)}-\frac{i e g^2 m_N}{F^2 \left(m_N^2-s_1\right)\left(M^2-s+s_1-t_2\right)}\notag\\
&-\frac{i \left(eg^2 m_N s-e g^2 m_N s_1\right)}{F^2 \left(m_N^2-s_1\right) \left(M^2-s+s_1-t_2\right)\left(2 M^2+2 m_N^2-2 s-t_1-t_2\right)}\notag\\
&+\frac{2 i e g^2 m_N}{F^2 \left(m_N^2-s_2\right) \left(2 M^2+2 m_N^2-2s-t_1-t_2\right)}\notag\\
&-\frac{i e g^2 m_N}{F^2 \left(M^2-s+s_2-t_1\right) \left(2 M^2+2 m_N^2-2s-t_1-t_2\right)}\notag\\
&+\frac{i \left(e g^2 m_N s-e g^2 m_N s_2\right)}{F^2\left(m_N^2-s_2\right) \left(M^2-s+s_2-t_1\right) \left(2 M^2+2 m_N^2-2 s-t_1-t_2\right)}\ ,\notag\\
\mathcal{M}^2_{10}=&-\frac{2 ic_4 e m_N}{F^2 \left(M^2-t_1\right)}+\frac{i ec_6 g^2 s_1}{2 F^2 \left(m_N^2-s_1\right) \left(M^2-s+s_1-t_2\right)}+\frac{i ec_6 g^2}{2 F^2\left(M^2-s+s_1-t_2\right)}\notag\\
&+\frac{i ec_6 g^2 s_1}{F^2 \left(m_N^2-s\right) \left(m_N^2-s_1\right)}-(s_1 \leftrightarrow s_2,t_1 \leftrightarrow t_2)\ ,\notag\\
\mathcal{M}^2_{11}=&\frac{16 i c_4 e m_N}{F^2 \left(M^2-t_1\right) \left(-2 M^2-2 m_N^2+2 s+t_1+t_2\right)}+(t_1 \leftrightarrow t_2)\ ,\notag\\
\mathcal{M}^2_{12}=&\frac{i c_6 e g^2 m_N}{2 F^2 \left(m_N^2-s_1\right) \left(-M^2+s-s_1+t_2\right)}+(s_1 \leftrightarrow s_2,t_1 \leftrightarrow t_2)\ .
\end{align}
\item Amplitudes $\mathcal{M}^3_j$:
\begin{align}
\mathcal{M}^3_{1}=&\frac{i c_6 e g^2}{4 F^2 \left(m_N^2-s_2\right)}-\frac{i c_6 e g^2}{4 F^2\left(-M^2+s-s_1+t_2\right)}\ ,\notag\\
\mathcal{M}^3_{2}=&\frac{i e c_4 \left(2 M^2+m_N^2-s-s_1+s_2-2 t_1\right)}{4 F^2 \left(-2 M^2-m_N^2+s+t_1+t_2\right)}-\frac{i e c_4\left(m_N^2-s-s_1+s_2\right)}{4 F^2 \left(m_N^2-s\right)}-\frac{i ec_6 g^2 m_N}{4 F^2 \left(M^2-s+s_2-t_1\right)}\notag\\
&+\frac{i e c_6g^2 m_N}{4 F^2\left(m_N^2-s_1\right)}-\frac{i ec_6\left( g^2-1\right)}{8 F^2 m_N}\ ,\notag\\
\mathcal{M}^3_{3}=&\frac{i c_6 e g^2 m_N}{2 F^2 \left(M^2-s+s_1-t_2\right) \left(2 M^2+m_N^2-s-t_1-t_2\right)}-\frac{ic_6 e g^2 m_N}{2 F^2 \left(m_N^2-s_1\right) \left(M^2-s+s_1-t_2\right)}\notag\\
&+\frac{i c_6 e g^2m_N}{2 F^2 \left(M^2-s+s_2-t_1\right) \left(2 M^2+m_N^2-s-t_1-t_2\right)}+\frac{i c_6 e g^2m_N}{2 F^2 \left(m_N^2-s_2\right) \left(M^2-s+s_2-t_1\right)}\notag\\
&+\frac{i c_6 e \left(g^2-1\right)}{4F^2 m_N \left(2 M^2+m_N^2-s-t_1-t_2\right)}-\frac{i c_6 e g^2 m_N}{2 F^2\left(m_N^2-s\right) \left(m_N^2-s_1\right)}\notag\\
&-\frac{i c_6 e g^2 m_N}{2 F^2 \left(m_N^2-s\right)\left(m_N^2-s_2\right)}+\frac{i c_6 e \left(g^2-1\right)}{4 F^2 m_N \left(m_N^2-s\right)}\ ,\notag\\
\mathcal{M}^3_{4}=&\frac{i ec_4 m_N}{F^2 \left(-2 M^2-m_N^2+s+t_1+t_2\right)}-\frac{i e c_4m_N}{F^2 \left(m_N^2-s\right)}+\frac{i c_6 eg^2}{4 F^2 \left(m_N^2-s_2\right)}\notag\\
&-\frac{i c_6 e g^2 (s-2 s_1-t_1+t_2)}{2 F^2 \left(M^2-s+s_1-t_2\right) \left(2M^2+m_N^2-s-t_1-t_2\right)}-\frac{i c_6 e g^2 s_1}{2 F^2 \left(m_N^2-s_1\right)\left(M^2-s+s_1-t_2\right)}\notag\\
&-\frac{i c_6 e g^2 (s-2 s_2+t_1-t_2)}{2 F^2\left(M^2-s+s_2-t_1\right) \left(2 M^2+m_N^2-s-t_1-t_2\right)}+\frac{i c_6 e g^2 s_2}{2 F^2\left(m_N^2-s_2\right) \left(M^2-s+s_2-t_1\right)}\notag\\
&-\frac{i c_6 e \left(7 g^2+1\right)}{4 F^2 \left(2M^2+m_N^2-s-t_1-t_2\right)}+\frac{i c_6 e g^2}{4 F^2 \left(M^2-s+s_1-t_2\right)}+\frac{i c_6 eg^2}{F^2 \left(M^2-s+s_2-t_1\right)}\notag\\
&-\frac{i c_6 e g^2 s_1}{2 F^2 \left(m_N^2-s\right)\left(m_N^2-s_1\right)}-\frac{i c_6 e g^2 s_2}{2 F^2 \left(m_N^2-s\right)\left(m_N^2-s_2\right)}-\frac{i c_6 e \left(3 g^2+1\right)}{4 F^2 \left(m_N^2-s\right)}\ ,\notag\\
\mathcal{M}^3_{5}=&-\frac{2 i e c_4\left(2 M^2+m_N^2-s-s_1+s_2-2 t_1\right)}{F^2 \left(m_N^2-s\right) \left(-2M^2-m_N^2+s+t_1+t_2\right)}+\frac{4 i ec_4 \left(M^2+m_N^2-s-s_1+s_2-t_1\right)}{F^2\left(t_2-M^2\right) \left(M^2+2 m_N^2-2 s-t_1\right)}\notag\\
&+\frac{4 i e c_4\left(M^2+m_N^2-s-t_1\right)\left(M^2+m_N^2-s-s_1+s_2-t_1\right)}{F^2 \left(m_N^2-s\right) \left(M^2+2 m_N^2-2 s-t_1\right) \left(-2M^2-2 m_N^2+2 s+t_1+t_2\right)}\notag\\
&-\frac{i e c_6g^2 m_N (t_1-t_2)}{F^2 \left(m_N^2-s\right) \left(m_N^2-s_1\right) \left(2 M^2+2 m_N^2-2s-t_1-t_2\right)}\notag\\
&+\frac{i ec_6 g^2 m_N}{F^2 \left(m_N^2-s_1\right) \left(2 M^2+2 m_N^2-2s-t_1-t_2\right)}+\frac{i e c_6g^2 m_N}{F^2\left(m_N^2-s_2\right) \left(2 M^2+2 m_N^2-2 s-t_1-t_2\right)}\notag\\
&+\frac{i ec_6 g^2 m_N}{F^2 \left(M^2-s+s_1-t_2\right) \left(2 M^2+2 m_N^2-2s-t_1-t_2\right)}\notag\\
&-\frac{i e c_6g^2 m_N (t_1-t_2)}{F^2 \left(m_N^2-s\right)\left(m_N^2-s_2\right) \left(2 M^2+2 m_N^2-2 s-t_1-t_2\right)}\notag\\
&+\frac{i e c_6g^2 m_N}{F^2\left(M^2-s+s_2-t_1\right) \left(2 M^2+2 m_N^2-2 s-t_1-t_2\right)}\notag\\
&+\frac{i e c_6\left(g^2 t_1-g^2t_2-t_1+t_2\right)}{2 F^2 m_N \left(m_N^2-s\right) \left(2 M^2+2 m_N^2-2 s-t_1-t_2\right)}\ ,\notag\\
\mathcal{M}^3_{6}=&-\frac{2 i ec_4 m_N}{F^2 \left(t_2-M^2\right)}+\frac{i e c_6g^2 \left(m_N^2+s_1\right)}{4 F^2 \left(m_N^2-s_1\right) \left(-M^2+s-s_1+t_2\right)}-\frac{i ec_6g^2 \left(M^2+2 m_N^2-s+s_2-t_1\right)}{4 F^2 \left(m_N^2-s_2\right) \left(M^2-s+s_2-t_1\right)}\ ,\notag\\
\mathcal{M}^3_{7}=&\frac{16 i c_4 e m_N}{F^2 \left(t_2-M^2\right) \left(M^2+2 m_N^2-2 s-t_1\right)}-\frac{8 i c_4 em_N}{F^2 \left(m_N^2-s\right) \left(-2 M^2-m_N^2+s+t_1+t_2\right)}\notag\\
&+\frac{16 i c_4 e m_N\left(M^2+m_N^2-s-t_1\right)}{F^2 \left(m_N^2-s\right) \left(M^2+2 m_N^2-2 s-t_1\right) \left(-2 M^2-2 m_N^2+2s+t_1+t_2\right)}\ ,\notag\\
\mathcal{M}^3_{8}=&\frac{i c_6 e g^2 m_N}{2 F^2 \left(M^2-s+s_1-t_2\right) \left(2 M^2+m_N^2-s-t_1-t_2\right)}+\frac{i c_6 e g^2m_N}{2 F^2 \left(m_N^2-s\right) \left(m_N^2-s_1\right)}\notag\\
&+\frac{ic_6 e g^2 m_N}{2 F^2 \left(M^2-s+s_2-t_1\right) \left(2 M^2+m_N^2-s-t_1-t_2\right)}+\frac{ic_6 e \left(g^2-1\right)}{4 F^2 m_N \left(2 M^2+m_N^2-s-t_1-t_2\right)}\notag\\
&+\frac{i c_6 e g^2 m_N}{2 F^2\left(m_N^2-s\right) \left(m_N^2-s_2\right)}-\frac{i c_6 e \left(g^2-1\right)}{4 F^2 m_N\left(m_N^2-s\right)}\ ,\notag\\
\mathcal{M}^3_{9}=&\frac{4 i e c_4\left(M^2+m_N^2-s-s_1+s_2-t_1\right)}{F^2 \left(M^2+2 m_N^2-2 s-t_1\right) \left(-2 M^2-2 m_N^2+2s+t_1+t_2\right)}-\frac{4 i e c_4\left(M^2+m_N^2-s-s_1+s_2-t_1\right)}{F^2 \left(t_2-M^2\right) \left(M^2+2m_N^2-2 s-t_1\right)}\notag\\
&+\frac{2 i e c_6g^2 m_N}{F^2 \left(m_N^2-s_1\right) \left(2 M^2+2 m_N^2-2 s-t_1-t_2\right)}-\frac{i ec_6 g^2 m_N}{F^2\left(m_N^2-s_2\right) \left(M^2-s+s_2-t_1\right)}\notag\\
&+\frac{i ec_6g^2 m_N \left(s-m_N^2\right)}{F^2 \left(m_N^2-s_1\right) \left(M^2-s+s_1-t_2\right) \left(2 M^2+2m_N^2-2 s-t_1-t_2\right)}\notag\\
&+\frac{i e c_6g^2 m_N}{F^2 \left(m_N^2-s_1\right)\left(M^2-s+s_1-t_2\right)}+\frac{2 i e c_6g^2 m_N}{F^2 \left(m_N^2-s_2\right) \left(2 M^2+2 m_N^2-2s-t_1-t_2\right)}\notag\\
&+\frac{i ec_6 g^2 m_N \left(m_N^2-s\right)}{F^2 \left(m_N^2-s_2\right)\left(M^2-s+s_2-t_1\right) \left(2 M^2+2 m_N^2-2 s-t_1-t_2\right)}\notag\\
&-\frac{i e c_6\left(g^2-1\right)}{F^2 m_N \left(2 M^2+2m_N^2-2 s-t_1-t_2\right)}\ ,\notag\\
\mathcal{M}^3_{10}=&\frac{2 i c_4 e m_N}{F^2 \left(t_2-M^2\right)}-\frac{2 i c_4 e m_N}{F^2 \left(m_N^2-s\right)}+\frac{ic_6 e g^2 m_N^2}{2 F^2 \left(m_N^2-s_1\right) \left(-M^2+s-s_1+t_2\right)}-\frac{i c_6 e}{2 F^2 \left(m_N^2-s\right)}\notag\\
&+\frac{i c_6 eg^2 m_N^2}{2 F^2 \left(m_N^2-s_2\right) \left(M^2-s+s_2-t_1\right)}-\frac{i c_6 e g^2\left(m_N^2 s+m_N^2 s_1-s s_1-s_1 s_2\right)}{2 F^2 \left(m_N^2-s\right) \left(m_N^2-s_1\right)\left(m_N^2-s_2\right)}\notag\\
&-\frac{i c_6 e g^2 m_N^2}{2 F^2 \left(m_N^2-s\right)\left(m_N^2-s_1\right)}-\frac{i c_6 e g^2 m_N^2}{2 F^2 \left(m_N^2-s\right)\left(m_N^2-s_2\right)}\ ,\notag\\
\mathcal{M}^3_{11}=&-\frac{16 i c_4 e m_N}{F^2 \left(M^2-t_2\right) \left(-2 M^2-2 m_N^2+2 s+t_1+t_2\right)}\ ,\notag\\
\mathcal{M}^3_{12}=&\frac{i c_6 e g^2 m_N}{2 F^2 \left(m_N^2-s_2\right) \left(M^2-s+s_2-t_1\right)}-\frac{i c_6 eg^2 m_N}{2 F^2 \left(m_N^2-s_1\right) \left(-M^2+s-s_1+t_2\right)}\ .
\end{align}
\item Amplitudes $\mathcal{M}^4_j$:
\begin{align}
\mathcal{M}^4_{1}=&\mathcal{M}^4_{2}=\mathcal{M}^4_{3}=\mathcal{M}^4_{5}=\mathcal{M}^4_{8}=\mathcal{M}^4_{9}=\mathcal{M}^4_{12}=0\ ,\notag\\
\mathcal{M}^4_{4}=&-\mathcal{M}^3_{4}(s_1 \leftrightarrow s_2,t_1 \leftrightarrow t_2)\ ,\quad \mathcal{M}^4_{6}=\mathcal{M}^3_{6}(s_1 \leftrightarrow s_2,t_1 \leftrightarrow t_2)\ , \notag\\\mathcal{M}^4_{7}=&-\mathcal{M}^3_{7}(s_1 \leftrightarrow s_2,t_1 \leftrightarrow t_2)\ ,\quad
\mathcal{M}^4_{10}=-\mathcal{M}^3_{10}(s_1 \leftrightarrow s_2,t_1 \leftrightarrow t_2)\ ,\notag\\ \mathcal{M}^4_{11}=&\mathcal{M}^3_{11}(s_1 \leftrightarrow s_2,t_1 \leftrightarrow t_2)
\end{align}
\item Amplitudes $\mathcal{M}^5_j$:
\begin{align}
\mathcal{M}^5_{1}=&\frac{(c_6+2c_7)e g^2}{4 F^2}\left(\frac{1}{M^2-s+s_2-t_1}+\frac{1}{m_N^2-s_2}\right)-(s_1 \leftrightarrow s_2,t_1 \leftrightarrow t_2)\ ,\notag\\
\mathcal{M}^5_{2}=&-\frac{c_4 e \left(2 M^2+m_N^2-s-s_1+s_2-2 t_1\right)}{4 F^2 \left(-2M^2-m_N^2+s+t_1+t_2\right)}-\frac{c_4 e \left(m_N^2-s-s_1+s_2\right)}{4 F^2 \left(m_N^2-s\right)}\ ,\notag\\
\mathcal{M}^5_{3}=&-\frac{e g^2 m_N (c_6+2 c_7)}{2 F^2 \left(M^2-s+s_1-t_2\right) \left(2M^2+m_N^2-s-t_1-t_2\right)}+\frac{e g^2 m_N (c_6+2 c_7)}{2 F^2 \left(m_N^2-s_1\right)\left(M^2-s+s_1-t_2\right)}\notag\\
&-\frac{e g^2 m_N (c_6+2 c_7)}{2 F^2 \left(M^2-s+s_2-t_1\right) \left(2M^2+m_N^2-s-t_1-t_2\right)}+\frac{e g^2 m_N (c_6+2 c_7)}{2 F^2 \left(m_N^2-s_2\right)\left(M^2-s+s_2-t_1\right)}\notag\\
&+\frac{e \left(1-g^2\right) (c_6+2 c_7)}{4 F^2 m_N \left(2M^2+m_N^2-s-t_1-t_2\right)}-\frac{e g^2 m_N (c_6+2 c_7)}{2 F^2 \left(m_N^2-s\right)\left(m_N^2-s_1\right)}\notag\\
&-\frac{e g^2 m_N (c_6+2 c_7)}{2 F^2 \left(m_N^2-s\right)\left(m_N^2-s_2\right)}+\frac{e \left(g^2-1\right) (c_6+2 c_7)}{4 F^2 m_N \left(m_N^2-s\right)}\ ,\notag\\
\mathcal{M}^5_{4}=&-\frac{e c_4m_N}{F^2 \left(-2 M^2-m_N^2+s+t_1+t_2\right)}-\frac{e c_4m_N}{F^2 \left(m_N^2-s\right)}-\frac{eg^2 s_1 (c_6+2 c_7)}{2 F^2 \left(m_N^2-s\right) \left(m_N^2-s_1\right)}\notag\\
&+\frac{e g^2 (c_6+2 c_7) (s-2 s_1-t_1+t_2)}{2 F^2 \left(M^2-s+s_1-t_2\right) \left(2M^2+m_N^2-s-t_1-t_2\right)}+\frac{e g^2 s_1 (c_6+2 c_7)}{2 F^2 \left(m_N^2-s_1\right)\left(M^2-s+s_1-t_2\right)}\notag\\
&+\frac{e g^2 (c_6+2 c_7) (s-2 s_2+t_1-t_2)}{2 F^2\left(M^2-s+s_2-t_1\right) \left(2 M^2+m_N^2-s-t_1-t_2\right)}+\frac{e g^2 s_2 (c_6+2 c_7)}{2F^2 \left(m_N^2-s_2\right) \left(M^2-s+s_2-t_1\right)}\notag\\
&+\frac{e \left(7 g^2+1\right) (c_6+2 c_7)}{4F^2 \left(2 M^2+m_N^2-s-t_1-t_2\right)}-\frac{e g^2 (c_6+2 c_7)}{4 F^2\left(M^2-s+s_1-t_2\right)}-\frac{e g^2 (c_6+2 c_7)}{4 F^2 \left(M^2-s+s_2-t_1\right)}\notag\\
&-\frac{e g^2 s_2(c_6+2 c_7)}{2 F^2 \left(m_N^2-s\right) \left(m_N^2-s_2\right)}-\frac{e \left(3 g^2+1\right) (c_6+2c_7)}{4 F^2 \left(m_N^2-s\right)}+\frac{e g^2 (c_6+2 c_7)}{4 F^2 \left(m_N^2-s_1\right)}+\frac{eg^2 (c_6+2 c_7)}{4 F^2 \left(m_N^2-s_2\right)}\ ,\notag\\
\mathcal{M}^5_{5}=&\frac{2 e c_4\left(2 M^2+m_N^2-s-s_1+s_2-2 t_1\right)}{F^2 \left(m_N^2-s\right) \left(-2M^2-m_N^2+s+t_1+t_2\right)}-\frac{4 e c_4\left(M^2+m_N^2-s-t_1\right)}{F^2 \left(m_N^2-s\right) \left(-2 M^2-2m_N^2+2 s+t_1+t_2\right)}\notag\\
&+\frac{e g^2 m_N (c_6+2 c_7) (t_2-t_1)}{F^2 \left(m_N^2-s\right) \left(m_N^2-s_1\right) \left(2M^2+2 m_N^2-2 s-t_1-t_2\right)}\notag\\
&-\frac{e g^2 m_N (c_6+2 c_7)}{F^2 \left(m_N^2-s_1\right)\left(2 M^2+2 m_N^2-2 s-t_1-t_2\right)}\notag\\
&-\frac{e g^2 m_N (c_6+2 c_7)}{F^2\left(M^2-s+s_1-t_2\right) \left(2 M^2+2 m_N^2-2 s-t_1-t_2\right)}\notag\\
&+\frac{e g^2 m_N (c_6+2 c_7)(t_2-t_1)}{F^2 \left(m_N^2-s\right) \left(m_N^2-s_2\right) \left(2 M^2+2 m_N^2-2s-t_1-t_2\right)}\notag\\
&+\frac{e g^2 m_N (c_6+2 c_7)}{F^2 \left(m_N^2-s_2\right) \left(2 M^2+2m_N^2-2 s-t_1-t_2\right)}\notag\\
&+\frac{e g^2 m_N (c_6+2 c_7)}{F^2 \left(M^2-s+s_2-t_1\right) \left(2M^2+2 m_N^2-2 s-t_1-t_2\right)}\notag\\
&+\frac{e \left(g^2-1\right) t_1 (c_6+2 c_7) (t_1-t_2)}{2 F^2m_N \left(m_N^2-s\right) \left(2 M^2+2 m_N^2-2 s-t_1-t_2\right)}\ ,\notag\\
\mathcal{M}^5_{6}=&\frac{(c6 + 2 c7) e g^2}{4 F^2}\left(\frac{2 s_1}{\left(m_N^2-s_1\right) \left(M^2-s+s_1-t_2\right)}+\frac{1}{M^2-s+s_1-t_2}+\frac{1}{m_N^2-s_1}\right)\notag\\
&-(s_1 \leftrightarrow s_2,t_1 \leftrightarrow t_2)\ ,\notag\\
\mathcal{M}^5_{7}=&\frac{8 c_4 e m_N}{F^2 \left(m_N^2-s\right) \left(-2 M^2-m_N^2+s+t_1+t_2\right)}\ ,\notag\\
\mathcal{M}^5_{8}=&-\frac{e g^2 m_N (c_6+2 c_7)}{2 F^2 \left(M^2-s+s_1-t_2\right) \left(2M^2+m_N^2-s-t_1-t_2\right)}+\frac{e \left(1-g^2\right) (c_6+2 c_7)}{4 F^2 m_N \left(m_N^2-s\right)}\notag\\
&-\frac{e g^2 m_N (c_6+2 c_7)}{2 F^2 \left(M^2-s+s_2-t_1\right)\left(2 M^2+m_N^2-s-t_1-t_2\right)}+\frac{e \left(1-g^2\right) (c_6+2 c_7)}{4 F^2 m_N \left(2M^2+m_N^2-s-t_1-t_2\right)}\notag\\
&+\frac{e g^2 m_N (c_6+2 c_7)}{2 F^2 \left(m_N^2-s\right)\left(m_N^2-s_1\right)}+\frac{e g^2 m_N (c_6+2 c_7)}{2 F^2 \left(m_N^2-s\right)\left(m_N^2-s_2\right)}\ ,\notag\\
\mathcal{M}^5_{9}=&-\frac{4 ec_4}{F^2 \left(-2 M^2-2 m_N^2+2 s+t_1+t_2\right)}+\frac{2 e g^2 m_N (c_6+2 c_7)}{F^2 \left(m_N^2-s_1\right) \left(2 M^2+2 m_N^2-2s-t_1-t_2\right)}\notag\\
&+\frac{e g^2 m_N (c_6+2 c_7)}{F^2 \left(M^2-s+s_1-t_2\right) \left(2 M^2+2m_N^2-2 s-t_1-t_2\right)}\notag\\
&+\frac{e g^2 m_N (c_6+2 c_7) (s_1-s)}{F^2\left(m_N^2-s_1\right) \left(M^2-s+s_1-t_2\right) \left(2 M^2+2 m_N^2-2 s-t_1-t_2\right)}\notag\\
&-\frac{e g^2m_N (c_6+2 c_7)}{F^2 \left(m_N^2-s_1\right) \left(M^2-s+s_1-t_2\right)}+\frac{2 e g^2 m_N(c_6+2 c_7)}{F^2 \left(m_N^2-s_2\right) \left(2 M^2+2 m_N^2-2 s-t_1-t_2\right)}\notag\\
&+\frac{e g^2m_N (c_6+2 c_7)}{F^2 \left(M^2-s+s_2-t_1\right) \left(2 M^2+2 m_N^2-2 s-t_1-t_2\right)}\notag\\
&+\frac{eg^2 m_N (c_6+2 c_7) (s_2-s)}{F^2 \left(m_N^2-s_2\right) \left(M^2-s+s_2-t_1\right) \left(2M^2+2 m_N^2-2 s-t_1-t_2\right)}\notag\\
&-\frac{e g^2 m_N (c_6+2 c_7)}{F^2 \left(m_N^2-s_2\right)\left(M^2-s+s_2-t_1\right)}+\frac{e \left(1-g^2\right) (c_6+2 c_7)}{F^2 m_N \left(2 M^2+2 m_N^2-2s-t_1-t_2\right)}\ ,\notag\\
\mathcal{M}^5_{10}=&\frac{e g^2 s_1 (c_6+2 c_7)}{2 F^2 \left(m_N^2-s_1\right) \left(M^2-s+s_1-t_2\right)}+\frac{eg^2 s_2 (c_6+2 c_7)}{2 F^2 \left(m_N^2-s_2\right) \left(M^2-s+s_2-t_1\right)}\notag\\
&+\frac{e g^2(c_6+2 c_7)}{2 F^2 \left(M^2-s+s_1-t_2\right)}+\frac{e g^2 (c_6+2 c_7)}{2 F^2\left(M^2-s+s_2-t_1\right)}-\frac{e g^2 s_1 (c_6+2 c_7)}{F^2 \left(m_N^2-s\right)\left(m_N^2-s_1\right)}\notag\\
&-\frac{e g^2 s_2 (c_6+2 c_7)}{F^2 \left(m_N^2-s\right)\left(m_N^2-s_2\right)}-\frac{e \left(3 g^2+1\right) (c_6+2 c_7)}{2 F^2 \left(m_N^2-s\right)}+\frac{eg^2 (c_6+2 c_7)}{2 F^2 \left(m_N^2-s_1\right)}+\frac{e g^2 (c_6+2 c_7)}{2 F^2\left(m_N^2-s_2\right)}\ ,\notag\\
\mathcal{M}^5_{11}=&0 \ ,\notag\\
\mathcal{M}^5_{12}=&\frac{e g^2 m_N (c_6+2 c_7)}{2 F^2 \left(m_N^2-s_1\right) \left(-M^2+s-s_1+t_2\right)}-(s_1 \leftrightarrow s_2,t_1 \leftrightarrow t_2)\ .
\end{align}
\item Amplitudes $\mathcal{M}^6_j$:
\begin{align}
\mathcal{M}^6_{1}=&0\ ,\notag\\
\mathcal{M}^6_{2}=&-\frac{e c_4\left(2 M^2+m_N^2-s-s_1+s_2-2 t_1\right)}{4 F^2 \left(-2 M^2-m_N^2+s+t_1+t_2\right)}-\frac{ec_4\left(m_N^2-s-s_1+s_2\right)}{4 F^2 \left(m_N^2-s\right)}-\frac{e g^2 m_N (c_6+2 c_7)}{4 F^2 \left(M^2-s+s_1-t_2\right)}\notag\\
&+\frac{e g^2 m_N (c_6+2c_7)}{4 F^2 \left(M^2-s+s_2-t_1\right)}+\frac{e g^2 m_N (c_6+2 c_7)}{4 F^2\left(m_N^2-s_1\right)}-\frac{e g^2 m_N (c_6+2 c_7)}{4 F^2 \left(m_N^2-s_2\right)}\ ,\notag\\
\mathcal{M}^6_{3}=&-\frac{c_6 e g^2 m_N}{2 F^2 \left(M^2-s+s_1-t_2\right) \left(2M^2+m_N^2-s-t_1-t_2\right)}-\frac{c_6 e g^2 m_N}{2 F^2 \left(m_N^2-s_1\right)\left(M^2-s+s_1-t_2\right)}\notag\\
&-\frac{c_6 e g^2 m_N}{2 F^2 \left(M^2-s+s_2-t_1\right) \left(2M^2+m_N^2-s-t_1-t_2\right)}-\frac{c_6 e g^2 m_N}{2 F^2 \left(m_N^2-s_2\right)\left(M^2-s+s_2-t_1\right)}\notag\\
&+\frac{c_6 e \left(1-g^2\right)}{4 F^2 m_N \left(2M^2+m_N^2-s-t_1-t_2\right)}-\frac{c_6 e g^2 m_N}{2 F^2 \left(m_N^2-s\right)\left(m_N^2-s_1\right)}\notag\\
&-\frac{c_6 e g^2 m_N}{2 F^2 \left(m_N^2-s\right)\left(m_N^2-s_2\right)}+\frac{c_6 e \left(g^2-1\right)}{4 F^2 m_N \left(m_N^2-s\right)}\ ,\notag\\
\mathcal{M}^6_{4}=&\frac{e c_4m_N}{F^2 \left(2 M^2+m_N^2-s-t_1-t_2\right)}-\frac{e c_4m_N}{F^2 \left(m_N^2-s\right)}\notag\\
&+\frac{c_6 e g^2 (s-2 s_1-t_1+t_2)}{2 F^2 \left(M^2-s+s_1-t_2\right) \left(2M^2+m_N^2-s-t_1-t_2\right)}-\frac{c_6 e g^2 s_1}{2 F^2 \left(m_N^2-s_1\right)\left(M^2-s+s_1-t_2\right)}\notag\\
&+\frac{c_6 e g^2 (s-2 s_2+t_1-t_2)}{2 F^2\left(M^2-s+s_2-t_1\right) \left(2 M^2+m_N^2-s-t_1-t_2\right)}-\frac{c_6 e g^2 s_2}{2 F^2\left(m_N^2-s_2\right) \left(M^2-s+s_2-t_1\right)}\notag\\
&+\frac{c_6 e \left(7 g^2+1\right)}{4 F^2 \left(2M^2+m_N^2-s-t_1-t_2\right)}-\frac{c_6 e g^2}{F^2 \left(M^2-s+s_1-t_2\right)}-\frac{c_6 eg^2}{F^2 \left(M^2-s+s_2-t_1\right)}\notag\\
&-\frac{c_6 e g^2 s_1}{2 F^2 \left(m_N^2-s\right)\left(m_N^2-s_1\right)}-\frac{c_6 e g^2 s_2}{2 F^2 \left(m_N^2-s\right)\left(m_N^2-s_2\right)}-\frac{c_6 e \left(3 g^2+1\right)}{4 F^2 \left(m_N^2-s\right)}\ ,\notag\\
\mathcal{M}^6_{5}=&-\frac{32 c_1 e t_1}{F^2 \left(M^2-t_1\right) \left(2 M^2+2 m_N^2-2 s-t_1-t_2\right)}+\frac{4 c_2 e(s-s_1-s_2+t_1+t_2)}{F^2 \left(M^2-t_2\right) \left(2 M^2+2 m_N^2-2 s-t_1-t_2\right)}\notag\\
&-\frac{2 c_2 e \left(2 s_1 s_2-2 s_1 t_1-s_2 t_1-s_2 t_2+t_1^2+t_1t_2\right)}{F^2 m_N^2 \left(M^2-t_1\right) \left(2 M^2+2 m_N^2-2 s-t_1-t_2\right)}+\frac{2c_2 e (s_1-t_2)}{F^2 m_N^2 \left(M^2-t_2\right)}-\frac{4 c_3 e}{F^2 \left(M^2-t_1\right)}\notag\\
&+\frac{4 c_3 e (4 s-2 s_1-2 s_2+3 t_1-3 t_2)}{F^2 \left(M^2-t_2\right) \left(2 M^2+2 m_N^2-2s-t_1-t_2\right)}+\frac{2 c_4 e (s_1+t_1)}{F^2 \left(m_N^2-s\right) \left(2 M^2+m_N^2-s-t_1-t_2\right)}\notag\\
&-\frac{2 c_4 et_1}{F^2 \left(m_N^2-s\right) \left(2 M^2+2 m_N^2-2 s-t_1-t_2\right)}+\frac{c_6 e g^2 m_N}{F^2 \left(m_N^2-s_1\right) \left(2 M^2+2m_N^2-2 s-t_1-t_2\right)}\notag\\
&+\frac{c_6 e g^2 m_N (t_2-t_1)}{F^2 \left(m_N^2-s\right) \left(m_N^2-s_1\right) \left(2 M^2+2m_N^2-2 s-t_1-t_2\right)}\notag\\
&+\frac{c_6 e g^2 m_N}{F^2 \left(M^2-s+s_1-t_2\right) \left(2 M^2+2m_N^2-2 s-t_1-t_2\right)}\notag\\
&+\frac{c_6 e \left(g^2-1\right) t_1}{2 F^2 m_N \left(m_N^2-s\right)\left(2 M^2+2 m_N^2-2 s-t_1-t_2\right)}-(s_1 \leftrightarrow s_2,t_1 \leftrightarrow t_2)\ ,\notag\\
\mathcal{M}^6_{6}=&\frac{c_6 e g^2 m_N^2}{2 F^2 \left(m_N^2-s_2\right) \left(M^2-s+s_2-t_1\right)}-(s_1 \leftrightarrow s_2,t_1 \leftrightarrow t_2)\ ,\notag\\
\mathcal{M}^6_{7}=&\frac{8 c_4 e m_N}{F^2 \left(m_N^2-s\right) \left(-2 M^2-m_N^2+s+t_1+t_2\right)}\ ,\notag\\
\mathcal{M}^6_{8}=&-\frac{c_6 e g^2 m_N}{2 F^2 \left(M^2-s+s_1-t_2\right) \left(2M^2+m_N^2-s-t_1-t_2\right)}+\frac{c_6 e g^2 m_N}{2 F^2 \left(m_N^2-s\right)\left(m_N^2-s_1\right)}\notag\\
&-\frac{c_6 e g^2 m_N}{2 F^2 \left(M^2-s+s_2-t_1\right) \left(2M^2+m_N^2-s-t_1-t_2\right)}-\frac{c_6 e \left(g^2-1\right)}{4 F^2 m_N \left(2M^2+m_N^2-s-t_1-t_2\right)}\notag\\
&+\frac{c_6 e g^2 m_N}{2 F^2 \left(m_N^2-s\right)\left(m_N^2-s_2\right)}-\frac{c_6 e \left(g^2-1\right)}{4 F^2 m_N \left(m_N^2-s\right)}\ .\notag\\
\mathcal{M}^6_{9}=&-\frac{32 c_1 e t_1}{F^2 \left(M^2-t_1\right) \left(2 M^2+2 m_N^2-2 s-t_1-t_2\right)}-\frac{32 c_1e}{F^2 \left(2 M^2+2 m_N^2-2 s-t_1-t_2\right)}\notag\\
&-\frac{2 c_2 e \left(2 s_1 s_2-2 s_1 t_1-s_2 t_1-s_2 t_2+t_1^2+t_1t_2\right)}{F^2 m_N^2 \left(M^2-t_1\right) \left(2 M^2+2 m_N^2-2 s-t_1-t_2\right)}+\frac{2 c_2 e(s_1+s_2-t_1-t_2)}{F^2 m_N^2 \left(2 M^2+2 m_N^2-2 s-t_1-t_2\right)}\notag\\
&-\frac{4 c_2 e(s-s_1-s_2+t_1+t_2)}{F^2 \left(M^2-t_1\right) \left(2 M^2+2 m_N^2-2 s-t_1-t_2\right)}-\frac{2c_2 e (s_2-t_1)}{F^2 m_N^2 \left(M^2-t_1\right)}+\frac{c_2 e}{F^2 m_N^2}\notag\\
&-\frac{4 c_3 (4 e s-2 e s_1-2 e s_2-3 e t_1+3 e t_2)}{F^2 \left(M^2-t_1\right) \left(2 M^2+2 m_N^2-2s-t_1-t_2\right)}+\frac{24 c_3 e}{F^2 \left(2 M^2+2 m_N^2-2 s-t_1-t_2\right)}\notag\\
&-\frac{4 c_3 e}{F^2\left(M^2-t_1\right)}+\frac{2 c_4 e}{F^2 \left(2 M^2+2 m_N^2-2 s-t_1-t_2\right)}\notag\\
&+\frac{2 c_6 e g^2 m_N}{F^2 \left(m_N^2-s_1\right) \left(2 M^2+2 m_N^2-2s-t_1-t_2\right)}+\frac{c_6 e g^2 m_N}{F^2\left(m_N^2-s_1\right) \left(M^2-s+s_1-t_2\right)}\notag\\
&-\frac{c_6 e g^2 m_N}{F^2 \left(M^2-s+s_1-t_2\right) \left(2 M^2+2 m_N^2-2s-t_1-t_2\right)}\notag\\
&+\frac{c_6 e g^2 m_N (s-s_1)}{F^2 \left(m_N^2-s_1\right)\left(M^2-s+s_1-t_2\right) \left(2 M^2+2 m_N^2-2 s-t_1-t_2\right)}\notag\\
&+\frac{c_6 e g^2 m_N}{F^2\left(m_N^2-s_1\right) \left(M^2-s+s_1-t_2\right)}+\frac{c_6 e \left(1-g^2\right)}{2 F^2 m_N \left(2M^2+2 m_N^2-2 s-t_1-t_2\right)}+(s_1 \leftrightarrow s_2,t_1 \leftrightarrow t_2)\ ,\notag\\
\mathcal{M}^6_{10}=&-\frac{2 c_4 e m_N}{F^2 \left(m_N^2-s\right)}-\frac{c_6 e g^2 s_1}{2 F^2 \left(m_N^2-s_1\right) \left(M^2-s+s_1-t_2\right)}-\frac{c_6 e g^2s_2}{2 F^2 \left(m_N^2-s_2\right) \left(M^2-s+s_2-t_1\right)}\notag\\
&-\frac{c_6 e g^2}{2 F^2\left(M^2-s+s_1-t_2\right)}-\frac{c_6 e g^2}{2 F^2 \left(M^2-s+s_2-t_1\right)}-\frac{c_6 e g^2s_1}{F^2 \left(m_N^2-s\right) \left(m_N^2-s_1\right)}\notag\\
&-\frac{c_6 e g^2 s_2}{F^2\left(m_N^2-s\right) \left(m_N^2-s_2\right)}-\frac{c_6 e \left(3 g^2+1\right)}{2 F^2 \left(m_N^2-s\right)}\ ,\notag\\
\mathcal{M}^6_{11}=&0\ ,\notag\\
\mathcal{M}^6_{12}=&-\frac{c_6 e g^2 m_N}{2 F^2 \left(m_N^2-s_1\right) \left(-M^2+s-s_1+t_2\right)}-(s_1 \leftrightarrow s_2,t_1 \leftrightarrow t_2)\ .
\end{align}
\end{itemize}

\bibliographystyle{utphys}
\bibliography{inspire_cite}

\providecommand{\href}[2]{#2}\begingroup\raggedright\begin{thebibliography}{10}

\bibitem{refprd341986}
S.~Capstick and N.~Isgur, ``{Baryons in a relativized quark model with
  chromodynamics},'' \href{http://dx.doi.org/10.1103/physrevd.34.2809}{{\em
  Phys. Rev. D} {\bfseries 34} (1986) 2809}.

\bibitem{refepja102001}
U.~Loring, B.~C. Metsch, and H.~R. Petry, ``{The Light baryon spectrum in a
  relativistic quark model with instanton induced quark forces: The Nonstrange
  baryon spectrum and ground states},''
  \href{http://dx.doi.org/10.1007/s100500170105}{{\em Eur. Phys. J. A}
  {\bfseries 10} (2001) 395--446},
  \href{http://arxiv.org/abs/hep-ph/0103289}{{\ttfamily arXiv:hep-ph/0103289}}.

\bibitem{refprd842011}
R.~G. Edwards, J.~J. Dudek, D.~G. Richards, and S.~J. Wallace, ``{Excited state
  baryon spectroscopy from lattice QCD},''
  \href{http://dx.doi.org/10.1103/PhysRevD.84.074508}{{\em Phys. Rev. D}
  {\bfseries 84} (2011) 074508},
  \href{http://arxiv.org/abs/1104.5152}{{\ttfamily arXiv:1104.5152 [hep-ph]}}.

\bibitem{Ma:2020hpe}
Y.~Ma, W.-Q. Niu, D.-L. Yao, and H.-Q. Zheng, ``{Dispersive analysis of low
  energy $\gamma N\to\pi N$ process and studies on the $N^*$(890) resonance},''
  \href{http://dx.doi.org/10.1088/1674-1137/abc169}{{\em Chin. Phys. C}
  {\bfseries 45} no.~1, (2021) 014104},
  \href{http://arxiv.org/abs/2005.10695}{{\ttfamily arXiv:2005.10695
  [hep-ph]}}.

\bibitem{Cao:2021kvs}
X.-H. Cao, Y.~Ma, and H.-Q. Zheng, ``{Dispersive analysis of low energy
  $\gamma^* N\rightarrow\pi N$ process},''
  \href{http://dx.doi.org/10.1103/PhysRevD.103.114007}{{\em Phys. Rev. D}
  {\bfseries 103} no.~11, (2021) 114007},
  \href{http://arxiv.org/abs/2101.12576}{{\ttfamily arXiv:2101.12576
  [nucl-th]}}.

\bibitem{Wang:2017agd}
Y.-F. Wang, D.-L. Yao, and H.-Q. Zheng, ``{New Insights on Low Energy $\pi N$
  Scattering Amplitudes},''
  \href{http://dx.doi.org/10.1140/epjc/s10052-018-6024-5}{{\em Eur. Phys. J. C}
  {\bfseries 78} no.~7, (2018) 543},
  \href{http://arxiv.org/abs/1712.09257}{{\ttfamily arXiv:1712.09257
  [hep-ph]}}.

\bibitem{Wang:2018gul}
Y.-F. Wang, D.-L. Yao, and H.-Q. Zheng, ``{On the existence of $N^*(890)$
  resonance in $S_{11}$ channel of $\pi N$ scatterings},''
  \href{http://dx.doi.org/10.1007/s11467-018-0877-9}{{\em Front. Phys.
  (Beijing)} {\bfseries 14} no.~2, (2019) 24501},
  \href{http://arxiv.org/abs/1810.07958}{{\ttfamily arXiv:1810.07958
  [hep-ph]}}.

\bibitem{Wang:2018nwi}
Y.-F. Wang, D.-L. Yao, and H.-Q. Zheng, ``{New insights on low energy $\pi N$
  scattering amplitudes: comprehensive analyses at $\mathcal{O}(p^3)$ level},''
  \href{http://dx.doi.org/10.1088/1674-1137/43/6/064110}{{\em Chin. Phys. C}
  {\bfseries 43} no.~6, (2019) 064110},
  \href{http://arxiv.org/abs/1811.09748}{{\ttfamily arXiv:1811.09748
  [hep-ph]}}.

\bibitem{Cao:2022zhn}
X.-H. Cao, Q.-Z. Li, and H.-Q. Zheng, ``{A possible subthreshold pole in
  S$_{11}$ channel from \ensuremath{\pi}N Roy-Steiner equation analyses},''
  \href{http://dx.doi.org/10.1007/JHEP12(2022)073}{{\em JHEP} {\bfseries 12}
  (2022) 073}, \href{http://arxiv.org/abs/2207.09743}{{\ttfamily
  arXiv:2207.09743 [hep-ph]}}.

\bibitem{refppnp512003}
B.~Krusche and S.~Schadmand, ``{Study of nonstrange baryon resonances with
  meson photoproduction},''
  \href{http://dx.doi.org/10.1016/S0146-6410(03)90005-6}{{\em Prog. Part. Nucl.
  Phys.} {\bfseries 51} (2003) 399--485},
  \href{http://arxiv.org/abs/nucl-ex/0306023}{{\ttfamily
  arXiv:nucl-ex/0306023}}.

\bibitem{refarxiv2207}
{\bfseries CBELSA/TAPS} Collaboration, T.~Seifen {\em et~al.}, ``{Polarization
  observables in double neutral pion photoproduction},''
  \href{http://arxiv.org/abs/2207.01981}{{\ttfamily arXiv:2207.01981
  [nucl-ex]}}.

\bibitem{refplb3631995}
A.~Braghieri {\em et~al.}, ``{Total cross-section measurement for the three
  double pion production channels on the proton},''
  \href{http://dx.doi.org/10.1016/0370-2693(95)01189-W}{{\em Phys. Lett. B}
  {\bfseries 363} (1995) 46--50}.

\bibitem{refplb5782004}
M.~Kotulla {\em et~al.}, ``{Double pi0 photoproduction off the proton at
  threshold},'' \href{http://dx.doi.org/10.1016/j.physletb.2003.10.056}{{\em
  Phys. Lett. B} {\bfseries 578} (2004) 63--68},
  \href{http://arxiv.org/abs/nucl-ex/0310031}{{\ttfamily
  arXiv:nucl-ex/0310031}}.

\bibitem{refplb6242005}
{\bfseries GDH, A2} Collaboration, J.~Ahrens {\em et~al.}, ``{Intermediate
  resonance excitation in the gamma p ---\ensuremath{>} p pi0 pi0 reaction},''
  \href{http://dx.doi.org/10.1016/j.physletb.2005.08.034}{{\em Phys. Lett. B}
  {\bfseries 624} (2005) 173--180}.

\bibitem{refepja342007}
{\bfseries GDH, A2} Collaboration, J.~Ahrens {\em et~al.}, ``{First measurement
  of the helicity dependence for the gamma p ---\ensuremath{>} p pi+ pi-
  reaction},'' \href{http://dx.doi.org/10.1140/epja/i2007-10491-5}{{\em Eur.
  Phys. J. A} {\bfseries 34} (2007) 11--21}.

\bibitem{refprl1032009}
{\bfseries Crystal Ball at MAMI, TAPS, A2} Collaboration, D.~Krambrich {\em
  et~al.}, ``{Beam-Helicity Asymmetries in Double Pion Photoproduction off the
  Proton},'' \href{http://dx.doi.org/10.1103/PhysRevLett.103.052002}{{\em Phys.
  Rev. Lett.} {\bfseries 103} (2009) 052002},
  \href{http://arxiv.org/abs/0907.0358}{{\ttfamily arXiv:0907.0358 [nucl-ex]}}.

\bibitem{refepja512015}
{\bfseries CBELSA/TAPS} Collaboration, V.~Sokhoyan {\em et~al.},
  ``{High-statistics study of the reaction $\gamma p\to p\;2\pi^0$},''
  \href{http://dx.doi.org/10.1140/epja/i2015-15187-7}{{\em Eur. Phys. J. A}
  {\bfseries 51} no.~8, (2015) 95},
  \href{http://arxiv.org/abs/1507.02488}{{\ttfamily arXiv:1507.02488
  [nucl-ex]}}. [Erratum: Eur.Phys.J.A 51, 187 (2015)].

\bibitem{refplb5512003}
{\bfseries GDH, A2} Collaboration, J.~Ahrens {\em et~al.}, ``{Helicity
  dependence of the gamma(pol.) p(pol.) --\ensuremath{>} n pi+ pi0 reaction in
  the second resonance region},''
  \href{http://dx.doi.org/10.1016/S0370-2693(02)03008-3}{{\em Phys. Lett. B}
  {\bfseries 551} (2003) 49--55}.

\bibitem{refplb7882019}
{\bfseries CLAS} Collaboration, E.~Golovatch {\em et~al.}, ``{First results on
  nucleon resonance photocouplings from the $\gamma p \to \pi^+\pi^-p$
  reaction},'' \href{http://dx.doi.org/10.1016/j.physletb.2018.10.013}{{\em
  Phys. Lett. B} {\bfseries 788} (2019) 371--379},
  \href{http://arxiv.org/abs/1806.01767}{{\ttfamily arXiv:1806.01767
  [nucl-ex]}}.

\bibitem{refplb8472023}
{\bfseries A2} Collaboration, D.~Ghosal {\em et~al.}, ``{Helicity dependent
  cross sections for the photoproduction of
  \ensuremath{\pi}0\ensuremath{\pi}\ensuremath{\pm} pairs from quasi-free
  nucleons},'' \href{http://dx.doi.org/10.1016/j.physletb.2023.138273}{{\em
  Phys. Lett. B} {\bfseries 847} (2023) 138273},
  \href{http://arxiv.org/abs/2308.15240}{{\ttfamily arXiv:2308.15240
  [nucl-ex]}}.

\bibitem{refprl1252020}
M.~Dieterle {\em et~al.}, ``{Helicity-Dependent Cross Sections for the
  Photoproduction of \ensuremath{\pi}$^0$ Pairs from Nucleons},''
  \href{http://dx.doi.org/10.1103/PhysRevLett.125.062001}{{\em Phys. Rev.
  Lett.} {\bfseries 125} no.~6, (2020) 062001},
  \href{http://arxiv.org/abs/2007.06079}{{\ttfamily arXiv:2007.06079
  [nucl-ex]}}.

\bibitem{refepja482012}
F.~Zehr {\em et~al.}, ``{Photoproduction of $\pi^0\pi^0$ and $\pi^0\pi^\pm$
  pairs off the proton from threshold to the second resonance region},''
  \href{http://dx.doi.org/10.1140/epja/i2012-12098-1}{{\em Eur. Phys. J. A}
  {\bfseries 48} (2012) 98}, \href{http://arxiv.org/abs/1207.2361}{{\ttfamily
  arXiv:1207.2361 [nucl-ex]}}.

\bibitem{ref2207.14079}
{\bfseries A2} Collaboration, S.~Garni {\em et~al.}, ``{Target and beam-target
  asymmetries for the $\gamma p \to \pi^0 \pi^0 p$ reaction},''
  \href{http://arxiv.org/abs/2207.14079}{{\ttfamily arXiv:2207.14079
  [hep-ex]}}.

\bibitem{refprl872001}
W.~Langgartner {\em et~al.}, ``{Direct observation of a rho decay of the
  D(13)(1520) baryon resonance},''
  \href{http://dx.doi.org/10.1103/PhysRevLett.87.052001}{{\em Phys. Rev. Lett.}
  {\bfseries 87} (2001) 052001}.

\bibitem{refplb6592008}
A.~V. Sarantsev {\em et~al.}, ``{New results on the Roper resonance and the
  P(11) partial wave},''
  \href{http://dx.doi.org/10.1016/j.physletb.2007.11.055}{{\em Phys. Lett. B}
  {\bfseries 659} (2008) 94--100},
  \href{http://arxiv.org/abs/0707.3591}{{\ttfamily arXiv:0707.3591 [hep-ph]}}.

\bibitem{refnpa6952001}
J.~C. Nacher, E.~Oset, M.~J. Vicente, and L.~Roca, ``{The Role of Delta(1700)
  excitation and rho production in double pion photoproduction},''
  \href{http://dx.doi.org/10.1016/S0375-9474(01)01110-1}{{\em Nucl. Phys. A}
  {\bfseries 695} (2001) 295--327},
  \href{http://arxiv.org/abs/nucl-th/0012065}{{\ttfamily
  arXiv:nucl-th/0012065}}.

\bibitem{refepja252005}
A.~Fix and H.~Arenhoevel, ``{Double pion photoproduction on nucleon and
  deuteron},'' \href{http://dx.doi.org/10.1140/epja/i2005-10067-5}{{\em Eur.
  Phys. J. A} {\bfseries 25} (2005) 115--135},
  \href{http://arxiv.org/abs/nucl-th/0503042}{{\ttfamily
  arXiv:nucl-th/0503042}}.

\bibitem{refrpj602017}
R.~R. Dusaev and M.~V. Egorov, ``{Double Photoproduction of Neutral Pions on a
  Proton and a Deuteron},''
  \href{http://dx.doi.org/10.1007/s11182-017-1040-8}{{\em Russ. Phys. J.}
  {\bfseries 60} no.~1, (2017) 26--36}.

\bibitem{refprl161966}
S.~D. Drell and A.~C. Hearn, ``{Exact Sum Rule for Nucleon Magnetic Moments},''
  \href{http://dx.doi.org/10.1103/PhysRevLett.16.908}{{\em Phys. Rev. Lett.}
  {\bfseries 16} (1966) 908--911}.

\bibitem{refprl731994}
M.~Benmerrouche and E.~Tomusiak, ``{Low-energy expansions for double pion
  photoproduction},'' \href{http://dx.doi.org/10.1103/PhysRevLett.73.400}{{\em
  Phys. Rev. Lett.} {\bfseries 73} (1994) 400--403}.

\bibitem{refnpa5801994}
V.~Bernard, N.~Kaiser, U.~G. Meissner, and A.~Schmidt, ``{Threshold two pion
  photoproduction and electroproduction: More neutrals than expected},''
  \href{http://dx.doi.org/10.1016/0375-9474(94)90910-5}{{\em Nucl. Phys. A}
  {\bfseries 580} (1994) 475--499},
  \href{http://arxiv.org/abs/nucl-th/9403013}{{\ttfamily
  arXiv:nucl-th/9403013}}.

\bibitem{refweinberg}
S.~Weinberg, ``{Phenomenological Lagrangians},''
  \href{http://dx.doi.org/10.1016/0378-4371(79)90223-1}{{\em Physica A}
  {\bfseries 96} no.~1-2, (1979) 327--340}.

\bibitem{refgasserandleutwyler}
J.~Gasser and H.~Leutwyler, ``{Chiral Perturbation Theory to One Loop},''
  \href{http://dx.doi.org/10.1016/0003-4916(84)90242-2}{{\em Annals Phys.}
  {\bfseries 158} (1984) 142}.

\bibitem{refscherer}
S.~Scherer and M.~R. Schindler,
  \href{http://dx.doi.org/10.1007/978-3-642-19254-8}{{\em {A Primer for Chiral
  Perturbation Theory}}}, vol.~830.
\newblock 2012.

\bibitem{refnpb3631991}
S.~Weinberg, ``{Effective chiral Lagrangians for nucleon - pion interactions
  and nuclear forces},''
  \href{http://dx.doi.org/10.1016/0550-3213(91)90231-L}{{\em Nucl. Phys. B}
  {\bfseries 363} (1991) 3--18}.

\bibitem{refap1581984}
J.~Gasser and H.~Leutwyler, ``{Chiral Perturbation Theory to One Loop},''
  \href{http://dx.doi.org/10.1016/0003-4916(84)90242-2}{{\em Annals Phys.}
  {\bfseries 158} (1984) 142}.

\bibitem{refap2832000}
N.~Fettes, U.-G. Meissner, M.~Mojzis, and S.~Steininger, ``{The Chiral
  effective pion nucleon Lagrangian of order p**4},''
  \href{http://dx.doi.org/10.1006/aphy.2000.6059}{{\em Annals Phys.} {\bfseries
  283} (2000) 273--302}, \href{http://arxiv.org/abs/hep-ph/0001308}{{\ttfamily
  arXiv:hep-ph/0001308}}. [Erratum: Annals Phys. 288, 249--250 (2001)].

\bibitem{refap3362013}
J.~M. Alarcon, J.~Martin~Camalich, and J.~A. Oller, ``{Improved description of
  the $\pi N$-scattering phenomenology in covariant baryon chiral perturbation
  theory},'' \href{http://dx.doi.org/10.1016/j.aop.2013.06.001}{{\em Annals
  Phys.} {\bfseries 336} (2013) 413--461},
  \href{http://arxiv.org/abs/1210.4450}{{\ttfamily arXiv:1210.4450 [hep-ph]}}.

\bibitem{refprd1002019}
G.~H. Guerrero~Navarro, M.~J. Vicente~Vacas, A.~N.~H. Blin, and D.-L. Yao,
  ``{Pion photoproduction off nucleons in covariant chiral perturbation
  theory},'' \href{http://dx.doi.org/10.1103/PhysRevD.100.094021}{{\em Phys.
  Rev. D} {\bfseries 100} no.~9, (2019) 094021},
  \href{http://arxiv.org/abs/1908.00890}{{\ttfamily arXiv:1908.00890
  [hep-ph]}}.

\bibitem{refpr601941}
W.~Rarita and J.~Schwinger, ``{On a theory of particles with half integral
  spin},'' \href{http://dx.doi.org/10.1103/PhysRev.60.61}{{\em Phys. Rev.}
  {\bfseries 60} (1941) 61}.

\bibitem{refjpg241998}
T.~R. Hemmert, B.~R. Holstein, and J.~Kambor, ``{Chiral Lagrangians and
  delta(1232) interactions: Formalism},''
  \href{http://dx.doi.org/10.1088/0954-3899/24/10/003}{{\em J. Phys. G}
  {\bfseries 24} (1998) 1831--1859},
  \href{http://arxiv.org/abs/hep-ph/9712496}{{\ttfamily arXiv:hep-ph/9712496}}.

\bibitem{refplb6832010}
H.~Krebs, E.~Epelbaum, and U.~G. Meissner, ``{Redundancy of the off-shell
  parameters in chiral effective field theory with explicit spin-3/2 degrees of
  freedom},'' \href{http://dx.doi.org/10.1016/j.physletb.2009.12.023}{{\em
  Phys. Lett. B} {\bfseries 683} (2010) 222--228},
  \href{http://arxiv.org/abs/0905.2744}{{\ttfamily arXiv:0905.2744 [hep-th]}}.

\bibitem{refplb7602016}
J.~Gegelia, U.-G. Mei\ss{}ner, and D.-L. Yao, ``{The width of the Roper
  resonance in baryon chiral perturbation theory},''
  \href{http://dx.doi.org/10.1016/j.physletb.2016.07.068}{{\em Phys. Lett. B}
  {\bfseries 760} (2016) 736--741},
  \href{http://arxiv.org/abs/1606.04873}{{\ttfamily arXiv:1606.04873
  [hep-ph]}}.

\bibitem{refprc862012}
T.~Bauer, J.~C. Bernauer, and S.~Scherer, ``{Electromagnetic form factors of
  the nucleon in effective field theory},''
  \href{http://dx.doi.org/10.1103/PhysRevC.86.065206}{{\em Phys. Rev. C}
  {\bfseries 86} (2012) 065206},
  \href{http://arxiv.org/abs/1209.3872}{{\ttfamily arXiv:1209.3872 [nucl-th]}}.

\bibitem{refpr1611988}
U.~G. Meissner, ``{Low-Energy Hadron Physics from Effective Chiral Lagrangians
  with Vector Mesons},''
  \href{http://dx.doi.org/10.1016/0370-1573(88)90090-7}{{\em Phys. Rept.}
  {\bfseries 161} (1988) 213}.

\bibitem{refijmpa252010}
D.~Djukanovic, J.~Gegelia, and S.~Scherer, ``{Path integral quantization for
  massive vector bosons},''
  \href{http://dx.doi.org/10.1142/S0217751X10049736}{{\em Int. J. Mod. Phys. A}
  {\bfseries 25} (2010) 3603--3619},
  \href{http://arxiv.org/abs/1001.1077}{{\ttfamily arXiv:1001.1077 [hep-th]}}.

\bibitem{refprl932004}
D.~Djukanovic, M.~R. Schindler, J.~Gegelia, G.~Japaridze, and S.~Scherer,
  ``{Universality of the rho-meson coupling in effective field theory},''
  \href{http://dx.doi.org/10.1103/PhysRevLett.93.122002}{{\em Phys. Rev. Lett.}
  {\bfseries 93} (2004) 122002},
  \href{http://arxiv.org/abs/hep-ph/0407239}{{\ttfamily arXiv:hep-ph/0407239}}.

\bibitem{Kawarabayashi:1966kd}
K.~Kawarabayashi and M.~Suzuki, ``{Partially conserved axial vector current and
  the decays of vector mesons},''
  \href{http://dx.doi.org/10.1103/PhysRevLett.16.255}{{\em Phys. Rev. Lett.}
  {\bfseries 16} (1966) 255}.

\bibitem{Riazuddin:1966sw}
Riazuddin and Fayyazuddin, ``{Algebra of current components and decay widths of
  rho and K* mesons},'' \href{http://dx.doi.org/10.1103/PhysRev.147.1071}{{\em
  Phys. Rev.} {\bfseries 147} (1966) 1071--1073}.

\bibitem{Basdevant:1970hm}
J.~L. Basdevant and J.~Zinn-Justin, ``{Yang-Mills fields and the pi pi
  interaction},'' \href{http://dx.doi.org/10.1103/PhysRevD.3.1865}{{\em Phys.
  Rev. D} {\bfseries 3} (1971) 1865}.

\bibitem{Guo:2007ff}
Z.~H. Guo, J.~J. Sanz~Cillero, and H.~Q. Zheng, ``{Partial waves and large N(C)
  resonance sum rules},''
  \href{http://dx.doi.org/10.1088/1126-6708/2007/06/030}{{\em JHEP} {\bfseries
  06} (2007) 030}, \href{http://arxiv.org/abs/hep-ph/0701232}{{\ttfamily
  arXiv:hep-ph/0701232}}.

\bibitem{refjhep052016}
D.-L. Yao, D.~Siemens, V.~Bernard, E.~Epelbaum, A.~M. Gasparyan, J.~Gegelia,
  H.~Krebs, and U.-G. Mei\ss{}ner, ``{Pion-nucleon scattering in covariant
  baryon chiral perturbation theory with explicit Delta resonances},''
  \href{http://dx.doi.org/10.1007/JHEP05(2016)038}{{\em JHEP} {\bfseries 05}
  (2016) 038}, \href{http://arxiv.org/abs/1603.03638}{{\ttfamily
  arXiv:1603.03638 [hep-ph]}}.

\bibitem{refpdg}
{\bfseries Particle Data Group} Collaboration, R.~L. Workman {\em et~al.},
  ``{Review of Particle Physics},''
  \href{http://dx.doi.org/10.1093/ptep/ptac097}{{\em PTEP} {\bfseries 2022}
  (2022) 083C01}.

\bibitem{refplb7632016}
J.~Gegelia, U.-G. Mei\ss{}ner, D.~Siemens, and D.-L. Yao, ``{The width of the
  $\Delta$-resonance at two loop order in baryon chiral perturbation theory},''
  \href{http://dx.doi.org/10.1016/j.physletb.2016.10.017}{{\em Phys. Lett. B}
  {\bfseries 763} (2016) 1--8},
  \href{http://arxiv.org/abs/1608.00517}{{\ttfamily arXiv:1608.00517
  [hep-ph]}}.

\bibitem{refplb4121997}
F.~Guerrero and A.~Pich, ``{Effective field theory description of the pion
  form-factor},'' \href{http://dx.doi.org/10.1016/S0370-2693(97)01070-8}{{\em
  Phys. Lett. B} {\bfseries 412} (1997) 382--388},
  \href{http://arxiv.org/abs/hep-ph/9707347}{{\ttfamily arXiv:hep-ph/9707347}}.

\bibitem{Pascalutsa:2002pi}
V.~Pascalutsa and D.~R. Phillips, ``{Effective theory of the delta(1232) in
  Compton scattering off the nucleon},''
  \href{http://dx.doi.org/10.1103/PhysRevC.67.055202}{{\em Phys. Rev. C}
  {\bfseries 67} (2003) 055202},
  \href{http://arxiv.org/abs/nucl-th/0212024}{{\ttfamily
  arXiv:nucl-th/0212024}}.

\bibitem{refplb5752003}
T.~Fuchs, M.~R. Schindler, J.~Gegelia, and S.~Scherer, ``{Power counting in
  baryon chiral perturbation theory including vector mesons},''
  \href{http://dx.doi.org/10.1016/j.physletb.2003.09.060}{{\em Phys. Lett. B}
  {\bfseries 575} (2003) 11--17},
  \href{http://arxiv.org/abs/hep-ph/0308006}{{\ttfamily arXiv:hep-ph/0308006}}.

\bibitem{refprl611988}
T.~N. Truong, ``{Chiral Perturbation Theory and Final State Theorem},''
  \href{http://dx.doi.org/10.1103/PhysRevLett.61.2526}{{\em Phys. Rev. Lett.}
  {\bfseries 61} (1988) 2526}.

\bibitem{Watson:1952ji}
K.~M. Watson, ``{The Effect of final state interactions on reaction
  cross-sections},'' \href{http://dx.doi.org/10.1103/PhysRev.88.1163}{{\em
  Phys. Rev.} {\bfseries 88} (1952) 1163--1171}.

\bibitem{refprd351987}
K.~L. Au, D.~Morgan, and M.~R. Pennington, ``{Meson Dynamics Beyond the Quark
  Model: A Study of Final State Interactions},''
  \href{http://dx.doi.org/10.1103/PhysRevD.35.1633}{{\em Phys. Rev. D}
  {\bfseries 35} (1987) 1633}.

\bibitem{Yao:2020bxx}
D.-L. Yao, L.-Y. Dai, H.-Q. Zheng, and Z.-Y. Zhou, ``{A review on partial-wave
  dynamics with chiral effective field theory and dispersion relation},''
  \href{http://dx.doi.org/10.1088/1361-6633/abfa6f}{{\em Rept. Prog. Phys.}
  {\bfseries 84} no.~7, (2021) 076201},
  \href{http://arxiv.org/abs/2009.13495}{{\ttfamily arXiv:2009.13495
  [hep-ph]}}.

\bibitem{Omnes:1958hv}
R.~Omnes, ``{On the Solution of certain singular integral equations of quantum
  field theory},'' \href{http://dx.doi.org/10.1007/BF02747746}{{\em Nuovo Cim.}
  {\bfseries 8} (1958) 316--326}.

\bibitem{rfprd832011}
R.~Garcia-Martin, R.~Kaminski, J.~R. Pelaez, J.~Ruiz~de Elvira, and F.~J.
  Yndurain, ``{The Pion-pion scattering amplitude. IV: Improved analysis with
  once subtracted Roy-like equations up to 1100 MeV},''
  \href{http://dx.doi.org/10.1103/PhysRevD.83.074004}{{\em Phys. Rev. D}
  {\bfseries 83} (2011) 074004},
  \href{http://arxiv.org/abs/1102.2183}{{\ttfamily arXiv:1102.2183 [hep-ph]}}.

\end{thebibliography}\endgroup

\end{document}